\RequirePackage{fix-cm}
\documentclass[12pt,a4paper,oneside]{book}

\usepackage{graphicx,xcolor,textpos}
\usepackage{helvet}
\usepackage{hyperref}
\hypersetup{
	colorlinks,
	linkcolor=blue
}
\graphicspath{ {figures/} }
\usepackage{array, float}
\usepackage{tcolorbox}
\usepackage{physics}
\usepackage{apacite}
\usepackage{amsmath}
\usepackage{amssymb}
\usepackage{mathtools}
\usepackage{siunitx}
\usepackage{natbib}
\usepackage[brazilian,english]{babel}	
\usepackage[nottoc]{tocbibind}
\usepackage{indentfirst}
\usepackage{enumitem}
\newlist{abbrv}{itemize}{1}
\setlist[abbrv,1]{label=,labelwidth=1in,align=parleft,itemsep=0.05\baselineskip,leftmargin=!}

\DeclareUnicodeCharacter{2212}{-}

\newenvironment{greybox}[1][]{\begin{tcolorbox}\noindent \rmfamily}{\medskip\end{tcolorbox}}
\newcommand{\id}{\textrm{d}}

\renewcommand{\cite}{\citep}

\definecolor{green}{RGB}{172,196,0}
\definecolor{bluetitle}{RGB}{29,141,176}
\definecolor{blueaff}{RGB}{0,0,128}
\definecolor{blueline}{RGB}{82,189,236}
\begin{document}
\thispagestyle{empty}
\newcommand{\form}[1]{\scalebox{1.087}{\boldmath{#1}}}
\sffamily
\begin{textblock}{191}(-24,-11)
\colorbox{green}{\hspace{139mm}\ \parbox[c][18truemm]{52mm}{\textcolor{white}{FACULTY OF SCIENCE}}}
\end{textblock}
\begin{textblock}{70}(-18,-19)
\textblockcolour{}
\includegraphics*[height=19.8truemm]{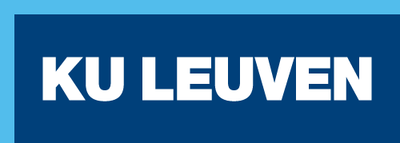}
\end{textblock}
\begin{textblock}{160}(-6,63)
\textblockcolour{}
\vspace{-\parskip}
\flushleft
\fontsize{40}{42}\selectfont \textcolor{bluetitle}{Statistical Mechanics of the Kompaneets Equation}\\[1.5mm]
\end{textblock}
\begin{textblock}{160}(8,153)
\textblockcolour{}
\vspace{-\parskip}
\flushright
\fontsize{14}{16}\selectfont \textbf{Guilherme Eduardo FREIRE OLIVEIRA}
\end{textblock}
\begin{textblock}{70}(-6,191)
\textblockcolour{}
\vspace{-\parskip}
\flushleft
Supervisor: Prof. Dr. Christian Maes\\[-2pt]
\textcolor{blueaff}{Instituut voor Theoretische Fysica, KU Leuven}\\[5pt]
\end{textblock}
\begin{textblock}{160}(8,191)
\textblockcolour{}
\vspace{-\parskip}
\flushright
Thesis presented in\\[4.5pt]
fulfillment of the requirements\\[4.5pt]
for the degree of Master of Science\\[4.5pt]
in Physics\\
\end{textblock}
\begin{textblock}{160}(8,232)
\textblockcolour{}
\vspace{-\parskip}
\flushright
Academic year 2020-2021
\end{textblock}
\begin{textblock}{191}(-24,248)
{\color{blueline}\rule{550pt}{5.5pt}}
\end{textblock}
\vfill
\newpage
\thispagestyle{empty}
\vspace*{\fill}
© Copyright by KU Leuven

Without written permission of the promoters and the authors it is forbidden to reproduce or adapt in any form or by any means any part of this publication. Requests for obtaining the right to reproduce or utilize parts of this publication should be addressed to KU Leuven, Faculteit Wetenschappen, Geel Huis, Kasteelpark Arenberg 11 bus 2100, 3001 Leuven (Heverlee), Telephone +32 16 32 14 01.

A written permission of the promoter is also required to use the methods, products, schematics and programs described in this work for industrial or commercial use, and for submitting this publication in scientific contests.

\newpage
\rmfamily
\setcounter{page}{0}
\pagenumbering{roman}

\chapter*{Acknowledgments}
\addcontentsline{toc}{chapter}{Acknowledgments}

Six years ago, I started my journey in Physics and I remember, as if it were yesterday, the fascination and enthusiasm with which I crossed the doors of the Institute for Exact Sciences of my beloved Federal University of Minas Gerais. Since then, so much has happened and I have had the tremendous opportunity to not only meet, but also work with incredible scientists, with whom I have made works which I am deeply proud of.

I believe that learning, evolving and improving is an essential part of life and that great moments should be marked with great joy and celebration. So, I would like to toast this work, which was so kindly elaborated and which marks such an important professional step for me, especially by thanking the people who were crucial in achieving this landmark.

Therefore, I would like to thank my dear mother, Cleide, whose tenderness was always present, giving me encouragement in moments of restlessness; to my dear father, Rogério, whose wisdom in our lasting conversations brought me light when lost; to my dear sister, Gabriela, whose companionship has always given me the certainty of never being alone; to my little goddaughter Beatriz, whose simplicity that only a baby can have filled my heart with joy and made simple what was once complicated; to my dear Julia, whose sweetness, support and affection lifted my heart, always making me to keep moving forward.

Finally, I could never let to thank my advisor, professor Christian Maes, whose brilliant and impeccable guidance was indispensable for the development of this work; to my colleague Kasper Meerts for many fruitful discussions; to my dear KU Leuven, a name I will carry forever and, on behalf of all my friends in Belgium, to Pavly and Lukas, who made me feel at home here.

The truth is that, if I were to thank every single one, maybe even in this paper that would not fit, so I apologize for all those who were important in this special period in my life, but which are not here.

To all of you, my most sincere thanks!

\chapter*{Agradecimentos}
\addcontentsline{toc}{chapter}{Agradecimentos}

Há seis anos atrás, comecei minha jornada na física e lembro-me, como se fosse ontem, do fascínio e entusiasmo com o qual atravessei as portas do Instituto de Ciências Exatas da minha querida Universidade Federal de Minas Gerais. De lá para cá muitas coisas aconteceram, e tive a tremenda oportunidade de, não só conhecer, mas também trabalhar com cientistas incríveis, com os quais desenvolvi trabalhos que me orgulho profundamente.

Acredito que, aprender, evoluir e aperfeiçoar é parte essencial da vida e que os momentos grandiosos devem ser marcados com muita alegria e festejo. Pois então, gostaria de brindar a este trabalho, que foi desenvolvido com tanto carinho e que marca uma etapa profissional tão importante para mim, sobretudo, agradecendo às pessoas que foram primordiais à conquista desse marco.

Gostaria, portanto, de agradecer à minha querida mãe, Cleide, cujo carinho sempre se fez presente, me dando alento em momentos de inquietação; ao meu querido pai, Rogério, cuja sabedoria em nossas conversas duradouras me trazia luz quando perdido; à minha querida irmã, Gabriela, cujo companheirismo sempre me deu a certeza de nunca estar sozinho; à minha afilhadinha Beatriz, cuja singeleza que só um bebê pode ter enchia meu coração de alegria e fazia tornar-se simples o que antes era complicado; à minha querida Júlia, cuja doçura, apoio e afeto elevava meu coração, me fazendo sempre continuar seguindo em frente.

Finalmente, eu não poderia nunca deixar de agradecer ao meu orientador, professor Christian Maes, cuja orientação brilhante e exímia fez-se essencial para o desenvolvimento deste trabalho; ao meu colega Kasper Meerts, pelas discussões frutíferas; à minha querida universidade KU Leuven, nome que carregarei para sempre e, em nome de todos meus amigos da Bélgica, ao Pavly e Lukas, que me fizeram sentir em casa aqui.

A verdade é que se fosse agradecer a todos, talvez nem neste papel caberia e, por isso, peço desculpas por todos aqueles que foram importantes neste período especial da minha vida e que não estão aqui.

A todos vocês, os meus mais sinceros agradecimentos!

\chapter*{Summary}
\addcontentsline{toc}{chapter}{Summary}

As an important subject in non-equilibrium Statistical Mechanics, we study in this thesis the relaxation to equilibrium of a photon gas in contact with an non-relativistic and non-degenerate electron bath. Photons and electrons interact via the Compton effect, establishing thermal equilibrium of radiation with matter as pointed out by A.S.~Kompaneets in \cite{kompa}. The evolution of the photon distribution function is then described by the eponymous partial differential equation, here viewed as the diffusion approximation to the relativistic Boltzmann equation that describes the system.

Being one of the few examples where this diffusion approximation can be performed in great detail, yielding the Bose-Einstein distribution as stationary solution, the Kompaneets equation also provides the description of the so-called Sunyaev-Zeldovich effect, which is the change of apparent brightness of the cosmic microwave background (CMB) radiation. 

There are many ways of deriving this equation, but one of them, which was proposed by Kompaneets in 1957 stands out for its directness and simplicity, explaining the reason why it is preferred by many references and included even in astrophysics textbooks. However, we point out in this work that there are some inconsistencies regarding this traditional derivation of the Kompaneets equation that were repeated by all the references we could find that follow the original framework of 1957, in such way that performing all the required calculations will lead you to the wrong equation.

These inconsistencies effectively break the conservation of photon-number that should happen at the level of the Boltzmann equation and we could not find any work which explicitly mentions or solves this problem. Remarkably enough, in his original work, Kompaneets does not mention the problem and manages to avoid it by invoking the strong and indirect argument that the equation should have the form of a continuity equation, with current vanishing for the Bose-Einstein distribution. References tend to repeat his argument, but here we show that there is no reason why this should work, \textit{i.e.}, we believe that the success of such procedure lies in a mathematical coincidence.

Therefore, this thesis will be divided in two parts: in the first we will be interested in how to deal with these inconsistencies, building the necessary basis in which the diffusion approximation to the Boltzmann equation is consistently performed, while also conversing with some history. In the second part, we will be interested in possible extensions to the famous equation and beyond reviewing some existing extensions, we will also show that a new setup involving a master equation of a random walk with suitable chosen transition rates in the photon reciprocal space furnishes not only Kompaneets equation but also a first generalization to a system of bosons under a possible driving. We believe that our framework may serve as an interesting point of departure to further extensions involving non-equilibrium conditions never, or mildly, considered in literature.

\chapter*{Resumo}
\addcontentsline{toc}{chapter}{Resumo}

Como um importante assunto em Mecânica Estatística de não-equilíbrio, nós estudamos nesta tese a relaxação ao equilíbrio de um gás de fótons em contato com um banho térmico de elétrons não-relativísticos e não-degenerados. Fótons e elétrons interagem através do efeito Compton, estabelecendo equilíbrio térmico da radiação com a matéria, como indicado por A.S.~Kompaneets em \cite{kompa}. A evolução temporal da função de distribuição dos fótons é, então, descrita pela epônima equação diferencial, aqui vista como a aproximação difusiva da equação de Boltzmann relativística do sistema.

Sendo um dos poucos exemplos no qual essa aproximação pode ser feita em grande detalhe, fornecendo a distribuição de Bose-Einstein como solução estacionária, a equação de Kompaneets também descreve o efeito Sunyaev-Zeldovich, responsável pela mudança do brilho aparente da radiação cósmica de fundo.

Existem muitas maneiras de encontrar essa equação, mas uma delas, a proposta por Kompaneets em 1957, destaca-se por ser direta e simples, explicando a razão pela qual é preferida por tantas referências e até mesmo incluída em livros-texto de astrofísica. Entretanto, apontamos nesse trabalho que existem algumas inconsistências acerca dessa tradicional derivação da equação, que são repetidas por todas as referências que seguem o trabalho original de 1957 que pudemos encontrar, de tal maneira que, ao desenvolver todos os cálculos, a equação encontrada está errada.

Essas inconsistências de fato quebram a conservação do número de fótons, que deve ocorrer no nível da equação de Boltzmann, e não conseguimos encontrar nenhum trabalho que explicitamente mencione ou resolva isso. Notavelmente, Kompaneets, em seu trabalho original, não apenas deixa de mencionar, como também consegue desviar do problema ao invocar o argumento forte e indireto de que a equação deve ter a forma de uma equação de continuidade, cuja corrente se anula para a distribuição de Bose-Einstein, e as referências tendem a repetir o seu argumento. No entanto, aqui mostramos que não existe nenhuma razão para que isso funcione, isto é, acreditamos que o sucesso desse procedimento reside em uma coincidência matemática.

Essa tese será dividida em duas partes: na primeira, lidaremos com essas inconsistências, construindo a base necessária para que a aproximação de difusão da equação de Boltzmann seja desenvolvida de forma consistente, ao mesmo tempo em que conversaremos com um pouco de história. Na segunda, estaremos interessados em possíveis extensões da famosa equação e, além de revisar algumas já existentes, mostraremos que considerando a equação mestra de uma caminhada aleatória com taxas de transição apropriadas no espaço recíproco dos fótons fornece, não apenas a equação de Kompaneets, mas também uma primeira generalização para bósons sob o efeito de um possível campo. Acreditamos que a nossa descrição pode servir como um interessante ponto de partida para outras extensões, envolvendo condições de não-equilíbrio nunca, ou pouco, consideradas. 
   
\chapter*{Vulgarising Summary}
\addcontentsline{toc}{chapter}{Vulgarising Summary}

The study of nature involves, in several aspects, the understanding of processes which are out of thermal equilibrium. As a matter of fact, this absence of equilibrium can come through many different ways, and, in order to better understand some ideas, let us consider a container with a partition in the middle and a gas that occupies one half of it while the other half is empty.
\begin{figure}[H]
	\centering
	\includegraphics[width=.6\linewidth]{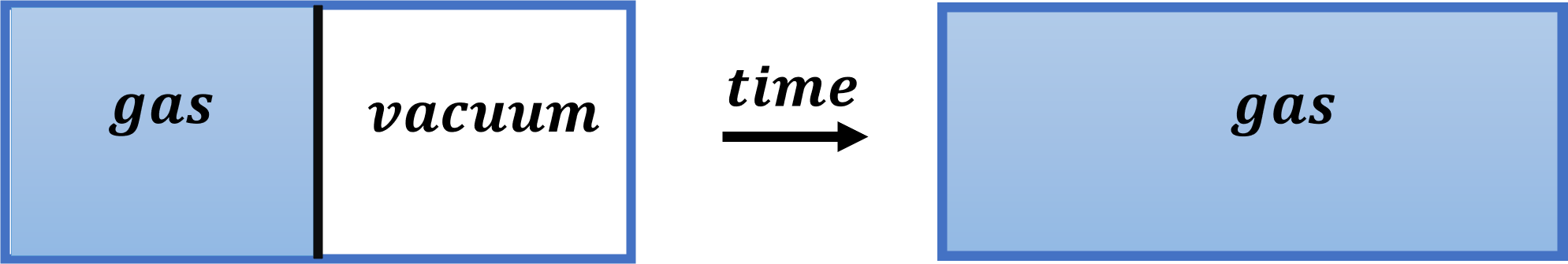}
\end{figure}
When we remove the partition, the gas will expand until it occupies the entire container, moment in which its state no longer is dependent on time. This is an instance of non-equilibrium, the so-called \textit{relaxation to equilibrium}: initially, at the moment we remove the partition, the gas is no longer in equilibrium (in fact, if it were, its state would not evolve in first place), but it expands to occupy the other half of the container, when equilibrium is established again.

In this thesis, we will exactly study this problem of relaxation to equilibrium and, unlike the previous example, we will investigate a gas mixture of photons and electrons. Here, the electrons play the same role as the container, which is to provide conditions for reaching equilibrium. On the other hand, we will be interested in the photons (which will be seen here, strangely as it may sound, as particles), studying how the time evolution of a central object, called the \textit{distribution function}, looks like.

The equation that describes this precise time evolution of the photon distribution function was proposed long time ago by Aleksandr Kompaneets \cite{kompa}, who was the first to point out that radiation (photons) requires contact with matter (here, the electrons) for the establishment of equilibrium. It is useful, then, to think of electrons as a reservoir, a sea, which is in equilibrium with a certain temperature. The photons will then interact (collide) with these electrons through an interaction we call the Compton effect, establishing after some time thermal equilibrium for radiation.

We are particularly interested in possible extensions of this equation and, therefore, this thesis is divided in two parts: in the first part, we propose how to solve a series of systematic inconsistencies on the derivation of the equation which appear in several references, including the original one. Traditionally, the problem is solved using a strong, indirect argument about the form of the equation, and here we show that this is not necessary, as long as we deal with these inconsistencies. In the second part, we show that, starting from an entirely different framework, it is possible to recover and extend the equation for some systems, and we believe this description to be useful for future generalizations.

\chapter*{Resumo de Divulgação}
\addcontentsline{toc}{chapter}{Resumo de Divulgação}

O estudo da natureza envolve, em vários aspectos, o entendimento sobre processos que estão fora de equilíbrio térmico. De fato, essa falta de equilíbrio pode vir através de diversas maneiras diferentes e, para entendermos a ideia, vamos supor um recipiente com uma divisória no meio e um gás que ocupa uma metade, enquanto a outra está vazia.
\begin{figure}[H]
	\centering
	\includegraphics[width=.6\linewidth]{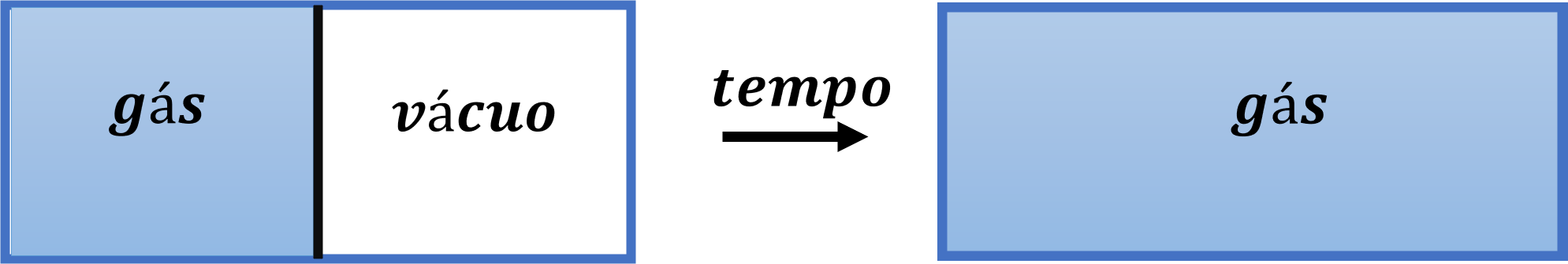}
\end{figure}
Quando abrirmos a divisória do meio, o gás irá expandir-se até ocupar todo o recipiente, momento no qual seu estado não irá mais depender do tempo. Este é um exemplo de um tipo de não-equilíbrio, a \textit{relaxação ao equilíbrio}: inicialmente, a partir do momento que retirarmos a divisória, o gás não estará mais em equilíbrio (pois se estivesse, seu estado não evoluiria) e evolui para ocupar a outra metade, estabelecendo equilíbrio novamente.

Nesta tese de mestrado, estudaremos exatamente este problema de relaxação ao equilí-\newline brio e, diferentemente do exemplo anterior, estudaremos uma mistura gasosa de fótons e elétrons. Aqui, os elétrons fazem o mesmo papel do recipiente, que é o de fornecer condições para equilíbrio ser atingido. Por outro lado, estaremos interessados nos fótons (que serão vistos, por mais estranho que pareça, como partículas) e queremos estudar como se dá a evolução temporal de um objeto central, chamado de \textit{função de distribuição}.

A equação que fornece essa precisa evolução temporal da distribuição do gás de fótons foi proposta, há muito tempo atrás, por Aleksandr Kompaneets \cite{kompa}, que foi o primeiro a apontar que radiação (os fótons) precisa do contato com a matéria (aqui, os elétrons) para que equilíbrio seja estabelecido. É útil, portanto, pensar nos elétrons como um reservatório, como um mar, que está em equilíbrio e que possui uma certa temperatura. Os fótons, então, interagem (colidem) com esses elétrons através de uma interação que chamamos de efeito Compton, estabelecendo após um certo tempo o equilíbrio térmico da radiação.

Estamos particularmente interessados em possíveis extensões dessa equação e, portanto, esta tese é dividida em duas partes: na primeira parte, propomos como resolver uma série de inconsistências sistemáticas, acerca da derivação da equação, que aparecem em diversas referências, incluindo a original. Tradicionalmente, o problema é resolvido empregando um argumento forte e indireto sobre a forma da equação e aqui mostramos que isso não é necessário, desde que lidemos com essas inconsistências. Na segunda parte, mostramos que é possível reencontrar e estender a equação para alguns sistemas partindo de uma descrição totalmente diferente, acreditamos que essa descrição é útil para futuras generalizações.

\listoffigures

\chapter*{List of Symbols and Abbreviations}
\addcontentsline{toc}{chapter}{List of Symbols and Abbreviations}
\begin{abbrv}
	\item[$m$]                     Particle mass
	\item[$m_e$]                   Electron mass
	\item[$n_e$]				   Electron density
	\item[$N$]                     Photon number
	\item[$c$]					   Speed of light in vacuum
	\item[$\omega$]                Photon frequency
	\item[$E$]                     Electron energy
	\item[$T$]					   Temperature or transition matrix (Ch.\ref{3}) 
	\item[$k_B$]                   Boltzmann's constant
	\item[$\beta$]				   Inverse temperature multiplied by Boltzmann's constant: $(k_BT)^{-1}$
	\item[$h$]					   Planck's constant
	\item[$\hbar$]                 Reduced Planck's constant: $\frac{h}{2\pi}$
	\item[$\gamma_v$]              Lorentz factor related to velocity $\mathbf{v}$
	\item[$\sigma$]                Total cross section
	\item[$\id\sigma,\frac{\id \sigma}{\id \Omega}$]             Differential cross section
	\item[$\Omega$]                Scattering solid angle
	\item[$\theta,\phi$]           Scattering angles
	\item[$\alpha$]                Photon-electron angle (Ch.\ref{3}, App.\ref{c}) or fine-structure constant (Ch.\ref{6})
	\item[$g$]                     Minus the determinant of the metric tensor
	\item[$\eta$]                  Minkowski metric tensor: $\mathrm{diag}(+ - - -)$
	\item[$\eta_{\mu\nu}$]         Minkowski metric tensor components
	\item[$\Lambda$]               Lorentz transformation matrix
    \item[$R$]					   Scale factor of the Universe
	\item[$\frac{\dot{R}}{R}$]     Hubble parameter
	\item[$\sigma_T$]              Total Thomson cross section
	\item[$D$]                     Diffusivity
	\item[$Z$]                     Atomic number
	\item[$\mathcal{F}$]           Flux (incident)
	\item[$\Delta$]                Dimensionless energy shift in the Compton effect
	\item[$x$]                 Dimensionless photon energy: $\frac{\hbar\omega}{k_B T}$	(Ch.\ref{4}), phase-space position four-vector (Ch.\ref{2}) or rescaled photon wave-vector (Ch.\ref{5})
	\item[$A$]                     Four-vectors
	\item[$\mathbf{A}$]			   Three-vectors
	\item[$A^{\mu}$]               Four-vector coordinates: $\mu=0,1,2,3$
	\item[$A^i$]                   Four-vector spatial coordinates: $i=1,2,3$
	\item[$\mu$]                   Phase-space point: $(x,p)$
	\item[$\Gamma$]                One-particle phase-space (Ch.\ref{2}) or Christoffel symbols of the Levi-Civita connection (Ch.\ref{6})
	\item[$\mathbf{v}$]             Electron (or particle) velocity vector
	\item[$\mathbf{p},p$]           Electron (or particle) three-momentum, four-momentum
	\item[$\mathbf{k},k$]           Photon three-momentum (or wave-vector), four-momentum
	\item[$\mathbf{\hat{n}}$]       Photon three-momentum unit vector
	\item[$\delta$]                Dirac's delta-function
	\item[$\delta^{(4)}$]            Four dimensional Dirac's delta-function
	\item[$f(t,\mathbf{x},\mathbf{p})$] Distribution function
	\item[$n(t,\mathbf{x},\mathbf{p})$] Occupation number distribution function
	\item[$j$]                          Current
    \item[$s,t,u$]                  Mandelstam variables
    \item[$I_l(x)$]                 Kompaneets' $l$-th integral
    \item[$f$]                      Driving (in the master equation)
    \item[$U$]                      Boson energy (in the master equation)
    \item[$W$]                      Transition rates (in the Boltzmann equation)
    \item[$w$]                      Transition rates (in the master equation)
    \item[$M$]                      Transition amplitude
    \item[$S$]                      Scattering matrix or S-matrix
    \item[CMB]                      Cosmic Microwave Background (radiation)
    \item[STF]                      Symmetric and Trace-Free (tensors) 
    \item[QFT]                      Quantum Field Theory
    \item[KN]                       Klein-Nishina
    \item[Th]                       Thomson
\end{abbrv}
\tableofcontents
\newpage


\setcounter{page}{0}
\pagenumbering{arabic}
\chapter{Introduction} \label{1}
Over the past decades, the interest in non-equilibrium phenomena has increased enormously. From bacteria motion to supernova explosions, it has become clear that non-equilibrium is an inherent aspect of nature and, as such, must not be neglected in the ultimate description of reality. The analysis of macroscopic equilibrium systems is well known in modern Statistical Mechanics, J.W.~Gibbs \cite{gibbs}, one of the pioneers, fundamentally classified these systems in his Theory of Ensembles, which quickly became the modern tool in a wide range of applications. However, no such generalized description exists for non-equilibrium systems, simply because their behavior is much richer, and one often has to proceed analyzing case by case. In this thesis, we are interested in one instance of non-equilibrium behavior, usually appearing in systems initially out of equilibrium, but that given enough time, \textit{relax} to it.

\textit{Relaxation to equilibrium} is a vast topic in Statistical Mechanics, but one recurring protagonist in its treatment is the so-called \textit{Boltzmann equation}, an integro-differential equation that gives the spatio-temporal evolution of a central object called the \textit{distribution function}. Although this equation is applicable to many systems, we are particularly interested here in the relaxation of a photon gas to equilibrium, forming the well-known Bose-Einstein distribution. 

As part of his research to build the soviet hydrogen bomb\footnote{As we mention in \cite{paper}, the famous \textit{Kompaneets equation} arose from Kompaneets' research for the nuclear program of URSS in 1949. After the equation turned out to be useless for their purpose, the results were declassified and published in 1957 \cite{peebles,longair}.}, the soviet physicist A.S.~Kompaneets was interested in the equilibrium properties of radiation, being one of the first to point out \cite{kompa} that, since Maxwell's equations are linear, radiation cannot reach thermal equilibrium alone, thus, needing to exchange energy with matter particles. By considering photons initially out of equilibrium but in contact with an electron bath in thermal equilibrium, Kompaneets writes down the Boltzmann equation of a gas mixture of photons and electrons interacting via Compton effect \cite{compton}, and proposes to carry an expansion up to second order in the photon energy shift. Integrating out the electron bath, Kompaneets finds an expression for the time evolution of the photons
\begin{equation}\label{ke}
	\omega^2\frac{\partial n}{\partial t}(t,\omega)= \frac{n_e\sigma_T 
		c}{m_e c^2}\frac{\partial }{\partial \omega}\omega^4\left\{k_B T 
	\frac{\partial n}{\partial \omega}(t,\omega) + 
	\hbar\left[1+n(t,\omega)\right]n(t,\omega)\right\}
\end{equation}
where $T$ is the temperature of the electron bath, $\sigma_T \approx \SI{0.66}{\barn}$ is the total Thomson cross section and $n_e, m_e$ are the electron density and mass, respectively. 

The above equation, which was named after him, expresses the time evolution of the (dimensionless) photon occupation number distribution function $n(t,\omega)$. Formally, Kompaneets procedure is what is called \textit{diffusion approximation} or \textit{Kramers-Moyal expansion} \cite{kramers} to the Boltzmann equation, yielding a Fokker-Planck version of this integro-differential equation.

The Kompaneets equation \eqref{ke} has the structure of a continuity equation in the photon number $$N\propto \int_0^\infty\id \omega\omega^2 n(t,\omega)$$ \textit{i.e.}, it is photon-number conserving\footnote{In fact, we should not expect differently, since Compton interaction is an elastic photon-electron scattering which preserves photon number.} with current
\begin{eqnarray}\label{cur}
	\frac{\partial n}{\partial t}(t,\omega)&=& \frac 1{\omega^2}\frac{\partial }{\partial \omega}\big(\omega^2\,j_t(\omega)\big)\nonumber\\
	j_t(\omega)&=& \frac{n_e \sigma_T c}{m_e c^2} \omega^2\left\{k_B T 
	\frac{\partial n}{\partial \omega}(t,\omega) + 
	\hbar\left[1+n(t,\omega)\right]n(t,\omega)\right\}
\end{eqnarray}
vanishing when $n(t,\omega)$ is given by the Bose-Einstein distribution
\[ n(t,\omega) = n_\text{eq}(\omega) = \frac{1}{\exp(\beta \hbar 
	\omega) - 1}.
\]

As we will see, for the derivation of \eqref{ke} it is assumed that electrons are non-relativistic ($k_B T \ll m_ec^2$) and that photons are soft, meaning that their energy is very small compared to the rest energy of the electron, but of the same order as the bath energy ($\hbar \omega \sim k_B T \ll m_e c^2$). Therefore, in that sense, the Kompaneets equation can also be regarded as the non-relativistic limit of the Boltzmann equation. 

Apart from being one of the few examples where the Kramers-Moyal expansion to the Boltzmann equation can be performed in great detail, \eqref{ke} not only provides a concrete example of relaxation to the Bose-Einstein distribution, but is also actively used in the study of astrophysical plasma, in the analysis of the \textit{Sunyaev-Zeldovich effect}, which is a distortion 
of the cosmic microwave background (CMB) radiation by Compton scattering of hot 
electrons during its passage through clusters of galaxies \citep{sunyaeveffect, sunyaev, burigana3}.

As we observed in \cite{paper}, the reader can easily realize that there exist many derivations of the Kompaneets equation in literature and that many authors have repeated or presented their best approach to this equation. However, we are motivated here by the appearance of systematic inconsistencies in many references, including Kompaneets' original paper \cite{kompa}. These inconsistencies mainly come from subtleties in performing the diffusion approximation to the Boltzmann equation correctly, also starting from an \textit{ab initio} consistent description. The derivation of \eqref{ke} as Kompaneets originally proposed is very didactically appealing and we believe that it is worth revisiting this traditional approach, so that part of this thesis will be devoted in clarifying these subtleties and inconsistencies that are rarely mentioned (and that a careful reader would stumble upon) even in textbook references.

Apart from talking to history we are also interested in possible extensions of \eqref{ke}, motivated by the interesting hypothesis \cite{arca} that the primordial plasma is away from equilibrium, yielding corrections to the Kompaneets equation due to the non-equilibrium nature of the electron bath. In that case, it is no longer true that the Bose-Einstein distribution is the solution of (now extended) \eqref{ke} and we should expect departures from it. In fact, the hypothesis in \cite{arca} is based on recent observations of deviations to the CMB spectrum in the low frequency regime that are yet not well understood \cite{arcade1, arcade2, edges}. As $n_{\text{eq}}(\omega)$ is used to obtain Planck spectrum (see for example Chapter \ref{4}), we should expect that corrections in \eqref{ke} yield also corrections to the stationary radiation spectrum. The other part of this thesis will then be devoted to the review of existing extensions, while proposing a framework for new possible extensions of the Kompaneets equation.

\section{A brief historical review}\label{review}

We reproduce here, in a slightly modified fashion, the historical review we have made in \cite{paper}.

In 1923, the famous physicist Wolfgang Pauli published a paper \cite{pauli} analyzing the conditions for thermal equilibrium of photons in an electron bath interacting via Compton scattering. Identifying what is called today \textit{detailed balance}, Pauli could retrieve the equilibrium distribution of radiation, known as the Planck spectrum, and was probably one of the first to lay the grounds for a description involving a master equation. However, Pauli did not manage to write the equation for the evolution of the photon distribution function, a task only performed a couple of years later by Kompaneets \cite{kompa}. In 1964, Dreicer more carefully elaborates the Fokker-Planck approximation to the Boltzmann equation of a photon-electron system, also not displaying the time evolution of the photon distribution \cite{dreicer}. Exactly one year later, Weymann writes a partial differential equation for the photon distribution function using Dreicer's formalism, but not showing any details of the calculation \cite{weymann}. Somewhat remarkable to note is that, although Weymann's and Dreicer's papers come after Kompaneets', they do not cite the latter work.

Not much later, in 1969, Sunyaev and Zeldovich concretely applies the equation found by Kompaneets to treat distortions of the CMB spectrum due to hot electrons, an effect named after them \cite{sunyaeveffect,sunyaev}. It did not take much time for physicists start looking into the first relativistic corrections to the Kompaneets equation, as the Sunyaev-Zeldovich effect would require the description of higher energy ranges. A first extension was done by Copper \cite{cooper} and later treated also by \cite{barbosa, itoh, itoh2, kohyama1, kohyama2, brown, kohyama3}. Normally, relativistic corrections are performed starting from the so-called \textit{manifestly covariant Boltzmann equation}. Although a more careful distinction and description of two equivalent versions of the kinetic equation will be given in due time, it is worth noting that the problems we will mention here do not happen with this manifestly covariant approach, making it simpler in some sense. On the other hand, it is also true that textbook references such as \cite{katz, rybicki} tend to avoid this description as it usually requires more background from the reader, such that revisiting problems with the traditional set up is worthwhile.

To derive \eqref{ke}, it is traditionally assumed that the electron bath is distributed according to the (equilibrium) Maxwell-Boltzmann statistics. However, \cite{barbosa} pointed out that one could obtain Kompaneets equation by using any isotropic distribution and a suitable definition of temperature, being one of the first to relax the condition of equilibrium to the electron bath. This was later also mentioned in \cite{ brown2, brown}. As far as we know, this was one of the first extensions of Kompaneets equation to the non-equilibrium case. In Chapter \ref{6} we will turn to less standard extensions, also making the connection of our framework to the already mentioned results of literature. There, it will become clear that, starting from Kompaneets' traditional approach to conclude the same as in the mentioned references, one extra constraint must be required on the electron distribution.

The condition that the photons are soft ($\hbar\omega\sim k_B T \ll m_ec^2$) may also be relaxed provided that we treat ($\hbar\omega \gg k_BT$), which is called \textit{down-Comptonization} regime, specifically. In fact, this regime should not be confused with the relativistic one because it is usually assumed that $k_BT\ll m_ec^2$ for the electron bath. Down-Comptonization first appeared in \cite{ross} in the so-called \textit{Ross-McCray equation}, where radiative transfer of X-ray photons is treated. The equation derived in \cite{ross}, however, does not yield the Bose-Einstein distribution as stationary solution and should be regarded only as an asymptotic limit of such extended Kompaneets equation. More recent and careful treatments can be found in \cite{liu, zhang}, where an extra term is found in \eqref{ke}.

\section{Problems with some derivations}

In order to understand the main problems with the traditional setup mentioned before, we must first understand the derivation of this equation as proposed originally by Kompaneets (since the details of the derivation itself will be made in Chapter \ref{4}, this section will be devoted to a more conceptual approach, such that details will be omitted sometimes for a better qualitative understanding). In his original paper, Kompaneets proposes to start from the Boltzmann kinetic equation for an electron-photon gas
\begin{equation}
	\label{komp-boltz}
	\frac{\partial n}{\partial t}(\omega) =  \int \id^3\mathbf{p} \,\id w\, 
	\left[n(\omega') f(\mathbf{p'}) (1+n(\omega)) - n(\omega) f(\mathbf{p}) 
	(1+n(\omega')) \right] 
\end{equation}
while performing an expansion up to second order in  $\Delta$, the photon energy shift $$ \Delta \coloneqq \frac{\hbar(\omega - \omega')}{k_B T} .$$ 

However, upon writing \eqref{komp-boltz} Kompaneets is vague about the format of the kinetic equation, specially because an expression of the rate $\id w$ is not explicitly given. The rate appearing in \eqref{komp-boltz} must be related to the cross section of the specific interaction, in such way that a careful examination of his paper suggests that Kompaneets is using the following expression for rate
\begin{equation}
	\label{ratewro}
	\id w = c\frac{\id \sigma^\text{Th}}{\id \Omega_\text{rest}}\,\id\Omega_\text{rest}
\end{equation}
where
\begin{equation}
	\label{thomson}
	\frac{\id \sigma^\text{Th}}{\id \Omega_\text{rest}} = \frac{3\sigma_T}{16\pi}\left(1 + \cos^2\theta_\text{rest}\right)
\end{equation}
is the Thomson scattering cross section evaluated in the rest frame of the electron. By performing the above-mentioned expansion, Kompaneets is left with an equation involving two integrals
\begin{align}
	\label{int_integrals}
	\frac{\partial n}{\partial t} = F\left(n, \frac{\partial n}{\partial \omega }\right)I_1(\Delta) +G\left(n,\frac{\partial n}{\partial \omega },\frac{\partial^2 n}{\partial \omega^2 }\right)I_2(\Delta^2) \
\end{align}
\textit{i.e.}, the first integral is proportional to the shift while the second is proportional to the shift squared. Above, $F$ and $G$ are some expression on $n$ and its derivatives, which are omitted for now. Kompaneets' strategy is then to compute the second integral only, while the first integral is fixed upon invoking the strong argument that \eqref{int_integrals} should be a continuity equation in the photon number. By using that the current should vanish in equilibrium for the Bose-Einstein distribution, the form of the current itself can be exactly found to be \eqref{cur} and the value of $I_1(\Delta)$ is completely fixed by this procedure. 

As far as we know, the value of $I_1(\Delta)$ was never computed without recurring to this argument and there is a strong tendency in literature to follow Kompaneets recipe to avoid the first integral. For example, \cite{katz, liu, rybicki, zhang} follow Kompaneets' set up, also using the Thomson cross section. We will highlight here that it is not possible to find Kompaneets equation by using the rate \eqref{ratewro}, while performing the diffusion approximation as Kompaneets originally proposed. As we will see, the problem lies precisely in the consistency of this particular description, for example, upon writing \eqref{thomson} we are fixing the electron rest frame, but the diffusion approximation as Kompaneets is proposing is not done in this frame, rather it is done in the frame where the electron distribution is isotropic, given by Maxwell-Boltzmann distribution\footnote{Hence, an observer seeing an electron gas distributed according to Maxwell-Boltzmann cannot use a cross section expressed in the electron rest frame, as there will be a probability of finding electrons with any velocity. Conversely, if we choose to express the cross section in the rest frame of the electron, we cannot use Maxwell-Boltzmann, as the scattering centers (the electrons) will be standing still.}. 

This also suggests that we must search for a covariant expression of the rate and \eqref{thomson} is no longer valid. We will also see that when it comes to that, an important prefactor must be add to \eqref{ratewro} to account correctly for the microscopic behavior of the Boltzmann equation. This prefactor is usually called M\o ller flux or M\o ller velocity factor and it is the relativistic kinematic correction that accounts for the flux of particles in the relativistic Boltzmann equation. The Thomson differential cross section will also be replaced by the covariant expression of the full relativistic Klein-Nishina differential cross section for Compton scattering.

The derivation of Kompaneets equation has already been called ``distinctly non-trivial" \cite{longair} and we hope to clarify a number of issues regarding its derivation in this thesis. On a more theoretical aspect, the problem we highlight is interesting and an important example of how inconsistencies can break conservation of photon number, \textit{i.e.}, the Boltzmann equation should be, from the start, photon number conserving, but if we would follow Kompaneets recipe while computing the integrals, we would be left with a non-conserving equation in the photon number. We also feel that these features are not fully explored in literature.

In the heart of the problem, Chapter \ref{2} will be devoted to the study of the Boltzmann relativistic equation, where we will see how to dialogue between both versions (covariant and the manifestly covariant) of the same kinetic equation. Chapter \ref{3} is aimed to the understanding of central objects appearing in the Boltzmann equation, these are transition rates, scattering matrices and scattering cross sections. In Chapter \ref{4} we will show how to consistently perform the diffusion approximation to the Boltzmann equation in two ways (i) as Kompaneets traditionally proposed and (ii) starting from the manifestly covariant formalism. Chapter \ref{5} approaches the Kompaneets equation from a new setup, which enables further extensions. There, a random walk in a bosonic reciprocal space is considered and we show that suitable chosen transition rates yield not only Kompaneets equation but also an extension to more general boson systems. An overview of less standard extensions to the Kompaneets equation is done in Chapter \ref{6}, while we will go through the conclusions of our work in Chapter \ref{7}. 

Finally, it is important to mention that this thesis project yielded the submitted paper \cite{paper} such that notation, some ideas and words contained in this work will be used, specially when it comes to Chapters \ref{4}, \ref{5} and \ref{6}. However, the discussion presented in this thesis is much more comprehensive to that in \cite{paper}, for example, some sections are transformed into chapters, where useful and extra details will be worked out, while new sections are created and subjects that are not addressed in \cite{paper} are added to enrich and illustrate our discussion (\textit{e.g.} Chapter \ref{2} and \ref{3}). To attain clarity and precision, we will mention whenever \cite{paper} is used explicitly.

\chapter{The relativistic Boltzmann equation}\label{2}

The Boltzmann equation is one of the pillars of Kinetic Theory. Under reasonable assumptions, it gives the spatio-temporal evolution of a central object called the \textit{distribution function}. When considering this equation, one usually has in mind a recipient containing a gas of several identical particles. These particles evolve according to some dynamics and collide among themselves, redistributing their momenta over the phase-space, while changing the form of their distribution function over time. Of course there may be more than one kind of particle and, in that case, one usually talks about a \textit{gas mixture}. This chapter will be devoted to the study of this equation, since it is essential to the correct derivation of the Kompaneets equation. As it is usually more natural to do so, we will start with the standard version of this equation, which we refer here as the covariant (or standard) relativistic Boltzmann equation. The second part we will develop the manifestly covariant formalism, while in the last section we will establish the connection between the two descriptions. It will become clear then what the precise meaning is and how \eqref{komp-boltz} can be correctly expressed. In order to simplify the notation, we will use the Einstein summation convention. The signature of the metric will be fixed to $\eta = \mathrm{diag}(+1, -1,-1,-1)$. Four-vectors will be denoted as $p$ while three-vectors will be boldfaced, $\mathbf{p}$.

This chapter is inspired by the nice works of \cite{ weert, kremer, lebellac, kolkata}, but does not follow any of these references in particular. It is true that the Boltzmann equation is a classical topic in Statistical Mechanics and, as such, it is often difficult to present this topic in a completely original manner. However, we have the perception that some topics are underdeveloped in literature, \textit{e.g.}, gas mixtures or the equivalence between the two descriptions of this equation, so that we will address these subjects here as well. Moreover, whenever possible we will search to clarify points that are sometimes not mentioned or left to the reader.

\section{On invariance, covariance and manifest covariance}\label{inv-cov-manc}

In the sections that follow, we will heavily use some jargons which are common in Theory of Relativity. Since sometimes these words can be misleading and somewhat confusing, considering also that some textbooks interchangeably use some definitions (\textit{e.g.}, covariance and invariance), we will briefly go over what we mean when some specific word is being used. 

It should be sufficient to keep in mind Special Relativity only. Hence, whenever we use the word transformation or some of its variations, we actually mean Lorentz transformations. Similarly, whenever reference frame is used, it should be understood as \textit{inertial} reference frame.

A quantity is said to be \textit{Lorentz invariant} when it remains \textbf{unchanged} under Lorentz transformations. This means that in any inertial reference frame\footnote{ Recall that Lorentz transformations are our way of connecting different frames of reference.} the quantity is given by the same (scalar) value. This is precisely the case of constants (\textit{e.g}, mass, charge or particle number) and scalar four-products like
\[A\cdot B\]
where $A$ and $B$ are four-vectors in Minkowski space. However, there might be quantities which are Lorentz invariant but that is not clear at first glance, this is the example of
\[\frac{\id \mathbf{A}}{A^0}\]
\textit{i.e}, a four-vector measure in three dimensional space divided by its time-component. In that case, we must prove that such quantities are indeed Lorentz invariant and we will do so for a couple of examples in next section. Since this is typically a property of scalar quantities, we will sometimes call them \textit{Lorentz scalars} to avoid cumbersome repetitions.

\textit{Lorentz covariance} refers to one of the principles of Relativity, that the laws of physics remain the same in any inertial reference frame. We expect, of course, that quantities like momentum or energy transform under the change of reference frame, nevertheless, the laws of physics, which are precisely some differential equation combining quantities that might transform, remain the same. This means the laws are so nice that, even though their building blocks change and transform, both sides of the equation transform in the same way, leaving the equation unchanged.

For example, we know that electric and magnetic fields do transform non-trivially under Lorentz transformations, however, Maxwell's equations are Lorentz covariant (or, simply, covariant), \textit{i.e.}, even though $\mathbf{E}$ and $\mathbf{B}$ transform, the equation is built in such way that both sides transform exactly in the same manner. In turn, this means that if the laws of Electromagnetism hold in one inertial frame, it holds in any inertial frame. Covariance is the basis of the principle of relativity.

As we will see, the relativistic Boltzmann equation is Lorentz covariant. In fact, the relativistic Boltzmann equation is an example of an equation which is not only covariant, but also invariant. This particularly happens because both sides are scalar quantities, however, we will refrain from using the term invariant for equations whenever possible.

Finally, we have seen that laws of physics are covariant, but sometimes quantities appearing there transform non-trivially (like in the case of Maxwell's equations). The natural question then is: can we write the same content of a given equation in a way that covariance is clear from the start? The answer is what we call \textit{manifest covariance}, \textit{i.e.}, we say that a covariant equation is written in a manifestly covariant way if all quantities appearing there are tensors and, as such, are clearly Lorentz covariant quantities at first glance. The textbook example is again Maxwell's equation, but now written in terms of the electromagnetic field tensor $F_{\mu\nu}$.	

Making connection to the next sections, there will be two representations of the relativistic Boltzmann equation, the covariant or standard, and the manifestly covariant one. As we have seen, the case of the Boltzmann equation is parallel to the example of Maxwell's equations, in which two different representations that have exactly same content exist. Indeed, the difference in being covariant or manifest covariant is just a matter of how we choose to write things.

\section{The standard representation of the relativistic Boltzmann equation}
\subsection{Elements of relativistic kinetic theory}
The kinematics of a relativistic particle of mass\footnote{For a photon, the mass is of course zero.} $m$ is characterized by a set of variables,
\[ x = \left(ct,\mathbf{x}\right) \ \ \ \ \ \ \ p=\left(\frac{E}{c}, \mathbf{p}\right)\]
which together completely determines the \textit{state} of the particle in the phase-space $\Gamma$, expressed by the combined set of coordinates
 \[ \mu = (x,p) \]
of course the four-momentum length is constrained by the well-known relation $p^2 = {(p^0)}^2 - \mathbf{p}^2=(mc)^2$.

Suppose that we now have a gas of N identical particles, then, the \textit{one-particle distribution function} $f$ is defined such that
\begin{equation}\label{distfunc}
	f(x,p) \id\mathbf{x} \id\mathbf{p} = f(t,\mathbf{x},\mathbf{p} ) \id\mathbf{x} \id\mathbf{p}
\end{equation} 
expresses the particle density in the phase-space volume $\id\mu = \id\mathbf{x}\id\mathbf{p}$. That is sometimes referred to as a \textit{coarse grained} description, where the volume is taken to be large enough compared to the microscopic scale but small enough to be treated as infinitesimal when compared to the macroscopic scale. One can think of the distribution function as the histogram of particles which have phase-space variables around $(\mathbf{x},\mathbf{p})$, therefore we can write informally
\[f(x,p) \approx \frac{1}{N} \{\# \ \mathrm{particles \ having \  }(\mathbf{x_i},\mathbf{p_i})\simeq (\mathbf{x},\mathbf{p}) \ \mathrm{at \ time \ }t \}\]
then, as the particle number $N$ becomes very large (or, $N\uparrow\infty$), the law of large numbers takes over, guaranteeing that, with probability $1$, $f(x,p)$ becomes truly the density in phase-space.

Upon writing \eqref{distfunc}, we are assuming that such description is possible, which of course does not have to be the case. The precise mathematical justification of it, which is related to formalizing the approximation above, is something important, but not subject of the present work. Yet, it is worth observing that to know the exact distribution of all particles in phase-space would be very difficult because we would have to consider $N$ (a number of order $10^{23}$) copies of $\Gamma$ and to define $f$ (which is now a function of $6N+1$ coordinates) such that it expresses the combined density of particles. Thus, the situation is enormously simplified if one considers a description using only the one-particle distribution function, seeking, then, a justification on the law of large numbers. The procedure in which the (classical) Boltzmann equation is derived from a N-body Hamiltonian dynamics is called the \textit{Boltzmann-Grad limit} \cite{scholarpedia}.

By integrating out the momentum, we are left with the density of particles
\begin{equation}\label{rho}
	\rho(t, \mathbf{x})=\int f(t,\mathbf{x},\mathbf{p} )  \id\mathbf{p}
\end{equation}
which corresponds to the following normalization\footnote{Some references choose to normalize the distribution function to unity. In that case, we must divide $f$ by total number of particles $N$.} for the distribution function
\begin{equation}\label{rho2}
N=\int f(t,\mathbf{x},\mathbf{p} ) \id\mathbf{x} \id\mathbf{p}
\end{equation}
if we have a dynamical quantity $Q(\mathbf{x},\mathbf{p})$ (\textit{e.g.}, energy) defined over phase-space we can define its average in a very natural way by using the one-particle distribution function
\begin{equation}
	\langle Q(t)\rangle\coloneqq \int Q(\mathbf{x},\mathbf{p}) f(t,\mathbf{x},\mathbf{p} )\id\mathbf{x} \id\mathbf{p} 
\end{equation}

The representation of the distribution function is also not unique. In fact, suppose we have some relation
\begin{align*}
	&\mathbf{x} = A \mathbf{a} \\
	&\mathbf{p} = B \mathbf{b}
\end{align*}
for some constants $A$ and $B$, then, we can change variables to find
\begin{equation}
	f_{ab}(t,\mathbf{a},\mathbf{b}) = \left(AB\right)^3f(t,\mathbf{x},\mathbf{p} )
\end{equation}
this is very convenient if we want to express the distribution function using wave vectors, for example, instead of momentum.

It is useful to keep track of invariant quantities, that is, quantities which do not change upon performing a Lorentz transformation. Therefore, let us begin by considering
\[\id t \id\mathbf{x} = \frac{1}{c}\id x,\] 
and claiming that this measure is Lorentz invariant, which in turn means that measuring volume and time in some frame while multiplying the result together yields a Lorentz scalar quantity, even though volume and time are separately not invariant. In order to check the claim, we observe that, in another inertial frame $K'$, this measure transforms as
\[\id x = |\det(J)|\id x'.\]
 
Under Lorentz transformations, four-vectors transform as
\[{V'}^\alpha = \Lambda^\alpha_{\, \mu}V^\mu\]
where $\Lambda^\alpha_{\, \mu}$ are the matrix elements related to the transformation in this particular basis. Thus, it is easy to realize that $J=\Lambda$, \textit{i.e.}, the Jacobian of the transformation is just the matrix of the Lorentz transformation, but since
\[|\det(\Lambda)|=1\]
for any Lorentz transformation \cite{carroll}, we readily have
\[\id x = \id x' \implies \id t \id\mathbf{x} = \id t'\id\mathbf{x'}\]
so that the measure in the whole Minkowski space is invariant\footnote{In fact, this is the same as stating that, multiplying measurements of volume and time as measured in some frame, yields a Lorentz scalar quantity.}. In fact, this holds for any four-measure.

Another invariant quantity which will be useful is 
\[\frac{\id \mathbf{p}}{p^0}\]
that is the momentum measure divided by the time component of the momentum. 

\begin{figure}[H]
	\centering
	\includegraphics[width=.4\linewidth]{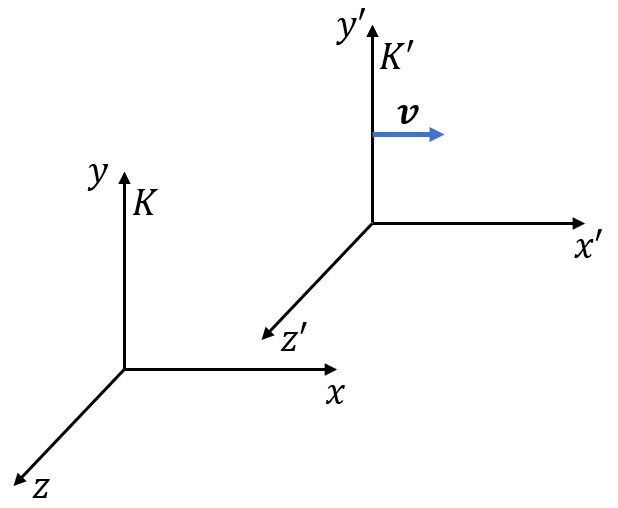}
	\caption{Frames $K$ and $K'$. Seen from $K$, the primed frame is moving with velocity $\mathbf{v}$ along the $x$-direction. This figure is inspired by \cite{kremer}.}
	\label{diffframes}
\end{figure}

Let us then follow \cite{kremer} and prove the result for a very general four-vector satisfying
\[A^\mu A_\mu=C\]
where $C$ is a constant. Denote $K$ as the frame where the components of vector $A$ is unprimed, while $K'$ is the frame where the four-vector $A$ is given by $\left({A^0}', \mathbf{A'}\right)$. We shall suppose now that $K'$ is moving with speed $|\mathbf{v}|$ in the $x$-direction as measured from $K$, see Figure \ref{diffframes}. This assumption is not needed and it is made here only to simplify the calculations. In fact, the Lorentz transformation matrix from a frame $K$ to a frame $K'$, moving with arbitrary velocity $\mathbf{v}$ as measured from $K$ can be found in Appendix \ref{a}.

Because the length of the four-vector $A$ is fixed, we can regard the time component as a function of the spatial components $A^0=A^0(\mathbf{A})$. The Lorentz transformation of the components of $A$ from $K$ to $K'$ is given by
\begin{equation}\label{Atransf}
{A^0}'=\gamma_v\left(A^0 - \frac{|\mathbf{v}|}{c}A^1\right), \ \ {A^1}'= \gamma_v\left(A^1 -\frac{|\mathbf{v}|}{c}A^0 \right),  \ \ {A^2}'=A^2, \ \ {A^3}'=A^3
\end{equation}
or, similarly, in matrix form
\begin{equation}\label{lorentztx}
	\begin{pmatrix}
		{A^0}' \\
		{A^1}' \\
		{A^2}' \\
		{A^3}'
	\end{pmatrix} = 
	\begin{pmatrix}
		\gamma_v & -\gamma_v\frac{|\mathbf{v}|}{c} & 0 & 0\\
		-\gamma_v\frac{|\mathbf{v}|}{c} & \gamma_v & 0 & 0\\
		0 & 0 & 1 & 0\\
		0 & 0& 0&1
	\end{pmatrix}
\begin{pmatrix}
	A^0\\
	A^1\\
	A^2\\
	A^3
\end{pmatrix}
\end{equation}
where
\[\gamma_v = \frac{1}{\sqrt{1-\frac{|\mathbf{v}|^2}{c^2}}}\]
is the Lorentz factor associated with $\mathbf{v}$.

Hence, if we change the reference system, the measure will change as
\begin{equation}\label{mesA}
\id{\mathbf{A'}} =|\det(J)| \id \mathbf{A}	
\end{equation}
where $J$ is the Jacobian of the transformation
\[J = \frac{\partial{\mathbf{A'}}}{\partial{\mathbf{A}}}\]
which by using \eqref{Atransf}	is given by
\begin{equation}
	J=
	\begin{pmatrix}
		\gamma_v\left(1 - \frac{|\mathbf{v}|}{c}\frac{\partial A^0}{\partial A^1}\right) &  - \gamma_v\frac{|\mathbf{v}|}{c}\frac{\partial A^0}{\partial A^2} &  - \gamma_v\frac{|\mathbf{v}|}{c}\frac{\partial A^0}{\partial A^3}\\
		0 & 1 & 0\\
		0 & 0 & 1
	\end{pmatrix}
\end{equation}
Observe that, differently than before, the Jacobian is not simply the Lorentz transformation matrix because we are, in fact, working in the three dimensional space (so not in the whole Minkowski space) and regarding time components as functions of the three-vectors. Therefore, we can write
\begin{equation}\label{detAJ}
	\det(J) = \gamma_v\left(1 - \frac{|\mathbf{v}|}{c}\frac{\partial A^0}{\partial A^1}\right)
\end{equation}

In order to calculate the partial derivative we use the relation $A^\mu A_\mu=C$, yielding
\[A_\mu\frac{\partial A^\mu}{\partial A^1}=0 \ \ \ \ \implies A_0\frac{\partial A^0}{\partial A^1} + A_1 = 0\]
which substituting back in \eqref{detAJ} gives
\begin{equation}
	\det(J) = \frac{1}{A_0}\gamma_v\left(A_0 + \frac{|\mathbf{v}|}{c}A_1\right)= \frac{{A^0}'}{A^0}
\end{equation}
where we used that $A^0=A_0$ and $A^1=-A_1$ because of our metric tensor. Hence,
\[\gamma_v\left(A_0 + \frac{|\mathbf{v}|}{c}A_1\right) \implies \gamma_v\left(A^0 - \frac{|\mathbf{v}|}{c}A^1\right)= {A^0}' \ \ \ \ (\mathrm{using \ \eqref{Atransf}}).\]

Now we substitute that back in \eqref{mesA}, to find
\begin{equation}\label{invmes2}
	\frac{\id{\mathbf{A'}}}{{A^0}'} =\frac{\id{\mathbf{A}}}{{A^0}}
\end{equation}
this proves the desired result and we note that 
\[\frac{\id \mathbf{p'}}{{p^0}'}=\frac{\id \mathbf{p}}{p^0}\]
is found by taking $A=p$.

The final observation we make here is that the distribution function as defined above is a Lorentz scalar quantity. In fact, since the number of observed particles is a Lorentz invariant quantity we have that
\[f(t,\mathbf{x},\mathbf{p})\id\mathbf{x}\id\mathbf{p}\]
is a Lorentz scalar. Now, let us suppose that we are in the rest frame  $K'$ (denoted here with primes) of the particle we are observing, which has four-momentum $p'$. The four-momentum in this frame is of course given by $p'=(mc, 0)$. Let us perform a Lorentz transformation to some other frame $K$ where the particle has four-momenta $p=(p^0,\mathbf{p})$ and is moving with velocity $\mathbf{v}$. By our last result,
\[\id\mathbf{p} = \frac{p^0}{mc}\id\mathbf{p'}= \gamma_v \id\mathbf{p'}\]
where $\gamma_v$ is the Lorentz factor. Similarly, the volume change is calculated using the invariance of the four-measure, yielding
\[\id x = \id x' \implies \id\mathbf{x}=\frac{\id \tau }{\id t}\id\mathbf{x'} = \frac{1}{\gamma_v}\id\mathbf{x'}\]
where we have used the proper time $\id\tau = \id t/\gamma_v$. Using all of this gives
\begin{equation}
	\id\mathbf{x}\id\mathbf{p} =\frac{1}{\gamma_v}\id\mathbf{x'}\, \gamma_v \id\mathbf{p'} = \id\mathbf{x'}\id\mathbf{p'}
\end{equation}
so that the phase-space measure is Lorentz invariant. Since the phase-space and number of particles are invariant, it follows that the distribution function is also a Lorentz invariant quantity. As a matter of fact and, as we will see, this will reflect in the covariance of the Boltzmann equation itself, meaning that we can perform calculations, approximations and express it in any inertial frame of reference.

\subsection{Free evolution of the distribution function}
Our task now is to find a compact expression to the evolution of $f$. Observe that the number of particles at time $t$ in the volume element $\id\mu(t)$ is given by
\begin{equation}
	N(t) = f(t,\mathbf{x},\mathbf{p} )\id\mu(t)
\end{equation}
we expect this number to change as the state of the particle evolves according to some dynamics. Then, we have for a small evolution in time
\begin{equation*}
	N(t+dt) = f(t+dt,\mathbf{x} + \id\mathbf{x}, \mathbf{p} + \id\mathbf{p} )\id\mu(t+dt)
\end{equation*}
which gives for $\id N = N(t+dt) - N(t)$,
\begin{equation}\label{dn}
	\id N =f(t+dt,\mathbf{x} + \id\mathbf{x}, \mathbf{p} + \id\mathbf{p} )\id\mu(t+dt) - f(t,\mathbf{x},\mathbf{p} )\id\mu(t)
\end{equation}
let us first examine the evolution of the phase-space volume. We have
\begin{equation*}
	\id\mu(t+dt) = |\det(J)| \id\mu(t)
\end{equation*}
where $J$ is the Jacobian of the transformation
\begin{equation*}
	J = \frac{\partial(\mathbf{x} + \id\mathbf{x}, \mathbf{p} + \id\mathbf{p})}{\partial(\mathbf{x},\mathbf{p})}
\end{equation*}

We calculate this Jacobian observing that
\begin{align}
	&\id\mathbf{x} =\frac{c\mathbf{p}}{p^0}\id t\\
	&\id\mathbf{p} = \mathbf{F} \id t
\end{align}
where we used 
\begin{equation}\label{force-vel}
	\mathbf{v} = \frac{c\mathbf{p}}{p^0} \ \ \ \ \ \ \mathrm{and} \ \ \ \ \ \ \frac{\id \mathbf{p}}{\id t} = \mathbf{F}
\end{equation}
above, $\mathbf{F}=\mathbf{F}(t, \mathbf{x},\mathbf{p})$ should be regarded as an external force (\textit{e.g.}, electromagnetic) acting on the particle.

Thus we have
\begin{equation*}
	J= \begin{pmatrix}
		1 & \id t \left[ c\mathbf{p}\cdot \frac{\partial }{\partial \mathbf{p}}\left(\frac{1}{p^0}\right) + \frac{3c}{p^0}\right]\\
		\id t \left[  \frac{\partial }{\partial \mathbf{x}}\cdot \mathbf{F} \right] & 1 + \id t \frac{\partial }{\partial \mathbf{p}}\cdot \mathbf{F}
	\end{pmatrix}
\end{equation*}
yielding
\begin{equation}
	\det(J) = 1 + \id t \frac{\partial }{\partial \mathbf{p}}\cdot \mathbf{F} + O(\id t^2)
\end{equation}
plugging that back in \eqref{dn} gives up to second order in time
\begin{equation}\label{dn2}
	\id N =\left[f(t+dt,\mathbf{x} + \id\mathbf{x}, \mathbf{p} + \id\mathbf{p} )\left(1 + \id t \frac{\partial }{\partial \mathbf{p}}\cdot \mathbf{F}\right) - f(t,\mathbf{x},\mathbf{p} )\right]\id\mu(t)
\end{equation}

Now, we Taylor expand the distribution function
\begin{equation*}
	f(t+dt,\mathbf{x} + \id\mathbf{x}, \mathbf{p} + \id\mathbf{p} ) = f(t,\mathbf{x}, \mathbf{p}) + \left(\frac{\partial f }{\partial t}(t,\mathbf{x}, \mathbf{p}) + \frac{c\mathbf{p}}{p^0}\cdot\frac{\partial f}{\partial \mathbf{x}}(t,\mathbf{x}, \mathbf{p}) + \mathbf{F}\cdot\frac{\partial f}{\partial \mathbf{p}}(t,\mathbf{x}, \mathbf{p}) \right)\id t +  O(\id t^2)
\end{equation*}
giving for the first parcel in \eqref{dn2}
\[f(t,\mathbf{x}, \mathbf{p}) + \left(\frac{\partial f }{\partial t}(t,\mathbf{x}, \mathbf{p}) + \frac{c\mathbf{p}}{p^0}\cdot\frac{\partial f}{\partial \mathbf{x}}(t,\mathbf{x}, \mathbf{p}) + \mathbf{F}\cdot\frac{\partial f}{\partial \mathbf{p}}(t,\mathbf{x}, \mathbf{p}) + f(t,\mathbf{x}, \mathbf{p})\frac{\partial }{\partial \mathbf{p}}\cdot \mathbf{F} \right)\id t +  O(\id t^2)\]
substituting that back in \eqref{dn2} yields the expression up to second order in time
\begin{equation}\label{dn3}
	\frac{\id N}{\id t} =\left[\frac{\partial f }{\partial t} + \frac{c\mathbf{p}}{p^0}\cdot\frac{\partial f}{\partial \mathbf{x}} + \frac{\partial }{\partial \mathbf{p}}\cdot \left(f\mathbf{F}\right)\right]\id\mu(t)
\end{equation}
if there is no collision among particles, this term should vanish. This only means that in absence of interactions, the distribution function will evolve freely as dictated by the dynamics of the system.

It is often convenient to express \eqref{dn3} in terms of the proper time $\id \tau$. Since $\id t$ is measured in the frame where the particle has velocity $\mathbf{v}$, the proper time is given by $\id t = \gamma_{v}\id \tau$ such that \eqref{dn3} can be written as
\begin{equation}\label{dn4}
	\frac{\id N}{\id \tau} =\gamma_{v}\left[\frac{\partial f }{\partial t} +\frac{c\mathbf{p}}{p^0}\cdot\frac{\partial f}{\partial \mathbf{x}} + \frac{\partial }{\partial \mathbf{p}}\cdot \left(f\mathbf{F}\right)\right]\id\mu(t)
\end{equation}

\subsection{Collision term}
Boltzmann idea was then to give an expression for the rate of change \eqref{dn3} involving interactions (or collisions) among particles. This suggests that we may decompose the change into two terms: a \textit{gain} term, expressing particles which are scattered into the volume element $\id\mu(t)$ and a \textit{loss} term, expressing particles initially in this volume element, but that scatters away from it. Hence
\begin{equation}
	\frac{\id N}{\id t} = \frac{\id N_+}{\id t} - \frac{\id N_-}{\id t}
\end{equation}
where first parcel in the r.h.s represents the gain term, while second parcel represents the loss term. In order to calculate these two terms, we shall use Boltzmann's Stoßzahlansatz\footnote{In German, this word means collision-number assumption \cite{kolkata}.}. This set of assumptions solves something called the \textit{Boltzmann hierarchy} \cite{spohn}, enabling a closed compact expression for the evolution of $f$.

\begin{greybox}
	\textbf{Boltzmann's Stoßzahlansatz.}
	\begin{itemize}
		\item The distribution function varies slowly in a time interval $\Delta t$ which is large compared to the duration of the collision $\delta t$ but small compared to the time in between collisions (``mean free time", $\tau$) 
		\begin{equation*}
			\delta t \ll \Delta t \ll \tau,
		\end{equation*}
		\textit{i.e.}, to leading approximation collisions are effectively \textit{instantaneous} and particles are only under each other's influence during the collision itself. 
		
		In that sense, the partial derivative should be actually understood as
		\begin{equation*}
			\frac{\partial f}{\partial t} \approx \frac{\Delta f}{\Delta t}
		\end{equation*}
		becoming meaningless to consider an infinitesimal time $\id t$. Then, in that case, Boltzmann equation should not be considered an exact equation \cite{lebellac}. However, we shall suppose that $\Delta t$ is still very small compared to the time we can measure, in such way that taking it to be infinitely small is justified.\footnote{In fact, the same reasoning applies to the coarse-grained description of the cell $\id \mu =\id \mathbf{x} \id \mathbf{p}$, in order to write things such as $\frac{\partial f}{\partial \mathbf{x}}, \frac{\partial f}{\partial \mathbf{p}}$ meaningfully.}
		
		The second inequality ($\Delta t \ll \tau$) serves not only to give meaning to the partial derivative with respect to time, but also to ensure that we can account for one collision episode solely. Otherwise (if $\Delta t \sim \tau$ for example), we would have to account for repeated binary scatterings, as particles collides more rapidly. 
		
		In particular, this reasoning cannot be true if the interactions considered are long-ranged (\textit{e.g.}, Coulomb interactions), since in that case the collision time would be actually infinite. Hence, we shall restrict ourselves to short-ranged, local interactions.  
	\end{itemize}
	\end{greybox}
	\begin{greybox}
	\begin{itemize}
		\item The probability of a scattering event involving more than two particles is very small and, thus, is neglected. Hence, it suffices to consider only binary collisions of particles. This is particularly a good approximation for a gas which is very dilute.
		\item Correlations among particles are neglected. In particular, this means that the two-particle correlation function can be factorized into a product of one-particle distribution functions, this is referred to as \textit{molecular chaos hypothesis}
		\begin{equation}\label{molecch}
			f^{(2)}(t,\mathbf{x_1},\mathbf{p_1}; \mathbf{x_2},\mathbf{p_2}) = f(t,\mathbf{x_1}, \mathbf{p_1})f(t,\mathbf{x_2}, \mathbf{p_2})
		\end{equation}
	\end{itemize}
\par As mentioned, Boltzmann's Stoßzahlansatz is a very good approximation to rarefied gases \cite{scholarpedia}.
\end{greybox}

\subsubsection*{Loss term}

\begin{figure}[H]
	\centering
	\includegraphics[width=0.3\linewidth]{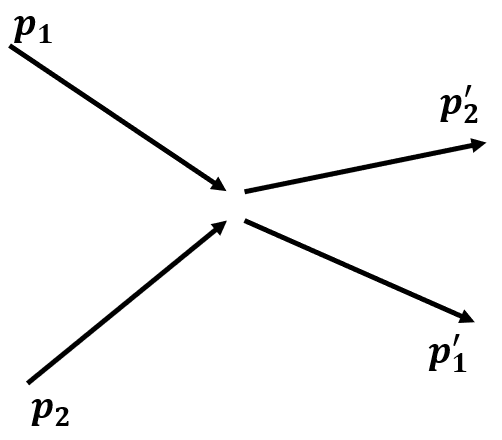}
	\caption{Loss term collisions as seen in laboratory frame. In this frame we see general collisions with scheme $\mathbf{p_1} + \mathbf{p_2} \to \mathbf{p'_1} + \mathbf{p'_2}$. Particle 1 momentum is fixed to $\mathbf{p_1}$ while particle 2 momentum varies within the possible range to account for every collision that leads to depopulate $\id \mathbf{x} \id \mathbf{p_1}$.}
	\label{labframe1}
\end{figure}

The loss term represents every collision scheme which depopulates $\id \mu=\id\mathbf{x}\id\mathbf{p} $. Invoking our first assumption, this leads to transitions between states expressed by pairs of momenta. Then, the loss term is given by collisions starting from states having at least one of the momenta given by $\mathbf{p}$, \textit{i.e.}, a general collision scheme given by
\begin{equation*}
	\mathbf{p_1} + \mathbf{p_2} \to \mathbf{p'_1} + \mathbf{p'_2}
\end{equation*}
where we relabeled $\mathbf{p}\to \mathbf{p_1}$ for notation convenience. Let us denote $\rho_1$ and $\rho_2$ as the density of particle 1 and 2 that participate in the collision in their own frame of reference, respectively. This means that in the frame where we see the collision scheme above (where the particles have initial velocities $\mathbf{v_1}$ and $\mathbf{v_2}$, see Figure \ref{labframe1}) we have
\begin{equation}\label{rhodist}
	\gamma_{v_1}\rho_1 =f(t, \mathbf{x}, \mathbf{p_1})\id \mathbf{p_1} \ \ \ \ \ \ \mathrm{and} \ \ \ \ \ \ \gamma_{v_2}\rho_2 = f(t, \mathbf{x}, \mathbf{p_2})\id \mathbf{p_2}
\end{equation}
Above equalities come from the definition of the distribution function and the observation that $\rho$, being a density, transforms as the inverse of the volume. Since the volume is contracted by a factor of $\gamma$, the density is expanded by the same factor.

Now, let us fix the reference frame in the particle 1. In this frame, the velocity before the collision of particle 1 is zero, while particle 2 has velocity given by $\mathbf{v^{\text{rel}}_{12}}$, the relative (relativistic) velocity with respect to particle 1. The density of particle 2 seen in this reference frame is also given analogously as we found before, but now with contraction related to the relative velocity, \textit{i.e},
\[ \gamma_{v^{\text{rel}}_{12}}\rho_2\] 
hence, the flux of particle 2 seen in the rest frame of particle 1 is given by
\begin{equation}\label{flux}
	\mathcal{F}_{12} = \gamma_{v^{\text{rel}}_{12}}\rho_2 |\mathbf{v^{\text{rel}}_{12}}|
\end{equation}

In order to count the number of collisions, we must introduce the \textit{scattering cross section}. The idea is the following: particle 2 is moving with velocity $\mathbf{v^{\text{rel}}_{12}}$ in the rest frame of 1. We define the $z$-axis in the direction of this vector and denote $\id \Omega_{12}$ as the solid angle having the $z$-axis in the direction of $\mathbf{v^{\text{rel}}_{12}}$. The likeness of the interaction to happen will be encoded in the differential cross section 
\begin{equation*}
	\id \sigma_{12} = \frac{\id \sigma}{\id \Omega_{\text{12}}}\id \Omega_{12}
\end{equation*}
 the subscript indicates that the solid angle is calculated (for now) in the rest frame of particle 1. The quantity $\id\sigma_{12}$ defines an area around the vector $\mathbf{v^{\text{rel}}_{12}}$, which can be thought as the area where the particle 2 will be scattered away by particle 1. If the area is large (then $\id\sigma$ is large) means that the interaction is more probable, while if the area is small, the interaction is less likely\footnote{In fact, $\id\sigma_{12}$ has units of area and it is constructed as such to represent the probability of having an interaction/collision. Imagine, for example, two billiard balls: let us ``stand" on one of the balls while the other comes to strike us. The differential cross section for this collision will be related to the (sectional) area we see from the ball coming to us, in such way that if this area is large, it will be more probable that it hit us.}.

The cross section defined in such way, when multiplied by the flux of incoming particles, gives the number of collisions per unit of time and  unit of target density (that is, particle 1) per unit of volume cell considered, or more precisely
\begin{equation*}
	\mathcal{F}_{12}\frac{\id \sigma}{\id \Omega_{\text{12}}}\id \Omega_{12} = \gamma_{v^{\text{rel}}_{12}}\rho_2 |\mathbf{v^{\text{rel}}_{12}}|\frac{\id \sigma}{\id \Omega_{\text{12}}}\id \Omega_{12}
\end{equation*} 
where we used \eqref{flux}. We multiply this quantity by the density of target particles $\rho_1$ and the volume element to obtain the total number of collisions per unit time in the volume, yielding
\begin{equation}\label{loss1}
	\frac{\id n_{-}}{\id\tau}= \rho_1\id\mathbf{x_{\text{rest}}}\, \gamma_{v^{\text{rel}}_{12}}\rho_2 |\mathbf{v^{\text{rel}}_{12}}|\frac{\id \sigma}{\id \Omega_{\text{12}}}\id \Omega_{12}
\end{equation} 
where $\id \tau$ is the proper time as measured in the rest frame of particle 1. Similarly, we recall that $\id\mathbf{x_{\text{rest}}}$ is also measured in the rest frame of 1.

Our task now is to express this quantity in terms of the distribution function\footnote{We are now using the molecular chaos hypothesis of Boltzmann's Stoßzahlansatz.}, using \eqref{rhodist}. We first rewrite \eqref{loss1} in the following way
\begin{equation}\label{loss2}
	\frac{\id n_{-}}{\id \tau}= \gamma_{v_1}\rho_1\, \gamma_{v_2}\rho_2\, \frac{\gamma_{v^{\text{rel}}_{12}}}{\gamma_{v_1}\gamma_{v_2}}\, |\mathbf{v^{\text{rel}}_{12}}|\, \frac{\id \sigma}{\id \Omega_{\text{12}}}\id \Omega_{12}\, \id\mathbf{x_{\text{rest}}}
\end{equation}
using \eqref{rhodist} this is written as
\begin{equation}\label{loss3}
	\frac{\id n_{-}}{\id \tau} =f(t, \mathbf{x}, \mathbf{p_1})\id \mathbf{p_1}\,f(t, \mathbf{x}, \mathbf{p_2})\id \mathbf{p_2}\, \frac{\gamma_{v^{\text{rel}}_{12}}}{\gamma_{v_1}\gamma_{v_2}}\, |\mathbf{v^{\text{rel}}_{12}}|\, \frac{\id \sigma}{\id \Omega_{\text{12}}}\id \Omega_{12}\, \id\mathbf{x_{\text{rest}}}
\end{equation}

Finally, it is convenient to simplify Equation \eqref{loss3} with the aid of the following identity
\begin{equation}\label{gammarelat}
\frac{\gamma_{v^{\text{rel}}_{12}}}{\gamma_{v_1}\gamma_{v_2}}\, |\mathbf{v^{\text{rel}}_{12}}| = \sqrt{(\mathbf{v_1} - \mathbf{v_2})^2 - \frac{1}{c^2}(\mathbf{v_1}\times\mathbf{v_2})^2} \eqqcolon v_{M12}
\end{equation}
which will be demonstrated in Appendix \ref{a}. This factor is called M\o ller velocity or, sometimes, M\o ller flux \cite{kremer,mirco,kolkata} and, as we have seen, it is necessary for consistent kinematic description of the relativistic Boltzmann equation. As we pointed out in \cite{paper}, this factor is commonly neglected in literature (see for example \cite{electronkompaneets, chen, tong}). It is worth noting that in the classical Boltzmann equation this factor is replaced by the relative velocity of the particles 
\[|\mathbf{v_1} - \mathbf{v_2}|\]
in that sense, the M\o ller velocity can also be thought as the relative speed\footnote{However, it should \textbf{not} be confused with the relativistic relative velocity (see Appendix \ref{a}). An evident contrast can easily be seen when calculating this quantity for a photon-electron scattering: while the modulus of the photon relative velocity is, of course, $c$, the M\o ller factor depends on the electron velocity and it is given by \eqref{molph-el}. In fact, as mentioned by \cite{weinberg} the M\o ller velocity can even exceed the speed of light.} which accounts correctly for the flux of particles in a relativistic treatment \cite{terrall}. In fact, when the velocities are co-linear (which is the case when we work in the center of momentum frame, for example), the second parcel inside the square root vanishes and we are left with the classical expression. 

Using the M\o ller velocity we rewrite \eqref{loss3} in a compact way
\begin{equation}\label{loss4}
	\frac{\id n_{-}}{\id t} =f(t, \mathbf{x}, \mathbf{p_1})f(t, \mathbf{x}, \mathbf{p_2})\,v_{M12}\, \frac{\id \sigma}{\id \Omega_{\text{12}}}\id \Omega_{12}\,  \id \mathbf{p_1}\, \id \mathbf{p_2}\, \id\mathbf{x_{\text{rest}}}	
\end{equation}

Now, expressing all quantities in the initial frame (depicted in Figure \ref{labframe1}), where particle 1 has initial velocity given by $\mathbf{v_1}$, yields
\begin{equation}\label{loss5}
	\frac{\id n_{-}}{\id \tau} = f(t, \mathbf{x}, \mathbf{p_1})f(t, \mathbf{x}, \mathbf{p_2})\,v_{M12}\, \id \sigma\,   \id \mathbf{p_2}\,
	\id \mathbf{p_1}\, \id\mathbf{x}	
\end{equation}
where we have used that $\id\mathbf{x_{\text{rest}}}=\gamma_{v_1}\id\mathbf{x}$ and $\id \tau =\frac{1}{\gamma_{v_1}} \id t$. At this point it is important to note that the cross section appearing in \eqref{loss} 
\[\id \sigma = \frac{\id \sigma}{\id \Omega}\id \Omega\]
must now be expressed in the frame of Figure \ref{labframe1}. From the discussion so far it is not clear that this can be done, \textit{i.e}, we do not know how differential cross sections transform, but the result will follow from the \textit{dynamical reversibility} condition of the differential cross section. We will develop that in great detail in Section \ref{dynrev-sec}, where will become clear that $\id \sigma$ is Lorentz invariant.

Equation \eqref{loss5} gives the total number of particles per unit of proper time that participates in the collision scheme  $(\mathbf{p_1},\mathbf{p_2})\to(\mathbf{p'_1},\mathbf{p'_2})$ which are scattered around the solid angle element $\id \Omega$, leading to depopulate the number of particles with momentum $\mathbf{p_1}$. However, in the Boltzmann equation, we should account for all possible collisions that depopulate state  $\mathbf{p_1}$. We can account for that by integrating over all possible incoming momenta for particle 2 and all possible scattering solid angle $\id \Omega$. This yields for the loss term in the Boltzmann equation

\begin{equation}\label{loss}
	\frac{\id N_{-}}{\id t} =\int_{\mathbf{p_2}}\int_{\Omega}f(t, \mathbf{x}, \mathbf{p_1})f(t, \mathbf{x}, \mathbf{p_2})\,v_{M12}\, \id \sigma\,   \id \mathbf{p_2}\,
	\id \mathbf{p_1}\, \id\mathbf{x}	
\end{equation}

\subsubsection*{Gain term}
Similarly, the gain term represents all possible collisions starting from some initial momenta, leading to populate $\id \mu=\id\mathbf{x}\id\mathbf{p_1}$ with particles of momentum $\mathbf{p_1}$. Analogously, this corresponds to the following collisions
\begin{equation*}
	\mathbf{p'_1} + \mathbf{p'_2} \to \mathbf{p_1} + \mathbf{p_2}
\end{equation*}

To calculate this term we proceed in a very similar fashion as we did for the loss term. However, we must work in the reference frame of particle 1, which has now a pre-collisional momentum giving by $\mathbf{p'_1}$. As we will see, this term is a bit more subtle than the former. Since now the collision promotes transitions $(\mathbf{p'_1},\mathbf{p'_2})\to(\mathbf{p_1},\mathbf{p_2})$, we should prime every quantity appearing in \eqref{loss}. As the collision still happens at position $\mathbf{x}$, the distribution functions should still be evaluated at this coordinate point, but care must be taken, as quantities such as the cell volume and the time are now measured in the pre-collision rest frame of 1, which is different than before. We can then write for the gain term

\begin{equation}\label{gain}
	\frac{dN_{+}}{dt'} =\int_{\mathbf{p'_2}}\int_{\Omega'}f(t, \mathbf{x}, \mathbf{p'_1})f(t, \mathbf{x}, \mathbf{p'_2})\,v'_{M12}\, \id \sigma' \,   \id \mathbf{p'_2}\,
	\id \mathbf{p'_1}\, \id\mathbf{x'}	
\end{equation}
where, similarly to the loss term, all quantities are expressed in the frame which particle 1 had initial velocity given by $\mathbf{v'_1}$.

The first observation we shall make is that the differential cross section appearing in \eqref{gain} does not need to be the same as the one in \eqref{loss}. The cross section appearing in \eqref{gain} is the cross section of the reverse collision process to that in \eqref{loss}. Stating that those two cross section are the same indicates that we have dynamical reversibility, \textit{i.e.}, that the chances of the process happening in one direction is the same as happening in the reverse direction. This does not need to be true and has profound implications to the Boltzmann equation. 

In fact, a more precise definition of dynamical reversibility will be given in next section. Here it is worth noting that it comes from the laws of the interactions we are considering, so, for example, if the interaction is electromagnetic we expect that reversibility holds and, in that case, the cross section is also Lorentz-invariant. In particular, this also means that the Boltzmann equation is particle-number preserving, leading to a Boltzmann equation which is also a master equation. This feature is hardly mentioned in literature mostly because in the vast majority of physical processes, dynamical reversibility is true. In the manifestly covariant formalism, this feature will be related to the unitarity of the scattering matrix.

The second observation we address is that we do not integrate the incoming momentum of particle 1, $\mathbf{p'_1}$. The reason is because conservation of energy-momentum bounds the incoming states $(\mathbf{p'_1},\mathbf{p'_2})$ to the outgoing states $(\mathbf{p_1},\mathbf{p_2})$, such that the set of momenta $(\mathbf{p_1},\mathbf{p_2},\mathbf{p'_1},\mathbf{p'_2})$ is not independent and our only degrees of freedom are expressed in the incoming momentum of particle\footnote{We choose to integrate the particle 2 just because our initial convention is to express the evolution of the distribution function of particle 1.} 2 and the scattering solid angle $\Omega'$. Making contrast with the standard, or (simply) covariant formalism, the manifestly covariant version treats the momenta $(\mathbf{p'_1},\mathbf{p'_2}, \mathbf{p_2})$ as (free) labels, we then integrate them in the Boltzmann equation and conservation of energy-momentum is taken care of by introducing a delta-function in the transition rates.
\subsection{Dynamical reversibility and time evolution of the distribution function}\label{dynrev-sec}

To express the gain and loss term in a compact way (the so-called \textit{Boltzmann collision functional}), we need to explore the symmetries of our interaction. This only means that we will express \eqref{gain} in quantities computed in the pre-collisional rest frame of particle 1, but now referring to the collision scheme $(\mathbf{p_1},\mathbf{p_2})\to(\mathbf{p'_1},\mathbf{p'_2})$, \textit{i.e.}, in the frame where it has initial momentum $\mathbf{p_1}$.

\begin{figure}[H]
	\centering
	\includegraphics[width=1\linewidth]{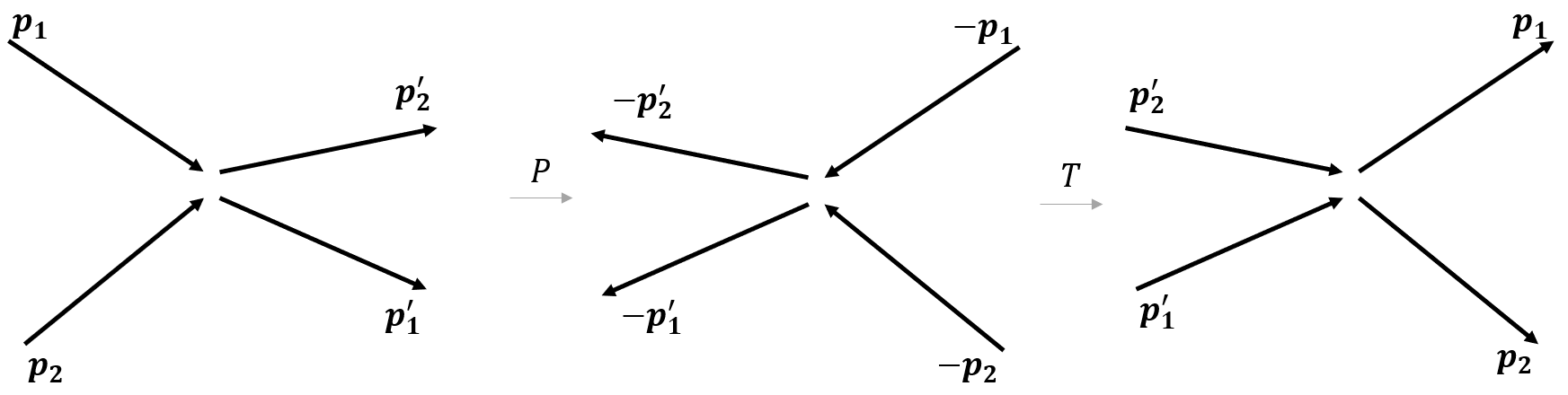}
	\caption{A sequence of parity ($P$) and time reversal ($T$) transformation leads to the inversion of the collision \textit{in} and \textit{out} states as viewed from the new reference frame. This figure is inspired by \cite{lebellac}.}
	\label{lorentz}
\end{figure}

More precisely, this corresponds to start with the collision
\begin{equation*}
	\mathbf{p'_1} + \mathbf{p'_2} \to \mathbf{p_1} + \mathbf{p_2}
\end{equation*}
and perform a Lorentz transformation to reverse the incoming states to the outgoing states. This Lorentz transformation is a composition of  \textit{parity} and \textit{time reversal} transformations (see Figure \ref{lorentz}).
In that new frame, the initial momentum of particle 1 is $\mathbf{p_1}$ and similarly for particle 2.

By doing that, one must express \eqref{gain} in this frame, where above collision is reversed (it is the same collision, however). Therefore, consider again the gain term
\begin{equation}\label{gaindyn}
	\id N_{+} =\int_{\mathbf{p'_2}}\int_{\Omega'}f(t, \mathbf{x}, \mathbf{p'_1})f(t, \mathbf{x}, \mathbf{p'_2})\,v'_{M12}\, \id \sigma' \,   \id \mathbf{p'_2}\,
	\id \mathbf{p'_1}\, \id\mathbf{x'}\,\id t'
\end{equation}
where we have written it in a slightly different manner. We have seen that the product $\id\mathbf{x'}\,\id t'$ is a Lorentz scalar quantity and we can write $$ \id\mathbf{x'}\,\id t' = \id\mathbf{x}\,\id t$$ so that 
\[v'_{M12}\, \id \sigma' \,   \id \mathbf{p'_2}\,
\id \mathbf{p'_1}\, \id\mathbf{x'}\,\id t' = v'_{M12}\, \id \sigma' \,   \id \mathbf{p'_2}\,
\id \mathbf{p'_1}\, \id\mathbf{x}\,\id t\]

In general, for relativistic particles we have $v'_{M12}\neq v_{M12}$ \cite{kremer}, but we can still rewrite the M\o ller velocity in a very suggestive way (see Appendix \ref{a})
\begin{equation}\label{rel-mol-4}
v_{M12} =  |\mathbf{v^{\text{rel}}_{12}}|\frac{p_1\cdot p_2}{p^0_1 p^0_2}
\end{equation}
hence
\[v'_{M12}\, \id \sigma' \,   \id \mathbf{p'_2}\,
\id \mathbf{p'_1}= |\mathbf{v^{\text{rel}'}_{12}}| p'_1\cdot p'_2\, \id \sigma' \,   \frac{\id \mathbf{p'_2}}{{p^0_2}'}\, \frac{\id \mathbf{p'_1}}{{p^0_1}'}\]
the scalar product and the relative velocity are also a Lorentz scalar quantity. In fact, the relative velocity can be written as
\begin{equation}\label{dotrel1}
	|\mathbf{v^{\text{rel}'}_{12}}| =c\sqrt{1-\frac{m^4c^4}{(p'_1\cdot p'_2)^2}}
\end{equation}
thus making clear its invariance. Above relation will be calculated in Appendix \ref{a} also.

Finally, we have seen that
\begin{align*}
	&\frac{\id \mathbf{p'_1}}{{p^0_1}'}= \frac{\id \mathbf{p_1}}{{p^0_1}}\\
	& \frac{\id \mathbf{p'_2}}{{p^0_2}'}= \frac{\id \mathbf{p_2}}{{p^0_2}}
\end{align*}
which enables us to write 
\[|\mathbf{v^{\text{rel}'}_{12}}|p'_1\cdot p'_2\, \id \sigma' \,   \frac{\id \mathbf{p'_2}}{{p^0_2}'}\,
\frac{\id \mathbf{p'_1}}{{p^0_1}'}= |\mathbf{v^{\text{rel}}_{12}}|p_1\cdot p_2\, \id \sigma' \,   \frac{\id \mathbf{p_2}}{p^0_2}\,
\frac{\id \mathbf{p_1}}{p^0_1}\]

The final ingredient we need is the invariance of the cross section. For this we define
\begin{greybox}
	\textbf{Dynamical reversibility (strong form)}. We call the strong form of dynamical reversibility when the \textit{differential} cross section of the reverse collision process is the same as the direct process, \textit{i.e.}
	\begin{equation}
		\id\sigma=\id\sigma'
	\end{equation}
This means that the interaction in place preserves Lorentz symmetry and, thus, is Lorentz invariant. In particular, this is why we omit the solid angle dependence sometimes. In fact, to be more precise we should write 
\begin{equation*}
	\id \sigma = \frac{\id \sigma }{\id \Omega }\id \Omega
\end{equation*}
where the solid angle is measured in the same frame of reference we express the Boltzmann equation in.
\end{greybox} 
It could be the case that the strong form of dynamical reversibility\footnote{Some references call this property detailed balance. We reserve this word, however, to apply it in the traditional set up of master equations.} does not hold when we consider more than one species of particles \cite{kolkata}, keeping that in mind we also define

\begin{greybox}
	\textbf{Dynamical reversibility (weak form)}. We call the weak form of dynamical reversibility when the \textit{total} cross section of the reverse collision process is the same as the direct process, \textit{i.e.}
	\begin{equation}
		\int_{\Omega}\id\sigma=\int_{\Omega'}\id\sigma'
	\end{equation}
\end{greybox}

As we will see, for the cross section we are interested, \textit{i.e.}, the Klein-Nishina cross section, the strong form of dynamical reversibility always hold. Using dynamical reversibility we have
\[\int_{\Omega'}v'_{M12}\, \id \sigma' \,   \id \mathbf{p'_2}\,
\id \mathbf{p'_1}\, \id\mathbf{x'}\,\id t' =\int_{\Omega} v_{M12}\, \id \sigma \,   \id \mathbf{p_2}\,
\id \mathbf{p_1}\, \id\mathbf{x}\,\id t\]
which is sufficient to rewrite \eqref{gaindyn} as
\begin{equation}
		\frac{\id N_{+}}{\id t} =\int_{\mathbf{p_2}}\int_{\Omega}f(t, \mathbf{x}, \mathbf{p'_1})f(t, \mathbf{x}, \mathbf{p'_2})\,v_{M12}\, \id \sigma\,   \id \mathbf{p_2}\,
	\id \mathbf{p_1}\, \id\mathbf{x}
\end{equation}

Finally, we combine \eqref{dn3}, \eqref{loss} and \eqref{gaindyn} to write
\begin{align}\label{prebol}
	\left\{\frac{\partial f_1 }{\partial t} + \frac{c\mathbf{p_1}}{p^0_1}\cdot\frac{\partial f_1}{\partial \mathbf{x}} + \frac{\partial }{\partial \mathbf{p_1}}\cdot \left(f_1\mathbf{F}\right)\right\} \,
	\id \mathbf{p_1}\, \id\mathbf{x} =\left\{\int_{\mathbf{p_2}}\int_{\Omega}\,v_{M12}\, \id \sigma\, \left(f_{1'}f_{2'} - f_1f_2 \right)  \id \mathbf{p_2}\right\} \,
	\id \mathbf{p_1}\, \id\mathbf{x}
\end{align}
with shorthand notation
\[f_i\coloneqq f(t, \mathbf{x}, \mathbf{p_i}) \ \ \ \ \mathrm{and} \ \ \ \ \ f_{i'}\coloneqq f(t, \mathbf{x}, \mathbf{p'_i})  \]
\begin{greybox}
	\textbf{Standard relativistic Boltzmann equation}. By looking \eqref{prebol} we can write
	\begin{equation}\label{boltz}
		\frac{\partial f_1 }{\partial t} + \frac{c\mathbf{p_1}}{p^0_1}\cdot\frac{\partial f_1}{\partial \mathbf{x}} + \frac{\partial }{\partial \mathbf{p_1}}\cdot \left(f_1\mathbf{F}\right) = \int_{\mathbf{p_2}}\int_{\Omega}\, \id \mathbf{p_2}\, v_{M12}\, \id \sigma\, \left(f_{1'}f_{2'} - f_1f_2 \right)
	\end{equation}

We will choose to call above equation the standard\footnote{We note here that in \eqref{boltz}, quantities are not expressed in a manifestly covariant way. As a matter of fact, the nomenclature is just a matter of how we choose to write things as we have seen in Section \ref{inv-cov-manc}.} relativistic Boltzmann equation in order to differentiate it from its other representation appearing in next section. However, sometimes we will, straightforwardly enough, simply call it \textit{the relativistic Boltzmann equation}. This equation gives the spatio-temporal evolution of the distribution function of a gas with many identical particles. Above, the prime momenta are implicitly related to the unprimed ones by energy-momentum conservation.
\end{greybox}
\begin{greybox}
This equation is Lorentz covariant (although not manifestly) and particle-number preserving by construction. In particular, the differential cross section appearing in \eqref{boltz} satisfies dynamical reversibility.

Being covariant, we can choose to compute quantities, express, solve or perform approximations to \eqref{boltz} in any inertial reference frame as long as we are consistent.
\end{greybox}

It is worth noting that some readers might find a factor of $1/2$ in front of \eqref{boltz} when looking into different references. This is because we are dealing with identical particles and one is counting collisions which leads to $(\mathbf{p_1},\mathbf{p_2})$ and $(\mathbf{p_2},\mathbf{p_1})$ as distinct\footnote{Of course that if the particles are identical these two states are the same and one must deal with the overcounting.}. This prefactor can be absorbed in the definition of cross section itself and we choose to do so. By doing that, the Boltzmann equation for a gas mixture has the same structure of \eqref{boltz}, also without the prefactor $1/2$.

\section{The manifestly covariant relativistic Boltzmann equation}\label{manif-cov-bol}

Now that we have seen how to derive the standard version of the relativistic Boltzmann equation, it is straightforward to generalize our result by rewriting it in a manifestly covariant way. For this, we shall look how to write quantities in a invariant way.

Let us begin with the phase-space measure, where we have seen previously that the quantity
\[\frac{\id \mathbf{p}}{p^0}\]
with $p^0 = \sqrt{\mathbf{p}^2 + (mc)^2}$, is a Lorentz scalar quantity, and thus, is invariant by Lorentz transformations. Since the distribution function is also Lorentz invariant, we must find a way of expressing the left hand side \eqref{dn3} in a manifestly covariant way.

We first rewrite \eqref{dn3} by using the relation $ \gamma_{v_1} = p^0_1/mc$
\begin{equation}\label{free}
		\frac{\id N}{\id t} =\frac{mc}{p^0_1}\gamma_{v_1}\left[\frac{\partial f_1 }{\partial t} + \frac{c\mathbf{p_1}}{p^0_1}\cdot\frac{\partial f_1}{\partial \mathbf{x}} + \frac{\partial }{\partial \mathbf{p_1}}\cdot \left(f_1\mathbf{F}\right)\right]\id\mu(t)
\end{equation}
the first two terms can be put in closed dot product
\begin{align}
	\gamma_{v_1}\left[\frac{\partial f_1 }{\partial t} + \frac{\mathbf{p_1}}{E_1}\cdot\frac{\partial f_1}{\partial \mathbf{x}}\right] &= \gamma_{v_1}\frac{\partial f_1 }{\partial t} +  \gamma_{v_1}\frac{c\mathbf{p_1}}{p^0_1}\cdot\frac{\partial f_1}{\partial \mathbf{x}}\nonumber \\
	&= \frac{ p^0_1\,  c}{mc}\frac{\partial f_1 }{\partial x^0} + \frac{\mathbf{p_1}}{m}\cdot\frac{\partial f_1}{\partial \mathbf{x}}\nonumber \\ \label{firstparcel}
	&= \frac{1}{m}p_1^{\mu}\frac{\partial f_1}{\partial x^\mu}
\end{align}
where we used Einstein's summation convention and the four-vectors $p_1= (p^0_1, \mathbf{p_1})$ and $x=(ct, \mathbf{x})$.

To work the last term, we define the Minkowski-four-force vector, defined in terms of the proper time
\begin{equation*}
	\frac{\id p_1}{\id \tau}=K 
\end{equation*}
since $p_1^2=(mc)^2$, we have
\begin{equation*}
 p_1 \cdot \frac{\id p_1}{\id \tau} = p_1 \cdot K = 0
\end{equation*}
above equation also guarantees that 
\begin{equation}\label{zeroc}
	K^0=\frac{\mathbf{p_1}}{p_1^0}\cdot \mathbf{K}
\end{equation}

By looking at our definition of the force vector $\mathbf{F}$ \eqref{force-vel}
\begin{equation}
	\mathbf{K} = \gamma_{v_1} \mathbf{F}
\end{equation}
the four-force above is defined in the whole Minkowski space. The four-momentum length is constraint by the mass, so that we have the zeroth component as a function of the momentum three vector $p_1^0=p_1^0(\mathbf{p_1})$, therefore, when differentiating $K$ with respect to $\mathbf{p_1}$ we have to use the chain rule, treating $p_1^0$ and $\mathbf{p_1}$ as independent variables\footnote{In fact, this transforms the three-divergence in a four-divergence. After the calculation is done, we can set $p^0=E/c$ again. As an example, suppose we have $f(x,y)$ with $y=y(x)$, then, the total variation with respect to $x$ is calculated by the bi-dimensional operator $\left(\frac{\partial }{\partial x} +\frac{\partial y}{\partial x}\frac{\partial }{\partial y}\right)$. This treats $(x,y)$ as independent variables (inside the sign of the partial derivative of course) and when the partial derivatives are done we use the relation $y=y(x)$.}
\[\frac{\partial}{\partial \mathbf{p_1}} \to \frac{\partial p_1^0}{\partial \mathbf{p_1}}\frac{\partial}{\partial p_1^0}+ \frac{\partial}{\partial \mathbf{p_1}}\]
where 
\[\frac{\partial p^0_1}{\partial \mathbf{p_1}}=\frac{\mathbf{p_1}}{p^0_1}\]
this yields for the last term in \eqref{free}
\begin{align}
	\gamma_{v_1}\frac{\partial }{\partial \mathbf{p_1}}\cdot \left(f_1\mathbf{F}\right) &= \gamma_{v_1}\left(\frac{\mathbf{p_1}}{p^0_1}\frac{\partial}{\partial p_1^0}+ \frac{\partial}{\partial \mathbf{p_1}} \right)\cdot \left(f_1\frac{mc\mathbf{K}}{p_1^0}\right)\nonumber \\
	&=\gamma_{v_1}\left(\frac{mc}{p^0_1}\frac{\partial }{\partial p_1^0}\left(f_1\frac{\mathbf{p_1}\cdot\mathbf{K}}{p_1^0}\right)+ \frac{mc}{p_1^0}\frac{\partial}{\partial \mathbf{p_1}}\cdot \left(f_1\mathbf{K}\right) \right)\nonumber \\
	&=\gamma_{v_1}\left(\frac{mc}{p^0_1}\frac{\partial }{\partial p_1^0}\left(f_1K^0\right)+ \frac{mc}{p^0_1}\frac{\partial}{\partial \mathbf{p_1}}\cdot \left(f_1\mathbf{K}\right) \right) \nonumber \\ \label{thirdp}
	&=\frac{\partial \left(f_1K^\mu\right) }{\partial p_1^\mu}
\end{align}
where in second line we used the independence inside the sign of the partial derivatives, while using \eqref{zeroc} in third line. We have also used the expression of the Lorentz factor in last line. Equation \eqref{free} is rewritten by using \eqref{firstparcel} and \eqref{thirdp} in a manifestly covariant way (in terms of the scalar four product) as 

\begin{equation}\label{free2}
	\frac{\id N}{\id t} =c\left[p_1^{\mu}\frac{\partial f_1}{\partial x^\mu} + m\frac{\partial \left(f_1K^\mu\right) }{\partial p_1^\mu}\right]\frac{\id\mu(t)}{p^0_1}
\end{equation}

To calculate the collision term, we now define the \textit{covariant transition rate per unit of volume}, such that the gain and loss terms are given by
\begin{align}
	\label{covgain}
	\id N_{+}=\int_{\mathbf{p_2},\mathbf{p'_1},\mathbf{p'_2}} \frac{\id \mathbf{p_2}}{p^0_2} \frac{\id \mathbf{p'_1}}{{p^0_1}'} \frac{\id \mathbf{p'_2}}{{p^0_2}'} f_1 f_2 W(p_1,p_2\to p'_1, p'_2)\,  \frac{\id \mathbf{p_1}}{p^0_1}\, \id \mathbf{x}\id x^0\\ \label{covloss}
	\id N_{-}=\int_{\mathbf{p_2},\mathbf{p'_1},\mathbf{p'_2}} \frac{\id \mathbf{p_2}}{p^0_2} \frac{\id \mathbf{p'_1}}{{p^0_1}'} \frac{\id \mathbf{p'_2}}{{p^0_2}'} f_{1'} f_{2'} W(p'_1,p'_2\to p_1, p_2)\,  \frac{\id \mathbf{p_1}}{p^0_1}\, \id \mathbf{x'}\id {x^0}'
\end{align}
where we integrated all the degrees of freedom apart from $\mathbf{p_1}$.

The number of scattered particles (left hand side of \eqref{covgain} and \eqref{covloss}) is invariant quantity. We also have seen that the product $ \id \mathbf{x}\id t$ or $ \id \mathbf{x'}\id t'$, together with the momentum measure divided by its time-component and the distribution function, are Lorentz invariant. This makes the transition rates $W$, as defined above, a Lorentz invariant quantity as well. Of course that, as we will see in Section \ref{cov-nonc}, this transition rate is related to the scattering cross section, motivating the definitions

\begin{greybox}
	\textbf{Dynamical reversibility (strong form)}. We call the strong form of dynamical reversibility when the transition rate per unit of volume of the reverse collision process is the same as the direct process, \textit{i.e.}
	\begin{equation}
		W(p_1,p_2\to p'_1, p'_2)=W(p'_1,p'_2\to p_1, p_2)
	\end{equation}
\end{greybox}
\begin{greybox}
	\textbf{Dynamical reversibility (weak form)}. We call the weak form of dynamical reversibility when the transition rate per unit of volume of the reverse collision process is the same as the direct process in the following sense
	\begin{equation}
\int_{\mathbf{p'_1},\mathbf{p'_2}} \frac{\id \mathbf{p'_1}}{{p^0_1}'} \frac{\id \mathbf{p'_2}}{{p^0_2}'}	W(p_1,p_2\to p'_1, p'_2)= \int_{\mathbf{p'_1},\mathbf{p'_2}} \frac{\id \mathbf{p'_1}}{{p^0_1}'} \frac{\id \mathbf{p'_2}}{{p^0_2}'} W(p'_1,p'_2\to p_1, p_2)
	\end{equation}
\end{greybox}

Since by construction our rate respects Lorentz symmetry, dynamical reversibility holds as well and we shall use it as before. In particular, it is related to the unitarity of the scattering matrix, a very general result from Quantum Field Theory (see, for example, \cite{kolkata} for a proof of this fact).

Exploring the invariance of  $ \id \mathbf{x}\id t$ and $ \id \mathbf{x'}\id t'$ analogously than before, we can write the collision term as
\begin{equation}\label{colterm}
\frac{\id N_{+}}{\id t} - \frac{\id N_{-}}{\id t}=\int_{\mathbf{p_2},\mathbf{p'_1},\mathbf{p'_2}} \frac{\id \mathbf{p_2}}{p^0_2} \frac{\id \mathbf{p'_1}}{{p^0_1}'} \frac{\id \mathbf{p'_2}}{{p^0_2}'}  W(p_1,p_2\to p'_1, p'_2)\left(f_{1'} f_{2'} - f_1 f_2\right)\, \id \mathbf{x} \frac{\id \mathbf{p_1}}{p^0_1}
\end{equation}
 As mentioned already, the manifestly covariant formalism, differently than before, treats all the other momenta as (dummy) labels which we have to integrate. Of course not all momenta are possible, the possible ones are given by combinations in which conservation of energy-momentum holds. Therefore, we conclude that, in the definition of the transition rates, there will be a delta-function, which guarantees four-momentum conservation. 

Finally we write using \eqref{free} and \eqref{colterm}
\begin{greybox}
\textbf{Manifestly covariant relativistic Boltzmann equation}. The equation below
\begin{equation}\label{covbol}
p_1^{\mu}\frac{\partial f_1}{\partial x^\mu} + m\frac{\partial \left(f_1K^\mu\right) }{\partial p_1^\mu}=\int_{\mathbf{p_2},\mathbf{p'_1},\mathbf{p'_2}} \frac{\id \mathbf{p_2}}{p^0_2} \frac{\id \mathbf{p'_1}}{{p^0_1}'} \frac{\id \mathbf{p'_2}}{{p^0_2}'}  W(p_1,p_2\to p'_1, p'_2)\left(f_{1'} f_{2'} - f_1 f_2\right)
\end{equation}
found by using \eqref{free} together with \eqref{colterm} is traditionally called the \textit{manifestly covariant relativistic Boltzmann equation}, which is written in way that explores Lorentz invariance of quantities, making covariance manifest.

The transition rates $W$ is related to the differential cross section (see Section \ref{cov-nonc}) and satisfies dynamical reversibility, ensuring particle-number conservation. In particular, the structure of this equation is similar to a master equation in the distribution function, where the transition rates represent a jump-process in phase-space, which is viewed here as a collision process among the particles.

As before, we note the possible appearance of the factor $1/2$ in front of the collision term (right hand side), related to fact we are treating identical particles. Of course this prefactor does not exist for the equation of a gas mixture and we can omit it in \eqref{covbol} by incorporating it in our definition of the rates.
\end{greybox}

Our last discussion for this section will show how the transition rates are defined by using the so-called scattering amplitude. There are two main ingredients in place: the transition amplitude and conservation of energy-momentum. The first ingredient is dealt with by using the scattering matrix. For example, suppose we start from an initial binary state $\bra{i}=\bra{p_1p_2}$\footnote{These are momenta states in what is called a Fock space. A Fock space is analogous to a Hilbert space, where the states are now particles, having some value of momentum.}, evolving to the (also binary) final state $\ket{f}=\ket{p'_1p'_2}$, the amplitude for this transition is given by the squared element of the transition matrix $\mathcal{M}$
\begin{equation}
	M(p_1,p_2\to p'_1,p'_2) = |\bra{p'_1p'_2}\mathcal{M}\ket{p_1p_2}|^2.
\end{equation}

As a matter of fact, this is a common procedure in QFT, where one finds the amplitude of a scattering process by considering the transition matrix $\mathcal{M}$. The exact precise treatment of such problem is important, but beyond the scope of the present work. Nevertheless, we will briefly scratch the surface of scattering theory in Chapter \ref{3}, presenting the main elements to treat such problems. Here, it suffices to know that such matrix exists and that it yields the transition probability per unit of time in going from an initial to a final momenta state.

Hence, considering the two points above we conclude that the transition rates should be proportional to
\begin{equation}\label{transrates}
 W(p_1,p_2\to p'_1, p'_2)	\propto M(p_1,p_2\to p'_1,p'_2) \delta^{(4)}(p_1 + p_2 - p'_1 - p'_2)
\end{equation}
the proportionality factor is a constant related to our convention in defining the distribution function (see next section). Following the definition from \cite{kolkata}, we can express the \textit{isotropic}\footnote{We note in Chapter \ref{3} that, since we consider unpolarized radiation, it is sufficient to consider isotropic transition amplitudes.} transition rates as

\begin{equation}\label{transrates2}
	W(p_1,p_2\to p'_1, p'_2)= \frac{1}{16(2\pi)^6}M(p_1,p_2\to p'_1,p'_2) (2\pi)^4\delta^{(4)}(p_1 + p_2 - p'_1 - p'_2)
\end{equation}

In Section \ref{cov-nonc} we will see how we go to a scattering cross section description by using the definition \eqref{transrates2}. This will make the link of the two representations of the Boltzmann equation.

\section{Degenerate gases and mixtures}

In this section we will treat two extensions of the Boltzmann equation. First we will account for degeneracy of particles, which will lead to the \textit{Boltzmann-Uehling-Uhlenbeck} equation \cite{boluehl}. This equation has a small modification in the collision functional to account for the quantum nature of the particle, that is, whether the particle is a fermion or a boson. The second extension will be related to gas mixtures, where we will account for collisions among different particles.

When going from a classical to a quantum description, one usually divides the momentum cell measure by the term $(2\pi \hbar)^3$
\[\id \mathbf{x}\id \mathbf{p} \to \id \mathbf{x}\frac{\id \mathbf{p}}{(2\pi\hbar)^3} \]
this term is only natural when counting the number of (quantum) states that fit inside a box with some volume $L^3$. In fact, this is a well-known result in Statistical Mechanics and we invite the reader to check, for example, \cite{lebellac}. 

This suggests that, for particles with no spin, $\id \mathbf{x}\frac{\id \mathbf{p}}{(2\pi\hbar)^3}$ is the number of available states in $\id \mathbf{x}\id \mathbf{p}$. If we want to describe particles having spins, the states grows by a number $g_s$, which is sometimes called \textit{degeneracy factor}. Thus, we have
\[ g_s\id \mathbf{x}\frac{\id \mathbf{p}}{(2\pi\hbar)^3} \]
 as the number of available states, where $g_s$ is given by (see for example \cite{kremer})
 \begin{equation*}
 g_s=\begin{cases}
 	2s +1 \ \ \ \ &\mathrm{if} \ m\neq 0 \\
  s \ \ \ \ \ &\mathrm{if} \ m=0
 \end{cases}
 \end{equation*}
where $s$ is the particle spin.

If we have a distribution function $f$ representing the particle density, we can write
\begin{align*}
	&f\id \mathbf{x}\id \mathbf{p} = \ \mathrm{density \ of \ particles}\\
	&f \frac{(2\pi\hbar)^3}{g_s}\id \mathbf{x}\id \mathbf{p} = \ \mathrm{density \ of \ occupied \ states}
\end{align*} 
\textit{i.e.}, the density of occupied states is the distribution function per number of states. Therefore, we can define 
\begin{equation}\label{ocup}
n(t, \mathbf{x}, \mathbf{p}) \coloneqq \frac{(2\pi\hbar)^3}{g_s} f(t, \mathbf{x}, \mathbf{p})
\end{equation}
which we call the \textit{occupation number distribution function}.

Representing the density of occupied states, we can now make the following statistical argument, to extend the collision term of the Boltzmann equation. Suppose, then, a collision $(\mathbf{p_1}, \mathbf{p_2})\to (\mathbf{p'_1}, \mathbf{p'_2})$, fermions will only make this transition if both states are unoccupied. Since $(1-n)$ represents the fraction of unoccupied states we should make the following change in the Boltzmann collision term
\[f_1f_2 \to f_1f_2\left(1- \frac{(2\pi\hbar)^3}{g_s} f_{1'}\right)\left(1- \frac{(2\pi\hbar)^3}{g_s} f_{2'}\right)\] 
and
\[f_{1'} f_{2'} \to f_{1'} f_{2'}\left(1- \frac{(2\pi\hbar)^3}{g_s} f_1\right)\left(1- \frac{(2\pi\hbar)^3}{g_s} f_2\right)\]
to account for Pauli exclusion principle. Likewise, for bosons we should replace the minus sign by a plus sign. We note here that this argument should not be taken too serious at this point. The complete argument should be sketched by working in symmetrized or antisymmetrized Fock spaces, identifying in there the transition rates corresponding to a jump process in reciprocal space. We shall do that in Chapter \ref{5}, where it will become clear that the rates should have an extra factor of $(1+\epsilon n)$ according to the quantum nature of the particle ($\epsilon=+1$ for bosons or $\epsilon=-1$ for fermions). Fortunately, the correction is the same as we find here, so that this qualitatively hand-waving argument is worthy at this point.
\begin{greybox}
	\textbf{Relativistic Boltzmann-Uehling-Uhlenbeck equation}. By replacing the collision term as we noted above, we can write the standard relativistic Boltzmann-Uehling-Uhlenbeck equation as
		\begin{align}
		\frac{\partial f_1 }{\partial t} + \frac{c\mathbf{p_1}}{p^0_1}\cdot\frac{\partial f_1}{\partial \mathbf{x}} + \frac{\partial }{\partial \mathbf{p_1}}\cdot \left(f_1\mathbf{F}\right)& =\nonumber \\
		 \int_{\mathbf{p_2}}\int_{\Omega}\, \id \mathbf{p_2}\, v_{M12}\, \id \sigma\,
		 \bigg{(}&f_{1'} f_{2'}\left(1+\epsilon \frac{(2\pi\hbar)^3}{g_s} f_1\right)\left(1+\epsilon \frac{(2\pi\hbar)^3}{g_s} f_2\right) \nonumber \\
		   -&f_1f_2\left(1+\epsilon \frac{(2\pi\hbar)^3}{g_s} f_{1'}\right)\left(1+\epsilon \frac{(2\pi\hbar)^3}{g_s} f_{2'}\right) \bigg{)}\label{boltzu}
	\end{align}
 	or, its manifestly covariant representation
 	\begin{align}
 		p_1^{\mu}\frac{\partial f_1}{\partial x^\mu} + m\frac{\partial \left(f_1K^\mu\right) }{\partial p_1^\mu}= \ \ \ \ \ \ \ \ \ \ \ \ \ \ \ \  \ \ \ \ &\nonumber \\
 		\int_{\mathbf{p_2},\mathbf{p'_1},\mathbf{p'_2}} \frac{\id \mathbf{p_2}}{p^0_2} \frac{\id \mathbf{p'_1}}{{p^0_1}'} \frac{\id \mathbf{p'_2}}{{p^0_2}'} W(p_1,p_2\to p'_1, p'_2)\bigg{(}&f_{1'} f_{2'}\left(1+\epsilon \frac{(2\pi\hbar)^3}{g_s} f_1\right)\left(1+\epsilon \frac{(2\pi\hbar)^3}{g_s} f_2\right)\nonumber \\
 		 -& f_1f_2\left(1+ \epsilon\frac{(2\pi\hbar)^3}{g_s} f_{1'}\right)\left(1+\epsilon \frac{(2\pi\hbar)^3}{g_s} f_{2'}\right)\bigg{)}\label{covbolu}
 	\end{align}	
 	where $\epsilon=1$ for bosons, while $\epsilon=-1$ for fermions
\end{greybox}

At this point it is worth making the observation that some careful readers may find versions of the relativistic Boltzmann equation which have the prefactor $\frac{g_s}{(2\pi\hbar)^3}$ in front of the right hand side of the equation (the collision term). This is related to the convention of expressing the time evolution of what we defined as the occupation number distribution function instead of the distribution function itself. So that, a straightforward calculation yields, for example 
\begin{align}
	\frac{\partial n_1 }{\partial t} + \frac{c\mathbf{p_1}}{p^0_1}\cdot\frac{\partial n_1}{\partial \mathbf{x}} + \frac{\partial }{\partial \mathbf{p_1}}\cdot \left(n_1\mathbf{F}\right) =	\frac{g_s}{(2\pi\hbar)^3}\int_{\mathbf{p_2}}\int_{\Omega}\, \id \mathbf{p_2}\, v_{M12}\, \id \sigma\,
	\big{(}&n_{1'}n_{2'}\left(1+ +\epsilon n_1\right)\left(1+ \epsilon n_2\right) \nonumber \\
	&-n_1n_2\left(1 +\epsilon n_{1'}\right)\left(1+ \epsilon n_{2'}\right) \big{)}\label{boltzuocu}
\end{align}
for the distribution $n$ defined as \eqref{ocup}. Following our programme of expressing the time evolution of the distribution function $f$ instead of $n$, we shall use the equation on $f$. This in turn will lead to simplifications when describing gas mixtures, as the degeneracy factor is different for different gases. Yet, we note here that some references denote $n$ as $f$, which can arise some confusions, for example \cite{boluehl, torresueh, kolkata} are using $f$ for what we call $n$, but as long as we are consistent and clear, this should not be a problem. As \cite{kremer} points out, when replacing $\frac{(2\pi\hbar)^3}{g_s} f\to f$ this new distribution function (which we denote by $n$) becomes the probability that the state is occupied rather than the probability density of particles in phase-space and one must also change \eqref{rho}, as well as \eqref{rho2}.

\subsection{Gas mixtures}

Now we shall turn our attention to gas mixtures. The  Boltzmann equation for mixtures was treated in a classical context, for example, in \cite{ross2}, while in a relativistic context by \cite{kremer}. Therefore, suppose we have a container with several different types of particles, for example, one can keep in mind a mixture of electrons and photons. If the particles only collide among themselves, one usually talks about an \textit{inert mixture}. On the other hand, if the particles not only collide, but also transform, one usually refers to as a \textit{reacting mixture}. Here, the reaction can be either chemical or nuclear in nature. Let us begin with inert mixtures with $k$ different species of particles. In that case, there will be two types of collisions
\begin{align*}
	&\mathbf{p_1}^i + \mathbf{p_2}^i \rightleftarrows \mathbf{p'_1}^i + \mathbf{p'_2}^i \ \ \ \ \mathrm{collisions \ among \ identical \  particles}\\
	&\mathbf{p_1}^i  + \mathbf{p_2}^j \rightleftarrows \mathbf{p'_1}^i + \mathbf{p'_2}^j \ \ \  \mathrm{collisions \ among \ different \  particles}
\end{align*}
for $i,j\in\{1,\dots,k\}$.

We will use superscripts to denote quantities referring to the i-th particle. For example, $\mathbf{p}^i$ denotes the momentum of the $i$-th particle, while $f^i(t,\mathbf{x},\mathbf{p})$ denotes its distribution. Then, the \textit{relativistic Boltzmann equation} for the $i$-th component ($i\in \{1,\dots ,k\}$) will have extra terms related to collisions among different particles and we can write
\begin{equation}\label{boltzinert}
	\frac{\partial f^i_1 }{\partial t} + \frac{c\mathbf{p_1}^i}{{p^0}^i_1}\cdot\frac{\partial f^i_1}{\partial \mathbf{x}} + \frac{\partial }{\partial \mathbf{p_1}^i}\cdot \left(f^i_1\mathbf{F}\right) = \sum_{j=1}^k C(f^i, f^j)
\end{equation}
where we used the shorthand notation for the collision functional
\begin{align}
	C(f^i, f^j) =  \int_{\mathbf{p_2}^j}\int_{\Omega^{ij}}\, \id \mathbf{p_2}^j\, v^{ij}_{M12}\, \id \sigma^{ij}\,
	\bigg{(}&{f_{1'}}^i{f_{2'}}^j\left(1+\epsilon^i \frac{(2\pi\hbar)^3}{g^i_s} f^i_1\right)\left(1+\epsilon^j \frac{(2\pi\hbar)^3}{g^j_s} f^j_2\right) \nonumber \\
	-&f^i_1f^j_2\left(1+\epsilon^i \frac{(2\pi\hbar)^3}{g^i_s} {f_{1'}}^i\right)\left(1+\epsilon^j \frac{(2\pi\hbar)^3}{g^j_s} {f_{2'}}^j\right) \bigg{)}\label{boltzuinert}
\end{align}

Since in next section we will show the equivalence of both descriptions, the choice of \eqref{boltzu} over \eqref{covbolu} is merely arbitrary. Above, particles $i$ and $j$ have respective degeneracy $g_i$ and $g_j$, being either bosons $\epsilon=+1$, fermions $\epsilon = -1$ or classical/non-degenerate $\epsilon=0$. The collision
\[\mathbf{p_1}^i  + \mathbf{p_2}^j \rightleftarrows \mathbf{p'_1}^i + \mathbf{p'_2}^j\]
have scattering cross section $\id \sigma^{ij}$ and the related M\o ller velocity is given by
\[v^{ij}_{M12} = \sqrt{(\mathbf{v_1}^i - \mathbf{v_2}^j)^2 - \frac{1}{c^2}(\mathbf{v_1}^i\times\mathbf{v_2}^j)^2} \]

Naturally, in light of last discussions, we assume that dynamical reversibility holds separately for every collision. Now, suppose that instead of this inert mixture, we have a reacting mixture, \textit{i.e.}, we consider a gas with the same $k$ components as before, but now, besides colliding, the $i$-th particle undergoes $q$ different (reversible) reactions
\begin{equation*}
	i + a_l \rightleftarrows b_l + c_l \ \ \ \ \mathrm{with \ }l\in\{1,\dots,q\}
\end{equation*}
where $a_l,b_l,c_l$ is one of the $k$ different species. The collision term in the Boltzmann equation will again have extra terms related to the reaction, thus, the natural extension of the collision functional is
\begin{equation}\label{reaccoll}
	C_{r-mix}(f^i) = \sum_{j=1}^k C(f^i,f^j) + \sum_{l=1}^q C^l_{\text{reac}}(f^i)
\end{equation}
where the last parcel is the collision term due to the $l$-th reaction
\begin{equation}
	C^l_{\text{reac}}(f^i) = \int_{\mathbf{p}^{a_l}}\int_{\Omega^{ia_l}}\, \id \mathbf{p}^{a_l}\, v^{ia_l}_{M}\, \id \sigma^{ia_l}_{\text{reac}}\,\left({f_{b_l}}{f_{c_l}}- f_if_{a_l}\right)\label{colisionreac}
\end{equation}
since the final states are always different particles, there is no enhancement or inhibition due to Quantum Mechanics in the reaction functional. The M\o ller velocity is of course given by
\[ v^{ia_l}_{M}= \sqrt{(\mathbf{v}^i - \mathbf{v}^{a_l})^2 - \frac{1}{c^2}(\mathbf{v}^i\times\mathbf{v}^{a_l})^2}\]
the reactive cross section $\id \sigma^{ia_l}_{\text{reac}} $ is the natural extension of the inert scattering cross section. As pointed out by \cite{kremer}, this cross section follows a form of dynamical reversibility given by
\begin{equation*}
\int_{\Omega^{ia_l}}\, v^{ia_l}_{M}\, \id \sigma^{ia_l}_{\text{reac}}\id \mathbf{p}^{i}\id \mathbf{p}^{a_l} = \int_{\Omega^{ia_l}}\, v^{b_lc_l}_{M}\, \id \sigma^{b_lc_l}_{\text{reac}}\id \mathbf{p}^{b_l}\id \mathbf{p}^{c_l} 
\end{equation*}
which connects the cross section of the forward reaction
\[i + a_l \rightarrow b_l + c_l \]
to that of the backward reaction
\[i + a_l \leftarrow b_l + c_l \]

As an example, we shall consider an inert mixture of electrons and photons.
\\
	\par\textbf{ \ \ \ \ Example: an electron-photon mixture}. Suppose we have an inert mixture of electrons and photons interacting via Compton effect with no external force acting on the system. If we consider that the cross section for photon-photon interaction is vanishingly small (which is a reasonable approximation up to leading order)\footnote{In fact, photon-photon scattering does exist but its cross section is, at best, of order $\approx 10^{-7}\, \mathrm{b}$, thus requiring an enormous flux of particles. In particular, note the difference in the order of magnitude when comparing with the Klein-Nishina cross section $\approx 10^{-1}\, \mathrm{b}$. Moreover, we expect photon-photon scattering to become more "relevant" for energies of order TeV, which considering the range of energies we are working ($\ll m_ec^2 \approx 0.5 \mathrm{MeV}$) is negligible \cite{silveira}.}, we can write the relativistic Boltzmann equation for the photon distribution function as
\begin{align*}
	\frac{\partial f_\gamma }{\partial t} + c\mathbf{\hat{n}}\cdot\frac{\partial f_\gamma}{\partial \mathbf{x}} & = \\
\int_{\mathbf{p}}\int_{\Omega}&\, \id \mathbf{p}\, v^{\gamma e}_{M}\, \id \sigma^{\gamma e}\,
	\bigg{(}f_{\gamma'} f_{e'}\left(1+ \frac{(2\pi\hbar)^3}{g_\gamma} f_\gamma\right)\left(1- \frac{(2\pi\hbar)^3}{g_e} f_e\right) \\
	-&f_\gamma f_e \left(1+ \frac{(2\pi\hbar)^3}{g_\gamma} f_{\gamma'}\right)\left(1- \frac{(2\pi\hbar)^3}{g_e} f_{e'}\right) \bigg{)}
\end{align*}
	where $\gamma$ and $e$ refer to the photons and electrons, respectively. Above, we have used $E_\gamma = \hbar\omega$ and slightly different notation for more clarity. We are now looking to the following collision scheme
	\[\mathbf{p} + \mathbf{k} \rightleftarrows \mathbf{p'} + \mathbf{k'}\]
	where the photon momentum is given by $\mathbf{k} =\mathbf{\hat{n}}\hbar\omega/c$
	
	For this system, the M\o ller velocity is given by
	\begin{align}
		v^{\gamma e}_{M} &= \sqrt{(\mathbf{v} - c\mathbf{\hat{n}})^2 - \frac{1}{c^2}(\mathbf{v}\times c\mathbf{\hat{n}})^2} \nonumber\\
		&= c \sqrt{\left(\frac{\mathbf{v}}{c}\right)^2 - 2\frac{\mathbf{v}}{c}\cdot\mathbf{\hat{n}} + 1 - \left(\frac{\mathbf{v}}{c}\times \mathbf{\hat{n}}\right)^2} \nonumber\\
		&= c \sqrt{\left(\frac{\mathbf{v}}{c}\right)^2 - 2\frac{\mathbf{v}}{c}\cdot\mathbf{\hat{n}} + 1 - \left(\frac{\mathbf{v}}{c}\right)^2 + \left(\frac{\mathbf{v}}{c}\cdot \mathbf{\hat{n}}\right)^2}\nonumber\\
		&=c \sqrt{ 1 - 2\frac{\mathbf{v}}{c}\cdot\mathbf{\hat{n}} + \left(\frac{\mathbf{v}}{c}\cdot \mathbf{\hat{n}}\right)^2}\nonumber\\ 
		&=c\left(1 - \frac{\mathbf{v}}{c}\cdot \mathbf{\hat{n}}\right) \label{molph-el}
	\end{align}
	
	If we now consider that the electrons can be treated as a non-degenerate gas, we can express the equation above as
	\begin{align}\label{prebkomp}
		\frac{\partial n_\gamma }{\partial t} + c\mathbf{\hat{n}}\cdot\frac{\partial n_\gamma}{\partial \mathbf{x}} =c\int_{\mathbf{p}}\int_{\Omega}\, \id \mathbf{p}\, \left(1-\frac{\mathbf{v}}{c}\cdot \mathbf{\hat{n}}\right)\, \id \sigma^{\gamma e}\,
		\left(n_{\gamma'} f_{e'}\left(1+ n_\gamma\right) - n_\gamma f_e \left(1+ n_{\gamma'}\right)\right)
	\end{align}
where we used definition \eqref{ocup} for the photon distribution function.

As we will see, \eqref{prebkomp} will be the starting point of the \textit{Kompaneets equation}. 

\section{Equivalence of the relativistic Boltzmann equation representations}\label{cov-nonc}

In this section we will show the equivalence of the two representations of the Boltzmann equation. Since we are most interested in this equation for an electron-photon system interacting via Compton effect, it is sufficient for us that we prove the equivalence for \eqref{prebkomp}. However, it is worth noting that our calculation is adaptable for any system. 

The equivalence of these two representations is done in \cite{kolkata} for a single particle gas. However, it is is readily extended by linearity for an arbitrary mixture by using ours and \cite{kolkata} result, \textit{i.e.}, the equivalence holds term by term in the collision functional \eqref{boltzuinert} or \eqref{reaccoll}.

Let us begin with the Boltzmann equation for a mixture of photons and electrons as calculated in last section, but now we use the covariant formalism.
\begin{align}\label{phot-eleccov}
k^{\mu}\frac{\partial n_\gamma}{\partial x^\mu} =\int_{\mathbf{p},\mathbf{k'},\mathbf{p'}} \frac{\id \mathbf{p}}{{p^0}} \frac{\id \mathbf{k'}}{{k^0}'} \frac{\id \mathbf{p'}}{{p^0}'}  W^{\gamma e}(p,k\to p', k')\left(n_{\gamma'} f_{e'}\left(1+ n_\gamma\right) - n_\gamma f_e \left(1+ n_{\gamma'}\right)\right)
\end{align}
where the four-vectors are given by
\begin{alignat*}{3}
&k = \left(\frac{\hbar \omega }{c}, \mathbf{k}\right)&&;  \ \ \ \ &&k' = \left(\frac{\hbar \omega' }{c}, \mathbf{k'}\right)\\
&p = \left(\frac{E}{c}, \mathbf{p}\right)&&;  \ \ \ \ &&p' = \left(\frac{E'}{c}, \mathbf{p'}\right)
\end{alignat*}
corresponding to the collision scheme
\[p+k\rightleftarrows p'+k'\]

In order to show the equivalence, we shall use definition \eqref{transrates2}, while computing the integrals in the outgoing degrees of freedom (primed indices). As we shall see in next chapter, the transition amplitude, represented in \eqref{transrates2} by the squared matrix element for an photon-electron scattering
\[M^{\text{KN}}(p,k\to p',k')= \left|\bra{p'k'}\mathcal{M}^{\text{KN}}\ket{pk}\right|^2\]
is related to the so-called Klein-Nishina scattering cross section by the following relation
\begin{align}\label{crosec}
M^{\text{KN}}(p,k\to p',k')	=16\pi \left(s-(m_ec)^2\right)^2\frac{\id \sigma }{\id t}^{\text{KN}}(s,t)
\end{align}
where $s$ and $t$ are called the Mandelstam variables defined as
\begin{align*}
	&s\coloneqq(p+k)^2 = 2p\cdot k + (m_ec)^2\\
	&t\coloneqq(k- k')^2=-2k\cdot k'\\
	&u\coloneqq (p'-k)^2 = -2p'\cdot k + (m_ec)^2 
\end{align*}
where we also added the definition of the Mandelstam variable $u$ (which is not independent from $s$ and $t$)\footnote{By using energy-momentum conservation $p+ k = p'+ k'$, there is one extra relation for each of the Mandelstam variables $s,t,u$ (see Chapter \ref{3}). However, we shall forget for the moment about energy-momentum conservation.}. These variables are clearly Lorentz invariant and completely describe the collision. Since these variables are also collision invariants, we expect that it is possible to express the full differential cross section in terms of them if dynamical reversibility holds. As a matter of fact, we shall assume that as well \cite{cross, cross2, kolkata}, check also Chapter \ref{3}.

By the definition of $t$, we have
\begin{equation*}
	 t = -2k^0 {k^0}'( 1 - \mathbf{\hat{n}}\cdot \mathbf{\hat{n}'})
\end{equation*}
where we defined the unit vector 
\[\mathbf{\hat{n}^{(')}} \coloneqq \frac{\mathbf{k}^{(')}}{{k^0}^{(')}} = \frac{c}{\hbar \omega }\mathbf{k}^{(')}\]
which enables the expression of the scattering angle $\cos \theta = \mathbf{\hat{n}}\cdot \mathbf{\hat{n}'}$. This yields for the derivative of $t$ with respect to $\theta$
\begin{align*}
	\id t &= 2k^0 {k^0}' \id \cos \theta\\
	&=\frac{1}{\pi }k^0 {k^0}' \id \Omega
\end{align*}

Above, we used the solid angle differential, which is given by $\id \Omega = 2\pi \id \cos \theta$ since the scattering respects azimuthal symmetry (that is indeed the case if we have unpolarized photons as we do). Hence, \eqref{transrates2} is given in terms of the cross section 
\begin{equation}
	W^{\gamma e}(p,k\to p',k')	=\frac{1}{4}\frac{ \left(s-(m_ec)^2\right)^2}{k^0 {k^0}'}\frac{\id \sigma^{\gamma e} }{\id \Omega }(s,t)\delta^{(4)}(p+k-p'-k')
\end{equation}

\begin{figure}[H]
	\centering
	\includegraphics[width=0.4\linewidth]{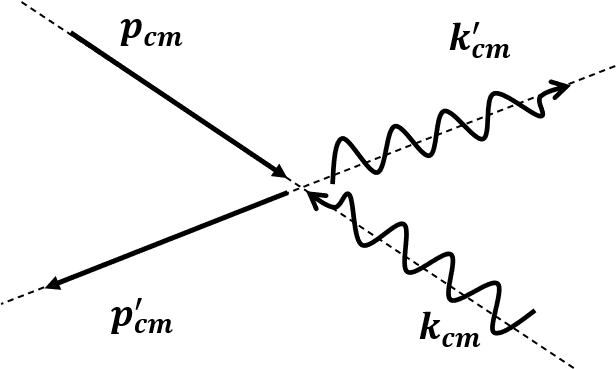}
	\caption{The center of momentum frame of the photon-electron scattering. In this frame the total momentum is zero, \textit{i.e.}, $\mathbf{p_{\text{cm}}}+\mathbf{k_{\text{cm}}} =0 = \mathbf{p'_{\text{cm}}}+\mathbf{k'_{\text{cm}}}$.}
	\label{comframe}
\end{figure}

We have the freedom to work in any intertial frame because of Lorentz covariance and the calculation is simplified if we work in the \textit{center of momentum frame}, see Figure \ref{comframe}. In this frame, the total momentum is always zero and the four vectors are given by
 \begin{alignat*}{3}
 	&k_{\text{cm}} = \left(\frac{\hbar \omega_{\text{cm}} }{c}, \mathbf{k}_{\text{cm}}\right)&&;  \ \ \ \ &&k'_{\text{cm}} = \left(\frac{\hbar \omega'_{\text{cm}} }{c}, \mathbf{k'}_{\text{cm}}\right)\\
 	&p_{\text{cm}} = \left(\frac{E_{\text{cm}}}{c}, -\mathbf{k}_{\text{cm}}\right)&&;  \ \ \ \ \ &&p'_{\text{cm}} = \left(\frac{E_{\text{cm}}'}{c}, -\mathbf{k'}_{\text{cm}}\right)
 \end{alignat*}
therefore, the relativistic Boltzmann equation yields
\begin{align}
	k_{\text{cm}}^{\mu}\frac{\partial n_\gamma}{\partial x_{\text{cm}}^\mu} =\int \frac{\id \mathbf{p}_{\text{cm}}}{{p^0}_{\text{cm}}} \frac{\id \mathbf{k'}_{\text{cm}}}{2{k^0}'_{\text{cm}}} \frac{\id \mathbf{p'}_{\text{cm}}}{2{p^0}'_{\text{cm}}}  \frac{ \left(s-(m_ec)^2\right)^2}{k^0_{\text{cm}} {k^0}'_{\text{cm}}}&\frac{\id \sigma^{\gamma e} }{\id \Omega_{\text{cm}} }(s,t)\delta^{(4)}(p_{\text{cm}}+k_{\text{cm}}-p'_{\text{cm}}-k'_{\text{cm}})\nonumber\\
	&\left(n_{\gamma'} f_{e'}\left(1+ n_\gamma\right) - n_\gamma f_e \left(1+ n_{\gamma'}\right)\right)
\end{align}
where \textit{all} quantities are computed in this particular frame of reference (including the distribution functions). We shall calculate the outgoing degrees of freedom, \textit{i.e.}, the following integral
\begin{align}
	I_{\text{out}} = \int\frac{\id \mathbf{k'}_{\text{cm}}}{2{k^0}'_{\text{cm}}} \frac{\id \mathbf{p'}_{\text{cm}}}{2{p^0}'_{\text{cm}}}  \frac{ \left(s-(m_ec)^2\right)^2}{k^0_{\text{cm}} {k^0}'_{\text{cm}}}&\frac{\id \sigma^{\gamma e} }{\id \Omega_{\text{cm}} }(s,t)\delta^{(4)}(p_{\text{cm}}+k_{\text{cm}}-p'_{\text{cm}}-k'_{\text{cm}})\nonumber\\ 
	&\left(n_{\gamma'} f_{e'}\left(1+ n_\gamma\right) - n_\gamma f_e \left(1+ n_{\gamma'}\right)\right)
\end{align}

Let us reduce the four-delta function by using the following identity, which holds in any frame of reference
\begin{equation}\label{3to4delta}
\frac{\id\mathbf{p'}_{\text{cm}}}{2{p^0}'_{\text{cm}}} = \int_{{p^0}'_{\text{cm}}}\id^4 p'_{\text{cm}}\delta({p'}^2_{\text{cm}} - (m_ec)^2)
\end{equation}
we integrate the four-momentum $p'$ while using the four-delta to set
\begin{equation*}
p'_{\text{cm}}=p_{\text{cm}}+k_{\text{cm}}-k'_{\text{cm}} \ \ \ \implies \ \ \ \  {p'}^2_{\text{cm}} - (m_ec)^2 = 2\left( p_{\text{cm}}\cdot k_{\text{cm}} - k'_{\text{cm}}\cdot(p_{\text{cm}} + k_{\text{cm}}) \right)
\end{equation*}
where we used that the unprimed electron is on its mass shell ${p}^2_{\text{cm}} = (m_ec)^2$

Then, we can rewrite the above integral as
\begin{align}
	I_{\text{out}} = \int\frac{\id \mathbf{k'}_{\text{cm}}}{2{k^0}'_{\text{cm}}}\frac{ \left(s-(m_ec)^2\right)^2}{k^0_{\text{cm}} {k^0}'_{\text{cm}}}\frac{\id \sigma^{\gamma e}}{\id \Omega_{\text{cm}} }(s,t) \delta(2( k'_{\text{cm}}\cdot(p_{\text{cm}} + k_{\text{cm}}) - p_{\text{cm}}&\cdot k_{\text{cm}}))\nonumber \\
	&\left(n_{\gamma'} f_{e'}\left(1+ n_\gamma\right) - n_\gamma f_e \left(1+ n_{\gamma'}\right)\right)
\end{align}

At this point, it is convenient to express things in a Lorentz invariant way in order to keep track of convenient quantities. Therefore, keeping that in mind, we shall express the total energy in the center of momentum frame using the Mandelstam $s$-variable 
\begin{equation}
s = (p_{\text{cm}}+k_{\text{cm}})^2=(p^0_{\text{cm}} + k^0_{\text{cm}})^2  \ \ \ \ \ \implies  p^0_{\text{cm}} + k^0_{\text{cm}}= \sqrt{s}
\end{equation}
as well as $k^0_{\text{cm}}$

\begin{equation}\label{s-frequency}
\frac{s-(m_ec)^2}{2}= p_{\text{cm}}\cdot k_{\text{cm}} =k^0_{\text{cm}} \sqrt{s} \ \ \ \ \ \implies k^0_{\text{cm}}= \frac{s-(m_ec)^2}{2\sqrt{s}}
\end{equation}
where we used in both expressions that $\mathbf{p}_{\text{cm}}=-\mathbf{k}_{\text{cm}}$.

Thus, the argument of the delta function is simplified by using the expression of the four-momenta together with calculations above, yielding
\begin{align*}
	&k'_{\text{cm}}\cdot(p_{\text{cm}} + k_{\text{cm}}) = {k^0}'_{\text{cm}}(p^0_{\text{cm}} + k^0_{\text{cm}}) = {k^0}'_{\text{cm}}\sqrt{s}\\
	&p_{\text{cm}}\cdot k_{\text{cm}}  = \frac{s-(m_ec)^2}{2}
\end{align*}
which enables the expression of the delta function
\begin{align}
	 \delta(2( k'_{\text{cm}}\cdot(p_{\text{cm}} + k_{\text{cm}}) - p_{\text{cm}}\cdot k_{\text{cm}}))&=\delta\left(2\sqrt{s}\left({k^0}'_{\text{cm}} - \frac{s-(m_ec)^2}{2\sqrt{s}}\right)\right)\nonumber \\
	 &=\frac{1}{2\sqrt{s}}\delta\left({k^0}'_{\text{cm}} - \frac{s-(m_ec)^2}{2\sqrt{s}}\right) \label{deltasimpl}
\end{align}
where we used the delta function identity
\[\delta(ax) = \frac{1}{|a|}\delta(x)\]

Finally, we plug that back in the integral, while using the expression of $k^0_{\text{cm}}$ to yield
\begin{align}
	I_{\text{out}} = \int\frac{\id \mathbf{k'}_{\text{cm}}}{{{k^0}'_{\text{cm}}}^2} \frac{\left(s-(m_ec)^2\right)}{2}\frac{\id\sigma^{\gamma e} }{\id \Omega_{\text{cm}} }(s,t) \delta\left({k^0}'_{\text{cm}} - \frac{s-(m_ec)^2}{2\sqrt{s}}\right)\left(n_{\gamma'} f_{e'}\left(1+ n_\gamma\right) - n_\gamma f_e \left(1+ n_{\gamma'}\right)\right)
\end{align}
we use now spherical coordinates to compute the final integral. By aligning the $z$-axis with the incoming photon direction, we have $\id \mathbf{k'}_{\text{cm}} = {{k^0}'_{\text{cm}}}^2 \id {k^0}'_{\text{cm}}\id \Omega_{\text{cm}}$. The integral over the delta function is trivial, yielding $1$ and imposing energy-momentum conservation, also implicitly determining the primed quantities inside the distribution function argument in terms of the electron-photon incoming states (unprimed labels), we have
\begin{align}
	I_{\text{out}} =\frac{\left(s-(m_ec)^2\right)}{2} \int_{\Omega_{\text{cm}}}\id \Omega_{\text{cm}} \frac{\id \sigma^{\gamma e} }{\id \Omega_{\text{cm}} }(s,t) \left(n_{\gamma'} f_{e'}\left(1+ n_\gamma\right) - n_\gamma f_e \left(1+ n_{\gamma'}\right)\right)
\end{align}
since
\[\frac{\left(s-(m_ec)^2\right)}{2} =p_{\text{cm}}\cdot k_{\text{cm}} \]
this yields for the Boltzmann equation
\begin{equation*}
k_{\text{cm}}^{\mu}\frac{\partial n_\gamma}{\partial x_{\text{cm}}^\mu} =\int_{\mathbf{p}_{\text{cm}}}\int_{\Omega_{\text{cm}}}\frac{\id \mathbf{p}}{{p^0_{\text{cm}}}}\,	(p_{\text{cm}}\cdot k_{\text{cm}})\, \id \Omega_{\text{cm}} \frac{\id\sigma^{\gamma e} }{\id \Omega_{\text{cm}} }(s,t) \left(n_{\gamma'} f_{e'}\left(1+ n_\gamma\right) - n_\gamma f_e \left(1+ n_{\gamma'}\right)\right)
\end{equation*}
this equation is expressed in the center of momentum frame. Therefore, we now explore invariance and observe that, because quantities are expressed in a manifestly covariant way (including the cross section, which is a Lorentz invariant quantity because of dynamical reversibility), we can transform the equation back to the original frame, where the collision is given by
\[p+k\rightleftarrows p'+k'\]
resulting in
\begin{equation}
	k^{\mu}\frac{\partial n_\gamma}{\partial x^\mu} =\int_{\mathbf{p}}\int_{\Omega}\frac{\id \mathbf{p}}{{p^0}}\,	(p\cdot k)\, \id \sigma^{\gamma e} \left(n_{\gamma'} f_{e'}\left(1+ n_\gamma\right) - n_\gamma f_e \left(1+ n_{\gamma'}\right)\right)
\end{equation}

Finally, we rearrange terms to find 
\begin{equation}
\frac{\partial n_\gamma }{\partial t} + c\mathbf{\hat{n}}\cdot\frac{\partial n_\gamma}{\partial \mathbf{x}} =c \int_{\mathbf{p}}\int_{\Omega}\, \id \mathbf{p}\, \frac{p\cdot k}{ p^0 k^0}\, \id \sigma^{\gamma e}\,
\left(n_{\gamma'} f_{e'}\left(1+ n_\gamma\right) - n_\gamma f_e \left(1+ n_{\gamma'}\right)\right)
\end{equation}
where we identify the M\o ller velocity (see Appendix \ref{a}, Equation \eqref{rel-mol-4-ap})
\[c\frac{p\cdot k}{ p^0 k^0} = c\frac{(p^0 k^0 - \mathbf{p}\cdot \mathbf{k})}{ p^0 k^0} = c\left(1 - \frac{\mathbf{v}}{c}\cdot \mathbf{\hat{n}}\right) \]
while using the relations
\[\frac{\mathbf{k}}{k^0 } =  \mathbf{\hat{n}} \ \ \ \ \ \mathrm{and } \ \ \ \ \ \frac{\mathbf{p}}{p^0 } =  \frac{\mathbf{v}}{c}\]

This proves the equivalence of the two representations of the relativistic Boltzmann equation in the context of a photon-electron inert mixture. As mentioned before, our calculation is adapted to work in more general systems, including in gases having a single particle component (see for example \cite{kolkata}).

\chapter{Scattering cross sections}\label{3}

In this chapter we shall see how to express scattering cross sections by using the scattering matrix formalism from Quantum Field Theory. There are many good books or reviews written about that and we do not intend here to explore this subject very deeply. The interested reader can find more detailed discussions, for example, in \cite{jauch,peskin, weinberg, cross, cross2, mirco}, these works will also be used to guide us throughout this chapter. It is also true that this subject is very vast and, while this is a very general formalism, which covers also many types of scattering processes, we will focus here in Compton scattering, giving rise to the so-called Klein-Nishina cross section. For us, it will be sufficient to consider (\textit{e.g.} for the photon) unpolarized states, hence, we shall not be concerned about complications arising from having polarization, which can be important for some applications. Therefore, quantities appearing here will be regarded as averaged over all polarization states. Finally, in last section we will be interested in how to express this particular cross section in a way that explores its Lorentz invariance. In this chapter, we shall also follow the convention of using natural units $\hbar=c=1$ in order to simplify notation. 

\section{From scattering matrices to cross-sections}

\begin{figure}[H]
	\centering
	\includegraphics[width=0.3\linewidth]{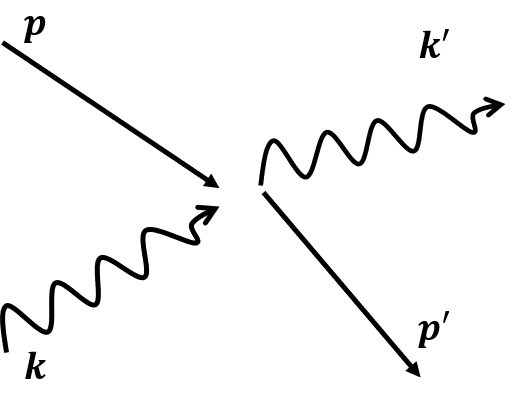}
	\caption{Photon-electron scattering as seen in the lab frame. In this frame, particles are allowed to have any value of (initial) momenta.}
	\label{labframe}
\end{figure}

Scattering experiments are our main source of knowledge to understand the processes that govern elementary particles. When dealing with them, we usually have in mind some set of incoming states, also called \textit{in} states, which interact, resulting in outgoing states, also called \textit{out} states. In the lab frame\footnote{We note here that it is a common procedure in many references to refer as the lab frame the frame where one of the particles (here we are thinking about binary states) is at rest. We shall differentiate this frame, by calling it rest frame. The lab frame will then be the frame where the collision is being performed in laboratory, and, thus, particles are allowed to have very general momenta, see Figure \ref{labframe} for example.} this corresponds to the general collisional scheme
\[p_1 + p_2 \to \{p_f\}=p'_1 + p'_2 \]  
where we restrict ourselves to binary collisions only. On one hand, we have seen in Chapter \ref{2} that the main protagonist which links measurable quantities in the laboratory with the probability of having collisions is the \textit{scattering cross section}, defined as
\begin{equation}\label{cr}
	\sigma = \frac{R_f}{\mathcal{F}}
\end{equation} 
where
\[R_f= \mathrm{number \ of \ scattering \ events \  with \ binary \ final \  state \ per \ unit \  volume \  per \ unit \  time }\]
and $\mathcal{F}$ is the incident flux of particles.

The cross section defined in such way is called the \textit{total cross section}, the word \textit{total} only means that we are looking at all possible collisions that lead to a set of binary states. The \textit{differential cross section} is then defined by looking at collisions that lead to the specific labeled out state $f=(p'_1,p'_2)$, \textit{i.e}
\begin{equation}\label{diffcr}
	\id \sigma = \frac{\id R_f}{\mathcal{F}}
\end{equation}
where
\[\id R_f= \mathrm{number \ of \ scattering \ events \  with \ final \ state \ }f=(p'_1,p'_2) \ \mathrm{per \ unit \  volume \  per \ unit \  time }\]

In fact, according to our reasoning, it is worth thinking about differentials above as differentials over the final momenta 
\[\frac{\id^2 \sigma }{\id \mathbf{p'_1} \id \mathbf{p'_2}}.\]

On the other hand, \textit{in} and \textit{out} states are linked by the so-called scattering or S-matrix. The idea is as follows: we start with the two particles infinitely far apart at time $t=-\infty$, so they are \textit{free particles}\footnote{For more detailed discussions and clarification of jargon, like \textit{asymptotically free particles}, we refer the reader to \cite{peskin, weinberg}.}, we bring them close together while they interact via some interaction described by your theory (\textit{e.g.}, electromagnetic interactions). Finally, after they interact, the particles move away from each other again, so that at $t=+\infty$ they are infinitely far apart and free again. Therefore, the overlap of \textit{in} and \textit{out} states is given by the elements of the $S$-matrix
\begin{equation}
S_{fi} = \lim_{t\to\infty}\bra{p'_1p '_2}U(t,-t)\ket{p_1p_2}
\end{equation}
where
\[U(t,-t)=\exp\left(-iH(2t)\right)\]
is the time evolution operator, see for example \cite{peskin}. The brackets representing the momenta state should be regarded, as we mentioned in Chapter \ref{2}, as truly quantum mechanical states in some Fock space of particles having some value of momentum. As mentioned in this reference, the scattering matrix has the structure of 
\[S= 1 + i T\]
where we have defined the matrix $T$, which represents the transition part of $S$, \textit{i.e}, the $S$-matrix has an identity component, which expresses that particles can miss each other, even when some interaction is taking place, whereas $T$ encodes the information related to have a transition mediated by the interaction. Since energy-momentum must be conserved there will always be a delta-function in the definition of $T$, such that we can define from $T$ the transition matrix $M$
\begin{equation}
\bra{p'_1p '_2}T\ket{p_1p_2} = (2\pi)^4\delta^{(4)}(p_1+ p_2 - p'_1 - p'_2)  \bra{p'_1p '_2}\mathcal{M}\ket{p_1p_2} 
\end{equation}

Now we must relate the elements of the matrix $M$ with the rate for having transitions $i\to f$, \textit{i.e}, scattering events $(p_1,p_2)\to(p'_1,p'_2)$. Using the definition of the differential cross section, we are looking into scattering events that leads to the final state $f=(p'_1,p'_2)$. Thus, in order to define a differential transition probability, we must use the infinitesimal volume element in momentum space $\id\mathbf{p}$, writing
\begin{equation}
	\id R_f = \frac{\id \mathbf{p'_1}}{(2\pi)^32E'_1}\frac{\id \mathbf{p'_2}}{(2\pi)^32E'_2}M(p_1, p_2 \to p'_1, p'_2) (2\pi)^4\delta^{(4)}(p_1+ p_2 - p'_1 - p'_2) 
\end{equation}
where we have defined the transition amplitude
\[M(p_1, p_2 \to p'_1, p'_2)\coloneqq |\bra{p'_1p '_2}\mathcal{M}\ket{p_1p_2} |^2\]

Above, the extra factors of $2\pi$ and $E$ comes from the normalization of the one-particle states (see for example \cite{mirco}), where
\[\bra{p}\ket{p'}=(2\pi)^3\, 2E\, \delta^{(3)}(\mathbf{p}-\mathbf{p'})\]
so that we always divide the measure by $(2\pi)^3\, 2E $ to have the result normalized.

According to \cite{cross, cross2, mirco}, the invariant incident flux of particles is given by\footnote{The reader may note that the expression given in \cite{peskin} (and others), for example, is slightly different from the one we give (there, $\mathcal{F}=4E_1E_2|\mathbf{v_1}-\mathbf{v_2}|$). As carefully discussed by \cite{mirco}, this flux is not Lorentz invariant and only holds for collinear velocities (this should require the use of the center of momentum frame or rest frames, for instance.). However, this expression is invariant under boosts along the $z$-direction.}\textsuperscript{,}\footnote{The task of deriving a Lorentz invariant expression for the flux which holds in any inertial reference frame is, as \cite{mirco} notes, not so straightforward and was first proposed by M\o ller in \cite{moller}.}
\begin{equation}\label{invflux}
	\mathcal{F} = 4\sqrt{(p_1\cdot p_2)^2-m^2_1m^2_2}
\end{equation}

This enables us to write for the differential scattering cross section
\begin{equation}\label{diffcr2}
\id \sigma=\frac{1}{4\sqrt{(p_1\cdot p_2)^2-m^2_1m^2_2}}\frac{\id \mathbf{p'_1}}{(2\pi)^32E'_1}\frac{\id \mathbf{p'_2}}{(2\pi)^32E'_2}M(p_1, p_2 \to p'_1, p'_2) (2\pi)^4\delta^{(4)}(p_1+ p_2 - p'_1 - p'_2) 
\end{equation}

The equation above is far from being trivial, but should not confuse the reader. Recall that we started from a suitable definition for the cross section, that is the \textit{transition rate per flux} \eqref{cr}, where we gave meaning to the transition rate $R_f$ and the incident flux $\mathcal{F}$ in terms of the incoming momenta and the transition part, $T$, of the $S$-matrix. We have also seen that the term \textit{differential} in the cross section definition is actually related to restricting our final state to some labeled pair of \textit{out} states $(p'_1,p'_2)$, where we must, then, drop the integration over the final states, giving meaning to $\id R_f$ in \eqref{diffcr}. The exact derivation of \eqref{diffcr2} is something important but not subject of the present work, however, we invite the reader to check \cite{peskin, weinberg, srednicki} for more details. The construction of \eqref{diffcr2} involves working in the language of Quantum Field Theory, where we can construct quantum states to the particles in some Fock space. It is worth noting, however, that once an equation such as \eqref{diffcr2} is established, we can start deriving expressions for the differential cross section, after calculating the transition amplitude $M$ of course. That is the subject of next section.

\section{Klein-Nishina cross section}

In this section we shall deal with scattering of photons and electrons, following mainly \cite{jauch, cross, cross2}. As usual, we denote the collision scheme as
\[p+k\to p'+k'\]
Let us then begin by recalling the definition made in Chapter \ref{2} of the  Mandelstam variables
\begin{align}
	s&\coloneqq(p+k)^2 = 2p\cdot k + m_e^2 \label{mandels}\\
	 &=(p'+k')^2= 2p'\cdot k' + m^2_e \nonumber\\
	t&\coloneqq(k- k')^2=-2k\cdot k' \label{mandelt}\\
	&=(p'-p)^2= - 2p\cdot p' + 2m^2_e \nonumber\\
	u&\coloneqq (p'-k)^2 = -2p'\cdot k + m_e^2  \label{mandelu}\\
	&=(p-k')^2=- 2p\cdot k' + m^2_e \nonumber
\end{align}
as we mentioned, these are Lorentz invariant quantities which explore the collisional degrees of freedom. For each Mandelstam variable, as we noted before, there is an extra relation obtained by using energy-momentum conservation (that is the second line in each of the definition). Since in the definition of the differential cross section, \eqref{diffcr2}, we have energy-momentum conservation guaranteed by the four-delta, we can use either expression for each of the Mandelstam variables. 

The transition amplitude $M$ can be calculated to leading order\footnote{Note that now we are using the jargon of Quantum Field Theory, where one usually performs something called \textit{perturbative expansion} to calculate these amplitudes to some desired order in the interaction coupling. The terms appearing in this expansion are represented by the so-called Feynman diagrams of the theory, see \cite{peskin, weinberg}.} by drawing the \textit{tree-level} Feynman diagrams. After taking the average over all polarization states of the photon (recall that we are disregarding polarization), the transition amplitude for Compton scattering can be written in terms of the Mandelstam variables (see \cite{cross, cross2}) as
\begin{align}
M^{\text{KN}}(s,u)=12\pi m^2_e\sigma_T\bigg{\{}\left(\frac{2m^2_e}{s-m_e^2} + \frac{2m^2_e}{u-m^2_e}\right)^2 + 2 \bigg{(}\frac{2m^2_e}{s-m_e^2} &+ \frac{2m^2_e}{u-m^2_e}\bigg{)} \nonumber\\
- &\frac{u-m_e^2}{s-m_e^2} - \frac{s-m_e^2}{u-m_e^2}\bigg{\}} \label{Mmand}
\end{align}
where $m_e$ is the mass of the electron and $\sigma_T$ is the total Thomson cross section as we have seen previously. Upon using the Mandelstam variables in expression above we already have in mind energy-momentum conservation. As matter of fact, recall that in definition \eqref{diffcr2} of differential cross section we have the delta-function to impose energy-momentum conservation, so that at the level of cross sections, this conservation always holds. Therefore, in the transition amplitude \eqref{Mmand}, we can use either expression in \eqref{mandels}, \eqref{mandelt} and \eqref{mandelu} for the Mandelstam variables. Expressing $s,u$-variables in terms of momenta we find
\begin{align}
	M^{\text{KN}}(p, k \to p', k')=12\pi m^2_e\sigma_T\bigg{\{}\left(\frac{m^2_e}{p\cdot k} - \frac{m^2_e}{p\cdot k'}\right)^2 + 2 \bigg{(}\frac{m^2_e}{p\cdot k} - \frac{m^2_e}{p\cdot k'}\bigg{)} + \frac{p\cdot k'}{p\cdot k } + \frac{p\cdot k}{p\cdot k'} \bigg{\}} \label{Mmand2}
\end{align}
or
\begin{align}
	M^{\text{KN}}(p, k \to p', k')=12\pi m^2_e\sigma_T\bigg{\{}\left(\frac{m^2_e}{p'\cdot k'} - \frac{m^2_e}{p'\cdot k}\right)^2 + 2 \bigg{(}\frac{m^2_e}{p'\cdot k'} - \frac{m^2_e}{p'\cdot k}\bigg{)} + \frac{p'\cdot k}{p'\cdot k' } + \frac{p'\cdot k'}{p'\cdot k} \bigg{\}} \label{Mmand3}
\end{align}
by using the second line in \eqref{mandels}, \eqref{mandelu}. From \eqref{Mmand2} and \eqref{Mmand3} we clearly see dynamical reversibility holding, since starting from \eqref{Mmand2} and performing the relabels to invert the \textit{in} and \textit{out} states: $p \to p'$ and $k \to k'$, we would end up with \eqref{Mmand3}, so that
\begin{equation}
		M^{\text{KN}}(p, k \to p', k') = 	M^{\text{KN}}(p', k' \to p, k)
\end{equation}
as expected. This, of course, leads to the dynamical reversibility of the transition rates as defined in \eqref{transrates2}. More generally, dynamical reversibility is a property inherited from the unitarity of the S-matrix, as we mentioned previously and as shown in \cite{kolkata}.

Finally, we can write the differential cross section as
\begin{equation}\label{knM}
\id \sigma^{\text{KN}}=\frac{1}{2(s-m_e^2)}M^{\text{KN}}(s,u)\frac{\id \mathbf{p'}}{(2\pi)^3 2E'}\frac{\id \mathbf{k'}}{(2\pi)^3 2\omega'} (2\pi)^4\delta^{(4)}(p+ k - p' - k')	
\end{equation}
where we have rewritten the flux\footnote{Since $p\cdot k >0$, we do not need to worry about a possible absolute value that may appear in the flux. In fact it is easy to see the inequality if we move to the center of momentum frame. In this frame it can be shown that $p_{\text{cm}} \cdot k_{\text{cm}} = (E_{\text{cm}} +\omega_{\text{cm}})\omega_{\text{cm}}$, which is a positive quantity. Now, since $s$ is a Lorentz invariant, it must be also positive in any other frame.} \eqref{invflux} in terms of the Mandelstam variable $s$.

Our task now is to write equation \eqref{knM} in terms of more physical quantities. Recall that our ultimate goal is to find an expression in terms of the scattering solid angle
\[\id \sigma^{\text{KN}} =\frac{\id \sigma}{\id \Omega}^{\text{KN}}\id \Omega\]

However, as we have seen it is useful to express things in a Lorentz invariant way, so that we not only explore the Lorentz invariance of the cross section but also have an easy point of departure to dialogue between different reference frames. The natural way of doing that is to use the Mandelstam variable $t$ instead of the scattering solid angle $\Omega$
\[\id \sigma^{\text{KN}} =\frac{\id \sigma}{\id t}^{\text{KN}}\id t\]
from its expression \eqref{mandelt}, $t$ is clearly related to the scattering solid angle, $\mathbf{k}\cdot\mathbf{k'}$.

To simplify \eqref{knM} we must eliminate the delta-function and we can do this by going to the total cross section
\begin{equation}\label{knM2}
\sigma^{\text{KN}}=\int\frac{\id \mathbf{p'}}{(2\pi)^3 2E'}\frac{\id \mathbf{k'}}{(2\pi)^3 2\omega'}\frac{1}{2(s-m_e^2)}M^{\text{KN}}(s,u) (2\pi)^4\delta^{(4)}(p+ k - p' - k')	
\end{equation}
where we compute some integrals while using the delta function to reduce the degrees of freedom of the measure. As we have seen, the total cross section is a Lorentz invariant quantity, so that we can perform this integral in any convenient frame of reference. Mimicking what we have done in Section \ref{cov-nonc}, we shall work in the center of momentum frame

Using again identity \eqref{3to4delta}, we can perform the integral over the electron momentum, yielding
\begin{equation}\label{knM3}
	\sigma^{\text{KN}}=\frac{1}{4(2\pi)^2}\int\frac{\id \mathbf{k'_{\text{cm}}}}{ \omega'_{\text{cm}}}\frac{M^{\text{KN}}(s,u)}{(s-m_e^2)} \delta(2( k'_{\text{cm}}\cdot(p_{\text{cm}} + k_{\text{cm}}) - p_{\text{cm}}\cdot k_{\text{cm}}))
\end{equation}
moving to spherical coordinates and aligning our $z$-axis in the direction of the incoming photon, we have $\id \mathbf{k'_{\text{cm}}} = \id\Omega_{\text{cm}}\, {\omega'_{\text{cm}}}^2\,  \id\omega'_{\text{cm}}$. On the other hand, the delta-function is simplified, yielding
\[ \delta(2( k'_{\text{cm}}\cdot(p_{\text{cm}} + k_{\text{cm}}) - p_{\text{cm}}\cdot k_{\text{cm}}))=\frac{1}{2\sqrt{s}}\delta\left(\omega'_{\text{cm}} - \frac{s-m_e^2}{2\sqrt{s}}\right)\]
where we used \eqref{deltasimpl} from Section \ref{cov-nonc}. Using all that in \eqref{knM3} and performing the integral over the radial direction enables the expression	
\begin{equation}\label{knM4}
	\sigma^{\text{KN}}=\frac{1}{16(2\pi)^2}\int\id \Omega_{\text{cm}}\frac{M^{\text{KN}}(s,u)}{s}
\end{equation}

Looking at \eqref{mandelt}, we have
\begin{align*}
	\id t &= 2\omega_{cm}\omega'_{cm} \id \cos \theta_{\text{cm}}\\
	&=\frac{1}{\pi }\omega_{cm}\omega'_{cm} \id \Omega_{\text{cm}}\\
	&=\frac{1}{\pi }\omega^2_{cm} \id \Omega_{\text{cm}}
\end{align*}
where we have used that $\id \cos \theta_{\text{cm}} =\frac{1}{2\pi }  \id \Omega_{\text{cm}}$ because of azimuthal symmetry and that
\[\omega_{cm}=\omega'_{cm}\]
because of conservation of energy-momentum. In fact we can check this by observing that energy-momentum conservation yields
\begin{align}
	p+k=p'+k' \implies &p'^2 = (p + k - k')^2 \nonumber\\
	 &m_e^2 = m_e^2 + 2p\cdot k - 2p\cdot k'- 2k\cdot k'\nonumber\\
\end{align}
so that
\begin{align}
	p+k=p'+k'  \implies &p\cdot k= p\cdot k' + k \cdot k'\label{conservation}
\end{align}
but it holds that
\[\mathbf{p_{\text{cm}}} + \mathbf{k_{\text{cm}}}= 0 =\mathbf{p'_{\text{cm}}} + \mathbf{k'_{\text{cm}}}\] 
in the center of momentum frame, so that we can write
\begin{align*}
	& p_{\text{cm}}\cdot k_{\text{cm}} = E_{\text{cm}}\omega_{\text{cm}} + \omega^2_{\text{cm}}\\
	& p_{\text{cm}}\cdot k'_{\text{cm}}= E_{\text{cm}}\omega'_{\text{cm}} + \mathbf{k_{\text{cm}}}\cdot\mathbf{k'_{\text{cm}}}\\
	&  k_{\text{cm}} \cdot k'_{\text{cm}} = \omega_{\text{cm}}\omega_{\text{cm}}' - \mathbf{k_{\text{cm}}}\cdot\mathbf{k'_{\text{cm}}} 
\end{align*}
yielding
\begin{equation}\label{compshiftcom}
\omega_{cm}=\omega'_{cm}
\end{equation}
in \eqref{conservation}. Substituting $\id \Omega_{\text{cm}}$ back in \eqref{knM4} we can write
\begin{equation}\label{knM5}
	\sigma^{\text{KN}}=\frac{1}{16\pi}\frac{1}{4}\int\id t \frac{M^{\text{KN}}(s,u)}{s\,\omega^2_{cm}} = \frac{1}{16 \pi}\int\id t \frac{M^{\text{KN}}(s,u)}{(s-m_e^2)^2}
\end{equation}
where we used that
\[\omega_{\text{cm}}=\frac{s-m_e^2}{2\sqrt{s}}\]
as we find using the definition of $s$ or looking at \eqref{s-frequency} in Section \ref{cov-nonc}. Finally, we invoke the \textit{Fundamental Theorem of Calculus} to write
\begin{equation}\label{knM6}
	\frac{\id\sigma}{\id t}^{\text{KN}} = \frac{1}{16 \pi (s-m_e^2)^2}M^{\text{KN}}(s,u)
\end{equation}
where the transition amplitude is given by \eqref{Mmand}.

This expression is Equation \eqref{crosec} in Section \ref{cov-nonc}, which was the starting point for showing the equivalence between both descriptions of the relativistic Boltzmann equation. Here, we have seen how to start from the natural definition of the differential cross section, while reducing the degrees of freedom with the four-delta to find an expression in terms of the Mandelstam variables for the Klein-Nishina differential cross section. The above expression explicitly exhibits the Lorentz invariance of $\id\sigma^{\text{KN}}$ (since all quantities are expressed in terms of $s,u,t$).

From \eqref{knM6} we can calculate the differential cross section in any inertial frame of reference\footnote{This is a common trick in Special Relativity, where we write things in a Lorentz invariant way to extend the definition of quantities to any inertial frame of reference.} and, in order to do that, we just need to express the Mandelstam variables in the frame we are interested. To see how that works, let us move to the electron rest frame\footnote{We emphasize again that some references call this the laboratory frame.}. In this frame, the observer sees the collision
\begin{figure}[H]
	\centering
	\includegraphics[width=0.3\linewidth]{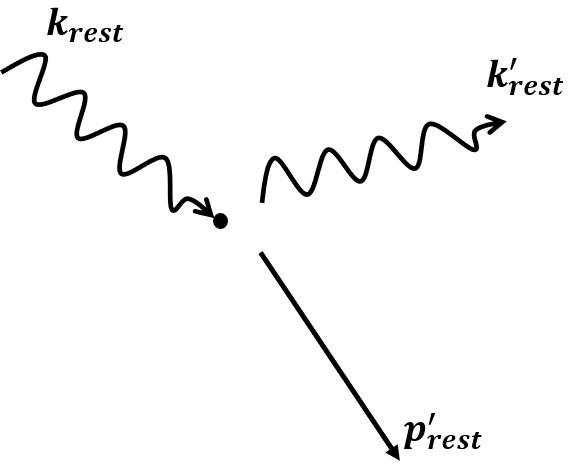}
	\caption{Photon-electron scattering as seen in the rest frame of the electron. In this frame, the initial electron momentum is zero $\mathbf{p_{\text{rest}}} =0$.}
	\label{restframe}
\end{figure}
Therefore we can write for the four-momenta
\begin{alignat*}{2}
	&k_{\text{rest}}=(\omega_{\text{rest}}, \mathbf{k_{\text{rest}}}) \ ; \ \  &&k'_{\text{rest}}=(\omega'_{\text{rest}}, \mathbf{k'_{\text{rest}}})\\
	&p_{\text{rest}}=(E_{\text{rest}}, 0)=(m_e,0) \ ; \ \  &&p'_{\text{rest}}=(E'_{\text{rest}}, \mathbf{p'_{\text{rest}}})
\end{alignat*}
yielding for the Mandelstam variables
\begin{alignat*}{2}
& s = 2E_{\text{rest}}\omega_{\text{rest}} +  m_e^2 &&\implies (s-m_e^2) = 2m_e\omega_{\text{rest}}\\
& t = -2\omega_{\text{rest}} \omega_{\text{rest}}'( 1 - \mathbf{\hat{n}_{\text{rest}}}\cdot \mathbf{\hat{n}'_{\text{rest}}}) \\
& u =-2E_{\text{rest}}\omega'_{\text{rest}} +m_ e^2 &&\implies (u - m_ e^2) = -2m_e\omega'_{\text{rest}}\\
\end{alignat*}
where $\mathbf{\hat{n}_{\text{rest}}} = \mathbf{k_{\text{rest}}}/\omega_{\text{rest}}$ (analogously for $ \mathbf{\hat{n}'_{\text{rest}}}$) and we define the rest frame scattering angle $\mathbf{\hat{n}_{\text{rest}}}\cdot \mathbf{\hat{n}'_{\text{rest}}} = \cos\theta_{\text{rest}}$. 

Thus, the transition amplitude \eqref{Mmand} can be written as
\begin{align}
	M^{\text{KN}}_{\text{rest}}=12\pi m^2_e\sigma_T\bigg{\{}\left(\frac{m_e}{\omega_{\text{rest}}} - \frac{m_e}{\omega'_{\text{rest}}}\right)^2 + 2 \bigg{(}\frac{m_e}{\omega_{\text{rest}}} - \frac{m_e}{\omega'_{\text{rest}}}\bigg{)} + \frac{\omega'_{\text{rest}}}{\omega_{\text{rest}} } + \frac{\omega_{\text{rest}}}{\omega'_{\text{rest}}}\bigg{\}} \label{Mrest}
\end{align}

To find the differential cross section in terms of the scattering angle we must change variables
\[t \to \Omega_{\text{rest}}\]
looking the $t$-variable expression we see that it depends on $\omega'_{\text{rest}}$ and, differently than in Section \ref{cov-nonc} and before, energy-momentum conservation is taking place now, so that $\omega'_{\text{rest}} = \omega'_{\text{rest}}(\Omega_{rest})$.	

Let us then start with \eqref{conservation} while expressing quantities in the rest frame of the electron to find a relation for $\omega'_{\text{rest}}$. Equation \eqref{conservation} can be written as
\begin{equation}\label{comptshiftrest}
	m_e\omega_{\text{rest}} = m_e\omega'_{\text{rest}} + \omega_{\text{rest}} \omega'_{\text{rest}}(1-\cos\theta_{\text{rest}}) \implies \omega'_{\text{rest}} = \frac{ \omega_{\text{rest}}}{1 + \frac{\omega_{\text{rest}}}{m_e}(1-\cos\theta_{\text{rest}})}
\end{equation}
this is the well-known Compton formula when the electron is initially at rest. Using that in $t$ we have
\[t = \frac{-2\omega^2_{\text{rest}}(1-\cos\theta_{\text{rest}})}{1 + \frac{\omega_{\text{rest}}}{m_e}(1-\cos\theta_{\text{rest}})}\]
which yields
\begin{equation}\label{trest}
\frac{\id t}{\id\cos\theta_{\text{rest}}} = \frac{2\omega^2_{\text{rest}}}{\left(1 + \frac{\omega_{\text{rest}}}{m_e}(1-\cos\theta_{\text{rest}})\right)^2} = 2{\omega'}^2_{\text{rest}}
\end{equation}

Since we are dealing with unpolarized photons, the scattering respects azimuthal symmetry
\[\id \Omega_{\text{rest}} = 2\pi \id \cos\theta_{\text{rest}}\]
and we write for $\id t$
\begin{equation}\label{trest1}
	\id t = \frac{1}{\pi}{\omega'}^2_{\text{rest}} \id \Omega_{\text{rest}}
\end{equation}

Using that and $s$ in \eqref{knM6} gives the differential cross section in terms of the scattering solid angle
\begin{equation}\label{knrest}
\frac{\id\sigma}{\id \Omega}^{\text{KN}}_{\text{rest}} = \frac{1}{16\pi^2 4m_e^2}\left(\frac{\omega'_{\text{rest}}}{\omega_{\text{rest}}}\right)^2M^{\text{KN}}_{\text{rest}}
\end{equation}

Finally, we observe that a slight rearrange of \eqref{comptshiftrest} yields
\begin{equation}\label{comptshiftrest2}
\frac{m_e}{\omega_{\text{rest} }} - \frac{m_e}{\omega'_{\text{rest}}} = (1-\cos\theta_{\text{rest}})
\end{equation}
and we recognize the first and second parcel appearing in \eqref{Mrest}. Thus, simplifying terms in \eqref{knrest} and \eqref{Mrest} we find the famous \textit{Klein-Nishina differential cross section} in the rest frame of the electron \cite{kleinnishina}
\begin{equation}\label{knrest1}
	\frac{\id\sigma}{\id \Omega}^{\text{KN}}_{\text{rest}} = \frac{3\sigma_T}{16\pi }\left(\frac{\omega'_{\text{rest}}}{\omega_{\text{rest}}}\right)^2\left[ \frac{\omega'_{\text{rest}}}{\omega_{\text{rest}} } + \frac{\omega_{\text{rest}}}{\omega'_{\text{rest}}} - \sin^2\theta_{\text{rest}}\right]
\end{equation}

When the scattering is non-relativistic ($\omega_{\text{rest}}\ll m_e$), the Compton shift is negligible ($\omega_{\text{rest}}\approx \omega'_{\text{rest}}$) and \eqref{knrest1} gives 
\begin{equation}\label{threst}
	\frac{\id\sigma}{\id \Omega}^{\text{Th}}_{\text{rest}} = \frac{3\sigma_T}{16\pi }\left(1+ \cos^2\theta_{\text{rest}}\right)
\end{equation}
this is the \textit{Thomson differential cross section} in the rest frame of the electron, usually referred in literature \cite{jauch} as the non-relativistic limit of the Klein-Nishina differential cross section, as we have seen here.

Now that we have witnessed how expression \eqref{knM6} works, it only remains to show how we can express the Klein-Nishina differential cross section in the very general frame we call laboratory frame, that is the frame where the observer sees the collision

The four-momenta are now given by
\begin{alignat*}{3}
	&k=(\omega, \mathbf{k}) &&; \ \  &&k'=(\omega', \mathbf{k'})\\
	&p=(E, \mathbf{p}) &&; \ \  &&p'=(E', \mathbf{p'})
\end{alignat*}
the generality of this frame allows particles to have any initial momenta (of course final momenta are bound by energy-momentum conservation) and will give rise to the covariant version (sometimes we will also call it the \textit{frame-independent} version) of the Klein-Nishina differential cross section, which holds in any inertial frame of reference.

As before we can express the Mandelstam variables as
\begin{alignat*}{2}
	& s = 2E\omega(1-\mathbf{v}\cdot\mathbf{\hat{n}}) +  m_e^2\\
	& t = -2\omega\omega'( 1 - \mathbf{\hat{n}}\cdot \mathbf{\hat{n}'}) \\
	& u = -2E\omega'(1-\mathbf{v}\cdot\mathbf{\hat{n'}}) +m_ e^2\\
\end{alignat*}
where we have  defined the scattering angle according to the following scheme
\begin{figure}[H]
	\centering
	\includegraphics[width=0.5\linewidth]{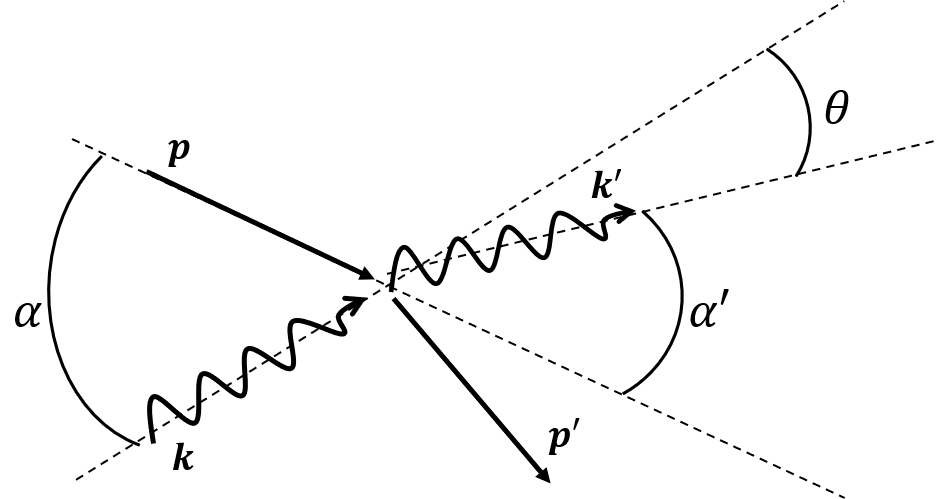}
	\caption{Collision scheme in the laboratory frame, highlighting the scattering angle $\theta$ and angles of the incoming electron momentum with the incoming photon, $\alpha$, and outgoing photon, $\alpha'$.}
	\label{collscheme}
\end{figure}
Thus, $\mathbf{\hat{n}}\cdot \mathbf{\hat{n}'}=\cos \theta$, with $\mathbf{\hat{n}}^{(')} = \mathbf{k}^{(')}/\omega^{(')}$ the unit vector along the direction of the photon movement.

The outgoing photon frequency is not independent from the incoming one, they are connected, of course, by conservation of energy-momentum. Hence, using \eqref{conservation}, we write
\begin{equation*}
	E\omega(1-\mathbf{v}\cdot\mathbf{\hat{n}}) = E\omega'(1-\mathbf{v}\cdot\mathbf{\hat{n'}}) + \omega\omega'( 1 - \mathbf{\hat{n}}\cdot \mathbf{\hat{n}'})
\end{equation*}
or, rearranging terms
\begin{equation}\label{shift1}
	\omega'= \frac{\omega(1-\mathbf{v}\cdot\mathbf{\hat{n}})}{1-\mathbf{v}\cdot\mathbf{\hat{n'}} + \frac{\omega }{E}( 1 - \mathbf{\hat{n}}\cdot \mathbf{\hat{n}'})}
\end{equation}

Restoring, for a brief moment, SI units we can express above equation as
\begin{equation}\label{shift}
	\omega' - \omega = \frac{c\mathbf{p}\cdot(\mathbf{\hat{n}'} - \mathbf{\hat{n}}) - \hbar\omega (1- 
		\mathbf{\hat{n}}\cdot\mathbf{\hat{n}'})}{\gamma_v m_ec^2 \left[1 - \mathbf{p}\cdot\mathbf{\hat{n}'}/\gamma_v m_ec + 
		(\hbar\omega/\gamma_v m_e c^2) (1 - \mathbf{\hat{n}} \cdot \mathbf{\hat{n}'})\right]}\, \omega
\end{equation}
where we used the initial momentum of the electron instead of the velocity. Of course that in SI units they are related by
\[\frac{\mathbf{v}}{c} = \frac{c \mathbf{p}}{\gamma_v m_e c^2}\]
\eqref{shift} is the \textit{Compton shift} equation \cite{longair}, less well-known to the case where the electron is not initially at rest. Expression \eqref{shift} will be important in Chapter \ref{4} and \ref{5}.

It is also interesting to note that \eqref{shift1} and \eqref{shift} both hold in any inertial reference frame, so that, in this sense, it is a \textit{frame-independent} expression. As a matter of fact, \eqref{shift1} and \eqref{shift} are both Lorentz invariant expressions, due to the invariance of \eqref{conservation} (recall that the four-product is clearly a Lorentz invariant quantity). Quantities with the same property as \eqref{shift1} and \eqref{shift}, that is the property of finding expressions for different frames by simply writing quantities in that particular frame, we sometimes refer to as \textit{frame-independent}\footnote{Of course that, formally, this is just a manifestation of Lorentz invariance or covariance. Nevertheless, frame-independence avoids the, sometimes confusing, jargon.}.  

So, for example, if we require $\mathbf{v}=0$, setting the Lorentz factor to $\gamma_v=1$, so that we are in the rest frame of the electron, \eqref{shift1} and \eqref{shift} reduce to \eqref{comptshiftrest}. Similarly we can do that for the center of momentum frame to find \eqref{compshiftcom}.

Now going back to \eqref{knM6} and using \eqref{shift1}, the $t$ variable can be written as
\begin{equation}\label{tlab}
t= \frac{-2\omega^2(1 - |\mathbf{v}|\cos\alpha)(1-\cos\theta)}{1-|\mathbf{v}|\cos\alpha' + \frac{\omega}{E}(1-\cos\theta)}
\end{equation}
where we have used the definition of the angle between the incoming electron-incoming photon $\alpha$, incoming electron-outgoing photon $\alpha'$ (see Figure \ref{collscheme}). To find the derivative with respect to the scattering angle, we must note that $\alpha'$ is not independent from $\theta$, in fact, if $\phi$ is the angle between the planes formed by $\mathbf{k},\mathbf{p}$ and $\mathbf{k},\mathbf{k'}$ we have (see \cite{jauch})
\[\cos \alpha'= \cos\alpha \cos \theta + \sin \alpha \sin\theta \cos \phi\]
using that in \eqref{tlab} and rearranging terms similarly as before enables the expression
\begin{equation}\label{tlab2}
	\frac{\id t}{\id \cos \theta}= 2\left[\frac{\omega(1 - |\mathbf{v}|\cos\alpha)}{1-|\mathbf{v}|\cos\alpha' + \frac{\omega}{E}(1-\cos\theta)}\right]^2 = 2{\omega'}^2
\end{equation}
leading to
\[\id t= \frac{1}{\pi} {\omega'}^2 \id\Omega.\]

Using $\id t$ and $s$ we can write for \eqref{knM6}
\begin{equation}\label{knlab}
	\frac{\id\sigma}{\id \Omega}^{\text{KN}} = \frac{1}{16 \pi^2 }\frac{1}{4E^2}\left(\frac{ \omega'}{\omega}\right)^2\frac{1}{(1-\mathbf{v}\cdot\mathbf{\hat{n}})^2}M^{\text{KN}}(s,u)
\end{equation}
while the transition amplitude is given substituting $s$ and $u$. For now, we write as before
\begin{equation}
M^{\text{KN}}=12\pi m^2_e\sigma_T\bigg{\{}\left(\frac{m^2_e}{p\cdot k} - \frac{m^2_e}{p\cdot k'}\right)^2 + 2 \bigg{(}\frac{m^2_e}{p\cdot k} - \frac{m^2_e}{p\cdot k'}\bigg{)} + \frac{p\cdot k'}{p\cdot k } + \frac{p\cdot k}{p\cdot k'} \bigg{\}} \label{Mmand22}
\end{equation}

By using \eqref{shift1} we can write the prefactor in front of the transition amplitude in \eqref{knlab} as 
\begin{equation}\label{prefator}
\frac{1}{16 \pi^2 }\frac{1}{4E^2}\left(\frac{ \omega'}{\omega}\right)^2\frac{1}{(1-\mathbf{v}\cdot\mathbf{\hat{n}})^2} = \frac{1}{16 \pi^2 }\frac{1}{4m_e^2}\left[\frac{1}{\gamma_v(1-\mathbf{v}\cdot \mathbf{\hat{n}'} + \frac{\omega}{E}(1-\mathbf{\hat{n}}\cdot \mathbf{\hat{n}'}))}\right]^2
\end{equation}

On the other hand, substituting the four-momenta and \eqref{shift1} in the transition amplitude gives for each parcel in \eqref{Mmand22}
\begin{align}
	&\frac{p\cdot k '}{p\cdot k} + \frac{p\cdot k }{p\cdot k'} = 1 + \frac{\omega(1-\mathbf{\hat{n}}\cdot \mathbf{\hat{n}'})}{E(1-\mathbf{v}\cdot\mathbf{\hat{n'}})} + \frac{1-\mathbf{v}\cdot\mathbf{\hat{n'}}}{1-\mathbf{v}\cdot\mathbf{\hat{n}} + \frac{\omega}{E}(1-\mathbf{\hat{n}}\cdot \mathbf{\hat{n}'})}\label{parcel1}\\
	&\frac{m_e^2}{p\cdot k} - \frac{m_e^2}{p\cdot k'} = - \frac{1-\mathbf{\hat{n}}\cdot \mathbf{\hat{n}'}}{\gamma_v^2(1-\mathbf{v}\cdot\mathbf{\hat{n}})(1-\mathbf{v}\cdot\mathbf{\hat{n'}})}\label{parcel2}
\end{align}

Replacing \eqref{parcel1} and \eqref{parcel2} in \eqref{knlab} results in
\begin{align}
M^{\text{KN}}=12\pi m^2_e\sigma_T\bigg{\{}1 + &\left[ 1-\frac{(1-\mathbf{\hat{n}}\cdot\mathbf{\hat{n}'})}{\gamma_v^2(1-\mathbf{v}\cdot\mathbf{\hat{n}})(1-\mathbf{v}\cdot\mathbf{\hat{n}'})} \right]^2\nonumber \\
 &\ \ \ \ \ \  \ \ \  \ +\frac{\omega^2(1-\mathbf{\hat{n}}\cdot\mathbf{\hat{n}'})^2}{E^2(1-\mathbf{v}\cdot\mathbf{\hat{n}'})(1-\mathbf{v}\cdot\mathbf{\hat{n}'} + \frac{\omega}{E}(1-\mathbf{\hat{n}}\cdot\mathbf{\hat{n}'}))}\bigg{\}}\label{knlab2}
\end{align}

Finally, we substitute \eqref{prefator} and \eqref{knlab2} in \eqref{knlab} to find
\begin{align}
	\frac{\id\sigma}{\id \Omega}^{\text{KN}} =	\frac{3\sigma_T}{16 \pi }\frac{1}{\gamma^2_v(1-\mathbf{v}\cdot \mathbf{\hat{n}'} + \frac{\omega}{E}(1-\mathbf{\hat{n}}\cdot \mathbf{\hat{n}'}))^2}\bigg{\{}&1 + \left[ 1-\frac{(1-\mathbf{\hat{n}}\cdot\mathbf{\hat{n}'})}{\gamma_v^2(1-\mathbf{v}\cdot\mathbf{\hat{n}})(1-\mathbf{v}\cdot\mathbf{\hat{n}'})} \right]^2\nonumber \\
	& +\frac{\omega^2(1-\mathbf{\hat{n}}\cdot\mathbf{\hat{n}'})^2}{E^2(1-\mathbf{v}\cdot\mathbf{\hat{n}'})(1-\mathbf{v}\cdot\mathbf{\hat{n}'} + \frac{\omega}{E}(1-\mathbf{\hat{n}}\cdot\mathbf{\hat{n}'}))}\bigg{\}}\label{knlab3}
\end{align}

Above expression is the covariant\footnote{In light of our discussion of Chapter \ref{2}, the full differential cross section, $\id \sigma$ is Lorentz invariant, while the differential cross section in terms of the solid angle, $\frac{\id \sigma}{\id \Omega}$, is ``only" covariant. This happens because we expect the scattering angle to transform non-trivially between different frames.} representation of the Klein-Nishina cross section. It is \textit{frame-independent} in the same sense as \eqref{shift1}, so that finding the expression for this differential cross section in any particular frame is as easy as writing a subscript in the quantities appearing in \eqref{knlab3}, while expressing them in the particular frame. To see how this works, let us move to the rest frame of the electron. Now, every quantity in \eqref{knlab3} must have the subscript ``rest". We know that $\mathbf{v}_{\text{rest}}=0$, $\gamma_{v}=1$, $E=m_e$ while $\id \Omega \to \id \Omega_{\text{rest}}$ and so on. This enables writing \eqref{knlab3} as
\begin{align}
	\frac{\id\sigma}{\id \Omega}^{\text{KN}}_{\text{rest}} =	\frac{3\sigma_T}{16 \pi }\frac{1}{(1 + \frac{\omega_{\text{rest}}}{m_e}(1-\cos\theta_{\text{rest}}))^2}\bigg{\{}1 + \cos^2\theta_{\text{rest}} +\frac{\omega_{\text{rest}}^2(1-\cos\theta_{\text{rest}})^2}{m^2_e(1 + \frac{\omega_{\text{rest}}}{m_e}(1-\cos\theta_{\text{rest}}))}\bigg{\}}\nonumber
\end{align}

Now we use \eqref{comptshiftrest} and \eqref{comptshiftrest2} to find
\begin{align}
	\frac{\id\sigma}{\id \Omega}^{\text{KN}}_{\text{rest}} =	\frac{3\sigma_T}{16 \pi }\left(\frac{\omega'_{\text{rest}}}{\omega_{\text{rest}}}\right)^2\bigg{\{}1 + \cos^2\theta_{\text{rest}} +\frac{\omega_{\text{rest}}\omega'_{\text{rest}}}{m^2_e}\left(\frac{m_e}{\omega_{\text{rest} }} - \frac{m_e}{\omega'_{\text{rest}}}\right)^2                                                                                    \bigg{\}}\label{knrest2}
\end{align}
where a rearranging is in order to find \eqref{knrest1}. Therefore, as we have checked, it is very straightforward to find the expression for the differential cross section in any inertial frame of reference, while starting from \eqref{knlab3}. Again, this is just a consequence of the Lorentz invariance of \eqref{knM6} in the same way that \eqref{shift1} is a consequence of the Lorentz invariance of \eqref{conservation}. As we state in \cite{paper}, expression \eqref{knlab3} for the Klein-Nishina cross section, to which we refer here sometimes as the frame-independent version, appears in \cite{barbosa}, but a derivation is not shown there. Jauch and Rohrlich \cite{jauch} give a derivation, but only express the differential cross section in terms of the scattering matrix. Since we could not find any other reference which contains the exact equation \eqref{knlab3} we have chosen to dedicate this chapter to this discussion. However, we also note here that once established the correct (and not so enlightening) expression above, we only need the first few orders of its expansion.

A dimensional analysis can be done in \eqref{knlab3} to find, in SI units, the expression
\begin{align}
	\label{klein-nishina1}
	\frac{\id\sigma}{\id \Omega}^{\text{KN}}= \frac{3 \sigma_T}{16 \pi}&\frac{1}{\gamma_v^2 \left(1- \mathbf{p}\cdot \mathbf{\hat{n}'}/\gamma_v m_ec + \frac{\hbar\omega}{\gamma m_ec^2} (1-\mathbf{\hat{n}}\cdot\mathbf{\hat{n'}}) \right)^2}\nonumber\\ 
	\Bigg\{ 1 & +\left[1 - \frac{(1-\mathbf{\hat{n}}\cdot\mathbf{\hat{n'}})}{\gamma_v^2( 1 - \mathbf{p}\cdot \mathbf{\hat{n}}/\gamma_v m_ec)(1 - \mathbf{p}\cdot\mathbf{\hat{n}'}/\gamma_v m_ec))}\right]^2 \nonumber \\  
	&\ \ \ \ \ \ \ \ \ +\frac{\left(\frac{\hbar\omega}{\gamma_v m_ec^2} (1-\mathbf{\hat{n}}\cdot\mathbf{\hat{n'}})\right)^2}{(1- \mathbf{p}\cdot \mathbf{\hat{n}'}/\gamma_v m_ec)(1- \mathbf{p}\cdot \mathbf{\hat{n}'}/\gamma_v m_ec + \frac{\hbar\omega}{\gamma_v m_ec^2} (1-\mathbf{\hat{n}}\cdot\mathbf{\hat{n'}}))} \Bigg\}
\end{align}
this is the expression for the Klein-Nishina differential cross section we will use in Chapter \ref{4}. As we will see, \eqref{klein-nishina1} is the correct and natural way of expressing the cross section which appears in the  relativistic Boltzmann equation. The reason is simply due to the fact that, in \eqref{klein-nishina1}, photons and electrons are allowed to have any initial value of momenta, while final momenta is obtained from the initial by energy-momentum conservation. Thus, to be viewed truly as a differential transition probability for repeated scattering processes (as is the case of the Boltzmann equation) we must describe any possible collision scheme\footnote{And not only the ones in which the electron is initially at rest, as it would be the case if we would use the expression for the differential cross section in the rest frame of the electron \eqref{knrest1}.} and \eqref{klein-nishina1} exactly accounts for that (see also the discussion in Section \ref{probcons} and Chapter \ref{7}).

\chapter{The Kompaneets equation}\label{4}
In this chapter we are going to turn our attention to the derivation of the Kompaneets equation, while showing how to perform the diffusion approximation as Kompaneets originally proposed consistently. In fact, this was already shown by us in \cite{paper}, so that many times we will use the notation, words and ideas from this paper. However, it is possible that the reader may find a slightly different exposition of the subject in this present work. In particular, our framework is more direct than that of \citep{kompa, dreicer,weymann,katz,rybicki} because we do not \emph{use} in the derivation itself that the Planck distribution is the stationary solution of the relativistic Boltzmann equation.

The diffusion approximation to the Boltzmann equation of an electron-photon gas to yield the Kompaneets equation can also be proven rigorously by requiring certain conditions on the possible differential cross sections, this was done by \cite{escobedo1}. However, Escobedo and Mischler do not explicitly link these conditions to the Thomson/Klein-Nishina cross section nor do they address the same problems we treat here.

\section{Diffusion approximation to the standard Boltzmann equation}\label{dif}

As originally proposed by Kompaneets \cite{kompa}, let us consider an inert mixture of photons and electrons, interacting via Compton effect. We shall assume that the occupation number distribution function of the photons is isotropic and homogeneous
\[n(t,\mathbf{x},\mathbf{k})=n(t,\omega)\]

Similarly, we also assume that the electrons distribution function is homogeneous and isotropic, given according to Maxwell-Boltzmann at temperature $T$
\begin{equation}\label{maxwell}
	f(t,\mathbf{x},\mathbf{p})\id^3\mathbf{p} = f_{\text{Eq}}(|\mathbf{p}|)\id^3\mathbf{p} = n_e (2\pi m_e k_B T)^{-3/2}\exp\left(-\frac{p_x^2 + p_y^2 + p_z^2}{2m_ek_BT}\right)\id^3\mathbf{p}
\end{equation}
where $n_e,m_e$ is the density of electrons and the electron mass, respectively. We also have chosen to drop the subscripts for a simpler notation.

Then, we can write the relativistic Boltzmann equation \eqref{prebkomp} of this system as
\begin{align}
	\frac{\partial n}{\partial t}(t,\omega) =c\int_{\mathbf{p}}\int_{\Omega}\, \id \mathbf{p}\, \left(1-\frac{\mathbf{v}}{c}\cdot \mathbf{\hat{n}}\right)\, &\id \sigma^{\gamma e}\, \nonumber\\
	&\left(n(t,\omega')f_{\text{Eq}}(|\mathbf{p}'|)	\left(1+ n(t,\omega)\right) - n(t,\omega) f_{\text{Eq}}(|\mathbf{p}|) \left(1+ n(t,\omega')\right)\right)\label{bol-komp}
\end{align}
where we already assumed non-degeneracy of the electron gas.

As we have seen, dynamical reversibility makes this a master-type equation, with the number of photons
\begin{equation}\label{photnumb}
	N=2\int \frac{\id\mathbf{k}}{(2\pi\hbar)^3} n(t,\mathbf{k})=\frac{1}{\pi^2c^3}\int \id\omega \,\omega^2 n(t,\omega)
\end{equation}
being conserved. Above, the factor of $2$ comes from the degeneracy factor of photons.

As the correct cross section, we shall use the general frame-independent expression of the Klein-Nishina differential cross section
\begin{equation}
	\id \sigma^{\gamma e} = \id\Omega\frac{\id \sigma}{\id \Omega}^{\text{KN}}(\mathbf{p},\mathbf{\hat{n}}, \Omega)
\end{equation}
where
\begin{align}
	\label{klein-nishina}
	\frac{\id\sigma}{\id \Omega}^{\text{KN}}(\mathbf{p},\mathbf{\hat{n}}, \Omega)= \frac{3 \sigma_T}{16 \pi}&\frac{1}{\gamma_v^2 \left(1- \mathbf{p}\cdot \mathbf{\hat{n}'}/\gamma_v m_ec + \frac{\hbar\omega}{\gamma m_ec^2} (1-\mathbf{\hat{n}}\cdot\mathbf{\hat{n'}}) \right)^2}\nonumber\\ 
	\Bigg\{ 1 & +\left[1 - \frac{(1-\mathbf{\hat{n}}\cdot\mathbf{\hat{n'}})}{\gamma_v^2( 1 - \mathbf{p}\cdot \mathbf{\hat{n}}/\gamma_v m_ec)(1 - \mathbf{p}\cdot\mathbf{\hat{n}'}/\gamma_v m_ec))}\right]^2 \nonumber \\  
	&\ \ \ \ \ \ \ \ \ +\frac{\left(\frac{\hbar\omega}{\gamma_v m_ec^2} (1-\mathbf{\hat{n}}\cdot\mathbf{\hat{n'}})\right)^2}{(1- \mathbf{p}\cdot \mathbf{\hat{n}'}/\gamma_v m_ec)(1- \mathbf{p}\cdot \mathbf{\hat{n}'}/\gamma_v m_ec + \frac{\hbar\omega}{\gamma_v m_ec^2} (1-\mathbf{\hat{n}}\cdot\mathbf{\hat{n'}}))} \Bigg\}
\end{align}
is the covariant expression of the Klein-Nishina differential cross section we have found and discussed in Chapter \ref{3}.

Upon writing \eqref{klein-nishina}, we have in mind this very general frame of reference (see Figure \ref{labframe}) in which the observer sees the collision scheme
\[\mathbf{p} + \frac{\hbar\omega}{c}\mathbf{\hat{n}} \rightleftharpoons \mathbf{p'} + \frac{\hbar \omega'}{c}\mathbf{\hat{n}'}\]
defining the scattering angle $\cos\theta = \mathbf{\hat{n}}\cdot \mathbf{\hat{n}'}$.

Since this frame of reference is the one where the observation (or experiment) is taking place, we use the convention of calling it the \textit{lab frame} (see Chapter \ref{3}). As we shall see, it would be inconsistent for this set up to use the cross section evaluated in the rest frame of the electron, for example.

Let us assume, exactly as phrased in \cite{paper} and as Kompaneets originally proposed, that: (i) the electrons are in thermal equilibrium at temperature $T$; (ii) the photons are soft, meaning that their energy is very small compared to the rest energy of the electron ($\hbar\omega \ll m_ec^2$), but of same order of the electron bath energy ($\hbar\omega \sim k_BT$) ; (iii) electrons are non-relativistic ($|\mathbf{p}|\ll m_ec$ or $k_BT\ll m_ec^2$) and non-degenerate. By combining (ii) and (iii) we conclude also that energy is transferred in small amounts only, permitting the continuum (or diffusion) approximation\footnote{In fact, the transfer of energy depends both on the incoming $\omega$ and $\mathbf{p}$ as dictated by the Compton formula \eqref{cs} and, to the lowest order, it is proportional to the product of both.}.

Points (ii) and (iii) suggest an expansion in terms of the energy shift. For this purpose we define the dimensionless energy shift
\[
\Delta \coloneqq \frac{\hbar(\omega' - \omega)}{k_B T}
\]
by looking at the Compton shift \eqref{shift}, we have
\begin{equation}
	\label{cs}
	\Delta(\omega, \mathbf{p}) = \frac{\hbar \omega}{k_B T}  \frac{c\mathbf{p}\cdot(\mathbf{\hat{n}'} - \mathbf{\hat{n}}) - \hbar\omega (1- 
		\mathbf{\hat{n}}\cdot\mathbf{\hat{n}'})}{\gamma m_ec^2 \left[1 - \mathbf{p}\cdot\mathbf{\hat{n}'}/\gamma m_ec + 
		(\hbar\omega/\gamma m_e c^2) (1 - \mathbf{\hat{n}} \cdot \mathbf{\hat{n}'})\right]} 
\end{equation}

It will be convenient to make the following natural change of variables
\begin{align*}
	&\omega \to x\coloneqq \frac{\hbar \omega }{k_B T}\\
	&\omega'\to x'\coloneqq \frac{\hbar \omega' }{k_B T}
\end{align*}
returning to the relativistic Boltzmann equation \eqref{bol-komp} and using the variables above, we can express the photon occupation number distribution function up to second order in the energy shift as
\begin{align}
	\label{pdist1}
	&n(x',t)( 1 + n(x,t)) = n(x,t)(1+n(x,t)) + (1+n(x,t))\frac{\partial n}{\partial x} \ \Delta + (1+n(x,t))\frac{\partial^2 n}{\partial x^2} \ \frac{\Delta^2}{2}\\
	\label{pdist2}
	&n(x,t)( 1 + n(x',t)) = n(x,t)(1+n(x,t)) +  n(x,t)\frac{\partial n}{\partial x} \ \Delta + n(x,t)\frac{\partial^2 n}{\partial x^2} \ \frac{\Delta^2}{2}
\end{align}

As we will see in next chapter, this diffusion approximation is nothing more than an instance of a continuous \textit{Kramers-Moyal} expansion. Now, by using conservation of energy in the electron distribution
\[E' = E - 
\Delta \ k_BT\]
we can express the distribution function evaluated in the outgoing momentum in terms of the incoming electron momentum and the energy shift only
\begin{equation}
	\label{edist}
	f_\text{Eq}(|\mathbf{p}'|) = f_\text{Eq}(|\mathbf{p}|)\left(1 + \Delta + \frac{\Delta^2}{2}\right)
\end{equation}

Let us now take \eqref{pdist1}--\eqref{pdist2} and \eqref{edist} to substitute back in \eqref{bol-komp}, obtaining a very concise expression for the spatio-temporal dynamics of the photon occupation number
\begin{align}
	\label{kwi}
	\frac{\partial n}{\partial t} =\left[\frac{\partial n}{\partial x} + n(1+n)\right]I_1(x) +\left[\frac{1}{2}\frac{\partial^2 n}{\partial x^2} + (1+n)\left(\frac{n}{2} + \frac{\partial n}{\partial x}\right)\right]I_2(x) \
\end{align}
where,
\begin{equation}
	I_\ell(x) = c\int \id^3\mathbf{p}\, \id\Omega \left(1 -\frac{ \mathbf{v}}{c}\cdot\mathbf{\hat{n}}\right) \frac{\id\sigma}{\id \Omega}^{\text{KN}}(\mathbf{p},\mathbf{\hat{n}}, \Omega) \, f_\text{Eq}(|\mathbf{p}|) \Delta^\ell
\end{equation}
the calculation of these integrals, with $\ell=1,2$, will be done in Appendix \ref{b}. These integrals will be referred, from now on, as the first ($I_1(x)$) and second ($I_2(x)$) Kompaneets' integrals, and they yield the result
\begin{align}
	\label{firi}
	&I_1(x)= \frac{n_e\sigma_T c\, k_BT}{m_ec^2}\;x(4-x)\\
	\label{seci}
	&I_2(x)= \frac{n_e\sigma_T c\, k_BT}{m_ec^2}\;2x^2
\end{align}

By plugging that back in \eqref{kwi} and performing some standard manipulations we are left with
\begin{equation}
	\label{kompfinal}
	x^2\frac{\partial n}{\partial t}(x,t) = \frac{n_e\sigma_T c\, k_BT}{m_ec^2}\frac{\partial }{\partial x}x^4\left\{\frac{\partial n}{\partial x}(x,t) +n(x,t)(1+n(x,t))\right\}
\end{equation}
which is the Kompaneets equation in terms of the dimensionless variable $x$. We could, of course, substitute back $x=\hbar\omega/k_BT$ to find 
\begin{equation}
	\omega^2\frac{\partial n}{\partial t}(t,\omega)= \frac{n_e\sigma_T 
		c}{m_e c^2}\frac{\partial }{\partial \omega}\omega^4\left\{k_B T 
	\frac{\partial n}{\partial \omega}(t,\omega) + 
	\hbar\left[1+n(t,\omega)\right]n(t,\omega)\right\}
\end{equation}
as desired.

\section{Diffusion approximation to the manifestly covariant Boltzmann equation}\label{covdiff}

In this section we shall demonstrate how to perform the diffusion approximation to the relativistic covariant Boltzmann equation. This section will follow closely the work of \cite{brown}. We will also follow Brown's convention of using natural units, thus, throughout this section $\hbar=c=k_B=1$. By doing that, the four-momenta participating in the collision are
\begin{align*}
	& k = ( \omega, \mathbf{k})\ ; \ \ k'= (\omega', \mathbf{k'})\\
	& p= (E, \mathbf{p}) \ ; \ \  p'= (E', \mathbf{p'})
\end{align*}
where we are looking at the scheme
\[p + k \rightleftharpoons p'+k'\]

We begin by writing the Boltzmann equation for an inert mixture of photons and electrons interacting via Compton effect, this is just equation \eqref{phot-eleccov}, where we assume isotropy and homogeneity as before, yielding

\begin{align}
	\omega\frac{\partial n}{\partial t}(t,\omega) =\int\frac{\id \mathbf{p}}{E} \frac{\id \mathbf{k'}}{\omega'} \frac{\id \mathbf{p'}}{E'}  W^{\gamma e}(p,k\to p', k')\big{(}&n(t,\omega') f_{Eq}(\mathbf{p'})\left(1+ n(t,\omega)\right)\nonumber\\
	 &- n(t,\omega) f_{Eq}(\mathbf{p}) \left(1+ n(t,\omega')\right)\big{)} \label{bol-kompc}
\end{align}

Upon integrating the electron bath, we can write this equation as a Boltzmann-master equation for the photons only

\begin{align}
	\omega\frac{\partial n}{\partial t}(t,\omega) =\int \frac{\id \mathbf{k'}}{\omega'}\left(W(k'\to k)n(t,\omega')\left(1+ n(t,\omega)\right)- W(k \to k') n(t,\omega)\left(1+ n(t,\omega')\right)\right) \label{bol-kompc2}
\end{align}
where the rates are now given by
\begin{align}
&W(k'\to k)= \int\frac{\id \mathbf{p}}{E}\frac{\id \mathbf{p'}}{E'}W^{\gamma e}(p,k\to p', k') f_{Eq}(\mathbf{p'})\label{in}\\
&W(k\to k')= \int\frac{\id \mathbf{p}}{E}\frac{\id \mathbf{p'}}{E'}W^{\gamma e}(p,k\to p', k') f_{Eq}(\mathbf{p})\label{out}
\end{align}

Following the discussion in Section 4 of our work \cite{paper}, we can slightly rewrite Equation \eqref{out} to find
\begin{equation}
	\label{out2}
	W(k\to k')= \int\frac{\id \mathbf{p}}{E}\frac{\id \mathbf{p'}}{E'}\, \, W^{\gamma e}(p,k\to p', k') f_{Eq}(\mathbf{p}')\frac{f_{Eq}(\mathbf p)}{f_{Eq}(\mathbf p')}
\end{equation}
and, if the electrons are in thermal equilibrium with inverse temperature $\beta = 1/T$, we verify that
\begin{equation}\label{conserve}
	\frac{f_{Eq}(\mathbf p)}{f_{Eq}(\mathbf p')} = e^{-\beta(E - E')} = 
	e^{-\beta(\omega' -\omega)} ,
\end{equation}
where the last equality follows from conservation of energy in the collisions.

Inserting \eqref{conserve} into \eqref{out2} yields the \textit{dynamical reversibility relation}\footnote{Sometimes also called the detailed balance relation for the transition rates as noted in Chapter \ref{2}.} for the photon transition rates \eqref{in}, \eqref{out}
\begin{equation}\label{microreversph}
\frac{W(k'\to k)}{W(k\to k')} = e^{-\beta( \omega - \omega')}
\end{equation}

We shall use now Definition \eqref{transrates2} of the transition rates to write
\begin{equation*}
	W^{\gamma e}(p,k\to p', k')= \frac{1}{16(2\pi)^6}M^{\text{KN}}(p,k\to p', k') (2\pi)^4\delta^{(4)}(p + k - p' - k')
\end{equation*}
where $M^{\text{KN}}(p,k\to p', k')$ is the Compton effect scattering amplitude, which we have seen in Chapter \ref{3} that it is given by
\begin{align}
	M^{\text{KN}}(p, k \to p', k')=12\pi m^2_e\sigma_T\bigg{\{}\left(\frac{m^2_e}{p\cdot k} - \frac{m^2_e}{p\cdot k'}\right)^2 + 2 \bigg{(}\frac{m^2_e}{p\cdot k} - \frac{m^2_e}{p\cdot k'}\bigg{)} + \frac{p\cdot k'}{p\cdot k } + \frac{p\cdot k}{p\cdot k'} \bigg{\}} \label{Mmand222}
\end{align}

This enables the expression of \eqref{in} and \eqref{out} in terms of the scattering amplitude
\begin{align*}
	&W(k'\to k)=\frac{1}{4(2\pi)^2} \int\frac{\id \mathbf{p}}{2E}\frac{\id \mathbf{p'}}{2E'}M^{\text{KN}}(p,k\to p', k') \delta^{(4)}(p + k - p' - k')f_{Eq}(\mathbf{p'})\\
	&W(k\to k')= \frac{1}{4(2\pi)^2}\int\frac{\id \mathbf{p}}{2E}\frac{\id \mathbf{p'}}{2E'}M^{\text{KN}}(p,k\to p', k') \delta^{(4)}(p + k - p' - k') f_{Eq}(\mathbf{p})
\end{align*}

At this point, it is convenient to rewrite equation \eqref{bol-kompc2} as 
\begin{align}
\frac{\partial n}{\partial t}(t,\omega) =\int {\omega'}^2 \id \omega'\left(\overline{W}(\omega'\to \omega)n(t,\omega')\left(1+ n(t,\omega)\right)- \overline{W}(\omega \to \omega') n(t,\omega)\left(1+ n(t,\omega')\right)\right) \label{bol-kompc3}
\end{align}
while also defining the isotropic transition rates
\begin{align}
	&\overline{W}(\omega '\to \omega)=\frac{1}{\omega\omega'} \int\id \Omega\,  W(k'\to k)\label{in4}\\
	&\overline{W}(\omega \to \omega')= \frac{1}{\omega\omega'}\int\id \Omega\, W(k\to k')\label{out4}
\end{align}
where in the integration of the photon outgoing momenta, we have aligned our $z$-axis in the direction of its incoming momentum, so that the solid angle $\id \Omega$ is actually the scattering solid angle\footnote{This is the same trick we used in Section \ref{cov-nonc}.}.

Making the usual assumptions that the photons are soft and the electrons are non-relativistic distributed according to \eqref{maxwell}, we can perform, similarly than before, the continuum Kramers-Moyal expansion in the energy shift of these rates. Since this expansion is highly non trivial, we shall address it in Appendix \ref{c}, yielding
\begin{align}
& \overline{W}(\omega '\to \omega)=\frac{n_e\sigma_T}{m_e}\left\{\left(m_e + \frac{7}{4}T\right)\frac{\delta(\omega'-\omega)}{\omega\omega'} - \delta'(\omega'-\omega) + T\delta''(\omega'-\omega) \right\} + O\left(\left(\omega' - \omega\right)^3\right)\\
&\overline{W}(\omega\to \omega')=\frac{n_e\sigma_T}{m_e}\left\{\left(m_e + \frac{7}{4}T\right)\frac{\delta(\omega-\omega')}{\omega\omega'} - \delta'(\omega-\omega') + T\delta''(\omega-\omega') \right\}+ O\left(\left(\omega - \omega'\right)^3\right)
\end{align}
where the primes denote the (formal) derivative of the delta-function with respect to shift $\omega'- \omega$.

Finally, we insert that back in Equation \eqref{bol-kompc3}. After computing the delta-function integrals (see Appendix \ref{c}), we find the Kompaneets equation expressed in natural units
\begin{equation}
	\omega^2\frac{\partial n}{\partial t}(t,\omega)=    \frac{n_e\sigma_T}{m_e}\frac{\partial }{\partial \omega}\omega^4\left\{T \frac{\partial n}{\partial \omega}(t,\omega) + \left[1+n(t,\omega)\right]n(t,\omega)\right\}
\end{equation}
a dimensional analysis is in order to easily retrieve \eqref{ke}.

\section{The structure of the Kompaneets equation}\label{struc}

Reproducing in this paragraph, as mentioned in our work \cite{paper}, the Kompaneets equation is composed by two terms inside the brackets in the right hand side: the first parcel is purely a diffusive term due to Doppler shift experienced by the photons in the reference frame of the electrons. The drift term (second parcel), describes stimulated emission and the Compton recoil. The so-called Comptonization, 
\textit{i.e.}, the redistribution of photon energies scattered by electrons consists of 
two terms indeed, both suppressed by the same factor of $m_ec^2$, the diffusion,
because the electron is moving so slowly (depending on $k_BT/m_ec^2$) that the 
Doppler shift is almost negligible, and the drift, because it is divided by the reduced Compton frequency $\omega_c = m_e c^2/\hbar \approx \SI{7.76e20}{\hertz}$, so that there is barely any recoil.

This equation has also a structure of a continuity equation in the photon number \eqref{photnumb}
\begin{equation}
	\frac{\id N}{\id t} \propto \int_0^\infty \id \omega \, \omega^2\frac{\partial n}{\partial t}(t,\omega) = \int_0^{\infty}\id \omega\, \, \frac{\partial }{\partial \omega}\{\omega^2j_t(\omega)\}=0
\end{equation}
so that the photon number is conserved as it should be (remember that Compton scattering preserves number of photons). Such behavior is not surprising since we start from a relativistic Boltzmann equation which is \textit{ab initio} constructed to preserve number of particles.

The current vanishes in equilibrium and it is easy to check that if we take
\[n(t,\omega)=n_{Eq}(\omega) = \frac{1}{\exp\left(\beta\hbar\omega\right) -1}\]
the current is identically zero
\[j_t(\omega)\equiv 0,\]
since in this case a straightforward calculation gives
\[k_BT\frac{\partial }{\partial \omega}n_{Eq}(\omega) = -\hbar\left[1+n_{Eq}(\omega)\right]n_{Eq}(\omega).\]
If we were to calculate, for example, higher order corrections to this equation, we would expect extra terms in the current. However, as \cite{brown} showed, these relativistic corrections would still retain the structure of a continuity equation, with all extra terms vanishing separately in equilibrium for the Bose-Einstein distribution.

As a matter of fact, the full current vanishing for the Bose-Einstein distribution is indeed a manifestation of \textit{detailed balance} in the Boltzmann equation, where we readily see that requiring 
\[	n_{Eq}(\omega')f_{\text{Eq}}(|\mathbf{p}'|)	\left(1+ n_{Eq}(\omega)\right) = n_{Eq}(\omega) f_{\text{Eq}}(|\mathbf{p}|) \left(1+ n_{Eq}(\omega')\right)\]
implies
\begin{equation}\label{detbal}
	\frac{n_{Eq}(\omega)}{\left(1+ n_{Eq}(\omega)\right)}=e^{\beta(E-E')}\frac{n_{Eq}(\omega')}{\left(1+ n_{Eq}(\omega')\right)}\implies \frac{n_{Eq}(\omega)}{\left(1+ n_{Eq}(\omega)\right)}=e^{\beta(\hbar\omega'-\hbar\omega)}\frac{n_{Eq}(\omega')}{\left(1+ n_{Eq}(\omega')\right)}
\end{equation}
which leads to the Bose-Einstein distribution naturally
\begin{equation}
	e^{\beta\hbar\omega}\frac{n_{Eq}(\omega)}{\left(1+ n_{Eq}(\omega)\right)}=\mathrm{constant} = e^{\beta\mu}=1
\end{equation}
where we used the chemical potential of photons $\mu=0$.

However, a more general argument in the context of a non-equilibrium electron bath should not rely on this fact, because in this case we cannot guarantee Equation \eqref{detbal}, such that a derivation which does not depend on the equilibrium solution from the start is worthwhile.

From the occupation number distribution function of photons we can calculate various important quantities, one of which is the \textit{spectral energy density} 
\[E_\gamma(t,\omega)\propto \hbar\left(\frac{\omega}{c}\right)^3\,n(t,\omega),\]
when the Bose-Einstein (equilibrium) distribution is reached, this spectral density will form the well-know black-body radiation spectrum. If we were to modify Kompaneets equation to yield a different equilibrium solution other than Bose-Einstein, the equilibrium spectral density will no longer be Planckian and we would naturally expect departures from the black-body spectrum. Such modification could be, for example, due to a non-equilibrium electron bath, which, in nature, should modify the spatio-temporal dynamics of the Kompaneets equation. See for example \cite{arca} for an interesting realization of this. For a more detailed discussion of various aspects of the Kompaneets equation, we invite the reader to also check \cite{practical}. 

As we also mention in \cite{paper}, the long time behavior of the Kompaneets equation is very interesting (see \textit{e.g} \cite{burigana1,burigana2} for a numerical code) and we will briefly mention here a feature often omitted in many references, that is the Bose-Einstein condensation. Suppose that we start with a gas having $N_0$ photons, that is, at $t=0$, $N(t=0)=N_0$. We observe now that the Bose-Einstein distribution is solely determined by the parameter $\beta$, \textit{i.e.}, the (inverse) temperature of the electron bath, so that, from the start, the number of photons that ``fits" under the Bose-Einstein curve is fixed by the temperature of the electron bath, and this number is given by
\[N_{BE}(\beta)=\frac{1}{\pi^2c^3}\int \id\omega \,\omega^2 n_{Eq}(\omega)\] 

Therefore, a simple reasoning leads us to conclude that if $N_0>N_{BE}(\beta)$, the remaining photons $N_0-N_{BE}(\beta)$ cannot disappear, since Kompaneets equation is number preserving. In fact, they will form a \textit{Bose-Einstein condensate} at $\omega=0$. For a very detailed and rigorous description of such interesting phenomenon, see \citep{escobedo1, levermore}. For us, it suffices to note here, in addition, that depending on the initial number of photons, the limiting photons as $t\uparrow\infty$ will have a Bose-Einstein distribution component ($n_\text{eq}(\omega)$), as well as a condensate (for $\omega = 0$).

\section{The importance of a consistent description}\label{probcons}

In this section we shall address the consistency problems in many derivations, \textit{e.g.}, \cite{katz, liu, rybicki, zhang} of the Kompaneets equation, including his original paper \cite{kompa}. These problems were already mentioned in Chapter \ref{1} but here we discuss it in a more detailed manner.

Before we start, it is worth emphasizing that these problems mainly (or only) occur in the standard description of the Boltzmann equation while working with the set up proposed by Kompaneets in 1957. To understand why is that, we first observe that problems in consistency have two sources: (i) the negligence of the M\o ller velocity factor and (ii) the consistency in expressing quantities (in particular the scattering cross section) in some inertial frame of reference. As we have seen, the M\o ller prefactor is hidden inside the transition rates in the manifestly covariant formalism and, the use of scattering matrices, usually comes with the preference of working manifestly covariant, so that none particular frame is chosen to express the transition rates. The use of these rates also visualizes the scattering process as a jump process, thus, we are in fact examining transitions
\[(k,p) \leftrightharpoons (k',p')\]
regarding momenta truly as states. This automatically deals with the problem of the frame of reference, not only because we are manifestly covariant and not choosing any particular frame of reference, but also because we will now have the actual transition amplitudes among these states.

On the other hand, the use of differential scattering cross sections introduces yet another problem, that is: in which frame of reference should we express it? The above-mentioned references use the electron rest frame, but this cannot be consistent with using Maxwell-Boltzmann for the distribution of electrons as we will find electrons with any possible velocity and consistency is lost. In order to be viewed as a transition probability for general transitions like in the instance of the scattering matrix, we should use a cross section written in a very arbitrary frame of reference, where photons and electrons are also allowed to have arbitrary momenta, \eqref{klein-nishina} being this expression. Then, we have the freedom to look at any possible collision and not only collisions in which the electrons are initially at rest, making this the correct description of the differential transition probability appearing in the Boltzmann equation.

As we have constructed to work as such, both descriptions of the Boltzmann equation hold in any inertial frame of reference. Thus, we have the freedom to solve, work and perform approximations to this equation as long as we express consistently the quantities appearing in \eqref{bol-komp} or \eqref{bol-kompc} in the frame of reference in which we are working. In the calculations done in this work, this is the lab frame, which is the frame where the electrons are distributed according to Maxwell-Boltzmann and where the observer sees multiple general photon-electron collisions of the type
\[\mathbf{p} + \frac{\hbar\omega}{c}\mathbf{\hat{n}} \rightleftharpoons \mathbf{p'} + \frac{\hbar \omega'}{c}\mathbf{\hat{n}'}\]

The reader should also note that it is not possible to go from the lab frame to a frame where \textit{all} electrons are standing still by making Lorentz transformations, so that if one decides to use the Klein-Nishina cross section evaluated in the electron rest frame, one must be very careful in how to express the electron distribution function. In any case, the set up as proposed by Kompaneets will fail and a new set up should be proposed, avoiding, if possible, repeating scatterings\footnote{We can see from the construction of the Boltzmann equation, that it deals with repeating scatterings among the particles. This type of description should be avoided if one decides to work in a frame where the electron is initially at rest simply because the first collision will make the electron move and, as a matter of logic, the next collision will no longer see this electron at rest.}. To our best knowledge, the only person who worked out the correct derivation of the Kompaneets equation while using a cross section evaluated in the electron rest frame in a completely original manner is the Nobel laureate J. Peebles in his book \cite{peebles}. Being a completely different approach, his derivation is not comparable to the one proposed by Kompaneets and that we address here.

To illustrate the discussion, let us consider the Boltzmann equation written in the following way
\begin{align}
	\frac{\partial n}{\partial t}(t,\omega) =c\int_{\mathbf{p}}\int_{\Omega}\, \id \mathbf{p}\,\id w\, \left(n(t,\omega')f_{\text{Eq}}(|\mathbf{p}'|)	\left(1+ n(t,\omega)\right) - n(t,\omega) f_{\text{Eq}}(|\mathbf{p}|) \left(1+ n(t,\omega')\right)\right)
\end{align}
where we have included the cross section and the M\o ller prefactor in the differential transition probability
\[\id w = \left(1-\frac{\mathbf{v}}{c}\cdot \mathbf{\hat{n}}\right)\, \id \sigma^{\gamma e}\]

Now, let us we repeat the whole diffusion approximation exactly as done in Section \ref{dif}, but now using the differential transition probability used by Kompaneets in 1957
\[	\id w = \id\Omega_{\text{rest}}\frac{\id \sigma}{\id \Omega}_{\text{rest}}^{\text{Th}}\]
where
\[\frac{\id \sigma}{\id \Omega}_\text{rest}^{\text{Th}} = \frac{3\sigma_T}{16\pi}\left(1+\cos^2\theta_{\text{rest}}\right)\]
is the Thomson differential cross section evaluated in the reference frame of the electron. So that we are not only expressing the cross section in a reference frame where the electron is at rest but also neglecting the M\o ller velocity factor.

The diffusion approximation can be carried in the same way as previously done and one will find \eqref{kwi}, but now with Kompaneets' first and second integral defined with the rate above. If those integrals are computed in the same way we do in Appendix \ref{b}, one will find

\begin{align}
	\label{firiw}
	&I_1(x)= \frac{n_e\sigma_T c\, k_BT}{m_ec^2}\;x(1-x)\\
	\label{seciw}
	&I_2(x)= \frac{n_e\sigma_T c\, k_BT}{m_ec^2}\;2x^2
\end{align}
instead of \eqref{firi} and \eqref{seci}. As we mentioned in \cite{paper}, to the best of our knowledge that first Kompaneets 
integral was never computed. As a matter of logic, as we have seen in the 
previous sections, it is not possible with solely the Thomson cross section 
(which ultimately leads to \eqref{firiw}) to find \eqref{kompfinal}, the Kompaneets equation.

Suppose now that instead of the rate above, we use
\[	\id w = \id\Omega_{\text{rest}}\frac{\id \sigma}{\id \Omega}_{\text{rest}}^{\text{KN}}\]
where
\[	\frac{\id\sigma}{\id \Omega}^{\text{KN}}_{\text{rest}} = \frac{3\sigma_T}{16\pi }\left(\frac{\omega'_{\text{rest}}}{\omega_{\text{rest}}}\right)^2\left[ \frac{\omega'_{\text{rest}}}{\omega_{\text{rest}} } + \frac{\omega_{\text{rest}}}{\omega'_{\text{rest}}} - \sin^2\theta_{\text{rest}}\right]\]
is the Klein-Nishina differential cross section but now evaluated in the rest frame of the electron as we have seen in Chapter \ref{3}.

By performing the diffusion approximation again with this rate, one finds
\begin{align}
	\label{firiw2}
	&I_1(x)= \frac{n_e\sigma_T c\, k_BT}{m_ec^2}\;x(5-x)\\
	\label{seciw2}
	&I_2(x)= \frac{n_e\sigma_T c\, k_BT}{m_ec^2}\;2x^2
\end{align}
these two results illustrate point (ii) above, where we mentioned that it is necessary to use a scattering cross section expressed in a consistent way (that is, in a covariant or frame-independent manner) to carry the diffusion approximation, \textit{i.e.}, we cannot use cross sections evaluated in the rest frame of the electron while using electrons distributed according to Maxwell-Boltzmann statistics.

To illustrate point (i) above, where we claim that the M\o ller velocity plays a crucial role for microscopic consistency, let us consider the following rate
\[	\id w = \id\Omega\frac{\id \sigma}{\id \Omega}^{\text{KN}}\]
where the differential cross section is given by the correct expression \eqref{klein-nishina}, while the M\o ller prefactor is neglected as well. Then, one finds in the diffusion approximation, the following values for the integrals
\begin{align}
	\label{firiw3}
	&I_1(x)= \frac{n_e\sigma_T c\, k_BT}{m_ec^2}\;x(3-x)\\
	\label{seciw3}
	&I_2(x)= \frac{n_e\sigma_T c\, k_BT}{m_ec^2}\;2x^2
\end{align}

So that only when the M\o ller prefactor is indeed added to the rate $\id w$ we can have the correct result, having the factor $4$ in the first integral\footnote{As a matter of fact, we note here that adding the M\o ller prefactor adds an extra factor of $\frac{n_e\sigma_T c\, k_BT}{m_ec^2}\;x$ to the first integral, so that using cross sections evaluated in the rest frame of the electron would still yield wrong result even when this factor is accounted on the rates. This observation only confirms point (ii) above.}. As we have seen in Chapter \ref{2}, this factor naturally appears when deriving the standard relativistic Boltzmann equation and accounts for the correct description of the flux of particles when seen in the rest frame of the electron. Moreover, in the lab frame, where we are observing the gases and seeing many collisions of photons and electrons, the electrons cannot be taken as they were at rest, simply because that is not what we are observing by using \eqref{maxwell}. Thus, we should address transitions having electrons and photons with any possible momenta in the Boltzmann equation and that is what \eqref{klein-nishina} accounts for.

As stated in \cite{paper}, it is very surprising, perhaps, 
that none of these effects are seen on $I_2(x)$, \textit{i.e.}, $I_2(x)$ is always given by the same value, whence making it possible to 
employ the indirect argument used traditionally (including in 
Kompaneets' original paper) to fix the value of $I_1(x)$, but this, however, is just a mathematical coincidence.

It is possible in the derivation of the Kompaneets equation to use the Thomson 
cross section together with the M\o ller factor (of course), because seen the Thomson cross section as the non-relativistic regime of the Klein-Nishina, it should satisfy, in the light of Chapter \ref{2},
dynamical reversibility as well. We would need, however, an expression analogous to \eqref{klein-nishina} for this cross section and we could not find such thing in literature. It is also true that transforming cross sections to arbitrary frames is a very difficult task, but since QFT enables a recipe to calculate cross sections in a manifestly covariant way, we can benefit from that to easily find complicated expressions such as \eqref{klein-nishina}, motivating its use instead of trying to generalize the Thomson cross section to an arbitrary frame.

As a matter of completeness, we shall demonstrate next that photon number can only be conserved if Kompaneets' integrals are given by \eqref{firi} and \eqref{seci}. For that goal, let us assume that the diffusion approximation yields 
\begin{align}
	\label{firiw4}
	&I_1(x)= \frac{n_e\sigma_T c\, k_BT}{m_ec^2}\;x(M-x)\\
	\label{seciw4}
	&I_2(x)= \frac{n_e\sigma_T c\, k_BT}{m_ec^2}\;2x^2
\end{align}
for some $M\in\mathbb{N}$. Then, we can rewrite Equation \eqref{kwi} in the following way
\begin{equation}
	x^2\frac{\partial n}{\partial t}(x,t) = \frac{n_e\sigma_T c\, k_BT}{m_ec^2}\, \frac{1}{x^{M-4}}\frac{\partial }{\partial x}x^M\left\{\frac{\partial n}{\partial x}(x,t) +n(x,t)(1+n(x,t))\right\}
\end{equation}
by looking \eqref{photnumb}, the rate in time of the photon number $N$ is proportional to the integral of the term on the left hand side so that
\begin{equation}
	\frac{\id N}{\id t} \propto \int^\infty_0\frac{\id x}{x^{M-4}}\frac{\partial }{\partial x}x^M\left\{\frac{\partial n}{\partial x}(x,t) +n(x,t)(1+n(x,t))\right\}
\end{equation}
integrating by parts and using that the current must vanish at infinity faster than any power of $x$, we have
\begin{equation}
		\frac{\id N}{\id t} \propto (M-4) \int^\infty_0 \id x\, x^3 j_t(x)
\end{equation}
where $j_t(\omega)$ is the current appearing the Kompaneets equation \eqref{cur}. Thus, we can only guarantee that the integral on the right side identically vanishes when $M=4$, as desired. This demonstrates that Kompaneets equation is photon number conserving only when the first Kompaneets' integral is indeed proportional to $x(4-x)$.

\chapter{Master-equation for a boson system}\label{5}

In this chapter we shall obtain the Kompaneets equation from a completely new set up. This will be done considering a random walk in the reciprocal space of the photon. The transition rates will be suitable chosen, in accordance with the discussion done in Section \ref{covdiff}. From our set up, it is easy to conclude that Kompaneets equation can be generalized for a more general boson gas, including also some forcing due to the environment where the bosons are performing the transitions. What follows is divided in two sections, in the first section we shall write down the corresponding master equation for our system while in the second section we will perform the Kramers-Moyal expansion to retrieve the generalized Kompaneets equation. This part follows closely our work \cite{paper}, so that throughout this whole chapter, the reader will find, almost integrally, the reproduction of the ideas and writings contained in Sections $2$ and $3$ of this work.

\section{Detailed balance with stimulated emission}
 Following \cite{paper}, let us consider a gas of photons (for now), where photons undergo a transition in the reciprocal space of wave vectors. The transition rates for this jump process arising within a quantum many-particle 
system are derived from one-particle Green's functions \citep{kadanoff}. Here we 
take photons with a symmetrized Fock space taking the tensor product 
over three-dimensional harmonic oscillators with wave vector ${\mathbf k}$  corresponding 
to frequency $\omega$.  An elementary transition is the annihilation of a 
photon with wave vector ${\mathbf k}$ while creating a photon with wave vector ${\mathbf k'}$.  
When the photons are in weak contact with a thermal bath at inverse temperature $\beta$, each transition creates a flux in reciprocal space with an expectation given by
\begin{equation}\label{ww}
	j({\mathbf k}\rightarrow {\mathbf k'})= b({\mathbf k},{\mathbf k'})\, 
	e^{\beta(\hbar\omega-\hbar\omega')/2}\,\big|\langle \text{f}| 
	a^\dagger_{\mathbf k'}a_{\mathbf k}|\text{i}\rangle\big|^2
\end{equation}
where $b({\mathbf k},{\mathbf k'}) = b({\mathbf k'},{\mathbf k})$ is 
symmetric and left unspecified for the moment as a parameter of dynamical 
activity. We have taken that
to lowest order  the matrix element will contain a single term annihilating and creating a photon $a^\dagger_{{\mathbf k'}}a_{\mathbf k}$.  Writing $n({\mathbf k})$ for the occupation/level at wave vector ${\mathbf k}$, we put
\[
|\text{i}\rangle = | \ldots n({\mathbf k})\ldots n({\mathbf k'})\ldots\rangle
\]
for the initial state.  The only non-zero matrix element will between that initial $|i\rangle$ and
the final state
\[
|\text{f} \rangle = | \ldots n({\mathbf k})-1\ldots n({\mathbf k'})+1\ldots\rangle
\]

Since on each of the Hilbert spaces $a|n\rangle = \sqrt{n}|n-1\rangle$ and  $a^\dagger|n\rangle = \sqrt{n+1}|n+1\rangle$, we find that
\[
\left|\langle \text{f}\,| a^\dagger_{{\mathbf k'}}a_{\mathbf k}|\text{i}\rangle\right|^2 = (1+n({\mathbf k'}))\,n({\mathbf k})\]

Therefore \eqref{ww} becomes
\begin{eqnarray}\label{www}
	j({\mathbf k}\rightarrow {\mathbf k'}) &=& b({\mathbf k},{\mathbf 
		k'})\,w({\mathbf k},{\mathbf k'}) \,n({\mathbf k})\\
	w({\mathbf k},{\mathbf k'}) &:=&  e^{\beta(\hbar\omega-\hbar\omega')/2}\,(1+n({\mathbf k'}))\nonumber
\end{eqnarray}

In above definition of the transition rates, we observe in connection with the discussion done in Section \ref{covdiff}, that the exponential prefactor is very natural to require, as it expresses dynamical reversibility of the transition rates \eqref{microreversph}. Thus, by adding this factor, we make sure that our set up is indeed compatible to the context of a Boltzmann-master equation, as we have seen previously.

In classic texts on Markov processes, \eqref{www} makes the sink term into ${\mathbf k'}$ from ${\mathbf k}$ and one would write a master equation for the probability of occupying the various wave vectors.  The source term is 
$j({\mathbf k'}\rightarrow {\mathbf k})$.  Ignoring however correlations between the occupations at different wave vectors, we can write the master equation directly for the (now expected) occupation numbers
\begin{equation}\label{meq}
	\frac{\partial}{\partial t} n_t({\mathbf k}) = \sum_{{\mathbf k'}} 
	b({\mathbf k},{\mathbf k'})\, \big[ 
	e^{\beta(\hbar\omega-\hbar\omega')/2}\,(1+n_t({\mathbf k'}))n_t({\mathbf k}) - 
	e^{\beta(\hbar\omega'-\hbar\omega)/2}\,(1+n_t({\mathbf k}))n_t({\mathbf 
		k'})\big]
\end{equation}

As such, the evolution equation \eqref{meq} does not need to be photon number-preserving, \textit{i.e}., it does not follow directly that
\begin{equation}\label{pa}
	\sum_{\mathbf k}\frac{\partial}{\partial t} n_t({\mathbf k}) = \sum_{{\mathbf 
			k},{\mathbf k'}} b({\mathbf k},{\mathbf k'})[w({\mathbf k'},{\mathbf k}) 
	n_t({\mathbf k'}) - w({\mathbf k},{\mathbf k'}) n_t({\mathbf k})] = 0
\end{equation}
unless $b({\mathbf k},{\mathbf k'})=b({\mathbf k'},{\mathbf k})$ is 
indeed symmetric. Only then, \eqref{meq} is a continuity equation.\\
Secondly, detailed balance requires that $j({\mathbf k}\rightarrow {\mathbf 
	k'}) - j({\mathbf k'}\rightarrow {\mathbf k})=0$ for all ${\mathbf k},{\mathbf 
	k'}$, or (for symmetric $b({\mathbf k},{\mathbf k'})$ always)
\begin{eqnarray}
	w({\mathbf k},{\mathbf k'})\,n({\mathbf k}) &=&  w({\mathbf k'},{\mathbf k})\,n({\mathbf k'})\;\;\text{ or }\nonumber\\
	e^{\beta(\hbar\omega-\hbar\omega')/2}\,(1+n({\mathbf k'}))\,n({\mathbf k}) &=&  e^{\beta(\hbar\omega'-\hbar\omega)/2}\,(1+n({\mathbf k}))\,n({\mathbf k'})\nonumber\\
	\implies e^{\beta\hbar\omega}\,\frac{n({\mathbf k})}{1+n({\mathbf k})}&=& \text{ constant\,} = e^{\beta\mu}  
\end{eqnarray}
which implies that in equilibrium
\[
n({\mathbf k}) = \frac 1{e^{\beta\hbar\omega} -1}
\]
by assuming that the chemical potential equals zero, $\mu=0$.  Note that we have only used \eqref{www} for an environment in thermal equilibrium where the temperature may refer to an electron gas or anything else.  The ``anything else'' would solely show in the prefactor $b({\mathbf k},{\mathbf k'})$ for the kinetics \eqref{meq}. In that sense. the evolution given by \eqref{meq} represents a general Kompaneets equation, before any diffusion approximation. Observe also that, in the same way, \eqref{bol-kompc2} is the continuum analog to the master equation \eqref{meq}, where the symmetric prefactor is recognized to be due to the electron bath and, thus, accounted inside the transition rates given in 
\eqref{bol-kompc2}. 

At this point it is useful to note that our framework is constructed as such to be compatible to the description at the level of the Boltzmann equation. Therefore, all the important (and necessary) ingredients to obtain the correct description of the Kompaneets equation such as dynamical reversibility of the rates and detailed balance are used to indicate that we are following the right direction.

\section{Discrete Kramers-Moyal expansion}\label{km}

In this section, we shall perform the diffusion approximation to the previous framework of the photon master equation, reproducing as it is done in Section 3 of our work \cite{paper}. First we observe that the previous section considers jumps in the space of wave vectors ${\mathbf k}$.  We associate in the present section an energy $U({\mathbf k})$ to the system and we expand the master equation \eqref{meq} for small energy changes ${\mathbf k}\rightarrow {\mathbf k'}$, that, as we have seen, is again the Kramers-Moyal or diffusion approximation.

Let us start in one dimension, where we consider a lattice mesh $\delta>0$ for $x\in \delta \mathbb{Z}$.  The $x = k_1$ denotes now the first component of the (rescaled) wave vector.  We imagine a walker hopping on that lattice of wave vectors, to nearest neighbor sites with transition rates
\begin{align}	
	w(x,x\pm\delta)= (1+n(x\pm\delta))B\left(x \pm 
	\frac{\delta}{2}\right)\,\exp\left\{-\frac{\beta}{2}\left(U(x\pm\delta)-U(x)\right)\pm\frac{\beta\delta}{2}f\left(x\pm\frac{\delta}{2}\right)\right\}\label{tr}
\end{align}

Here, $n$ is the instantaneous number of walkers; its presence in the rates 
represents the stimulated emission. The function $B>0$ is an inhomogeneous 
activity rate and  $\beta$ is the inverse temperature of a medium enabling the 
hopping.  There is also a driving force\footnote{Note we are in reciprocal 
	space here so that $\delta$ is an inverse length and $f$ is measured in 
	multiples of $\hbar\,c$.} $f$ and a potential $U$ which are added following the 
condition of local detailed balance at fixed environment inverse temperature 
$\beta = (k_BT)^{-1}$  \citep{ldb}. At this moment we do not dwell on the 
physical meaning of the driving $f$ and we do not restrict ourselves to photons but to bosonic systems more generally; see also Chapter \ref{6}.
The rate \eqref{www} is a special case of \eqref{tr}, where  $f\equiv0$ and the photon energy $U=\hbar\omega$. Abusing notation, we also incorporate the symmetric activity $b({\mathbf k},{\mathbf k'})$ in the rates as the prefactor $B$.

For fixed $\delta$ the master equation as in \eqref{meq} becomes
\begin{eqnarray}
	\frac{\partial n_t}{\partial t}(x) &+&  j_t(x,x+\delta) - j_t(x-\delta,x) =0 \;\;\text{ for}\\
	j_t(x,x+\delta) &=& n_t(x) w(x,x+\delta) -n_t(x+\delta)w(x+\delta,x) 
	\nonumber\\
	j_t(x-\delta,x) &=& n_t(x-\delta)w(x-\delta,x) -n_t(x) w(x,x-\delta) \nonumber 
\end{eqnarray}

We expand this last equation to second order in $\delta$; see Appendix \ref{d}.
The result is
\begin{align}\label{mm}
	\frac{\partial}{\partial t}n_t = \delta^2\bigg{\{} &\left(\beta B g' + \beta B'g\right)(1+n)n +\left(\beta B g+B'\right)n' + 2\beta B g n n' + Bn''
	\bigg{\}} 
\end{align}
with $g(x) \coloneqq U'(x)-f(x)$. That can be written more explicitly as a continuity equation,
\begin{equation}
	\frac{\partial n_t}{\partial t}(x) = \delta^2\frac{\partial}{\partial x}\bigg{\{} B(x)\,\bigg(\frac{ \partial n_t}{\partial x}(x)+ \beta g(x)\big(1 + n_t(x)\big)n_t(x) \bigg)
	\bigg{\}} 
\end{equation}
in which we recognize the structural elements of the Kompaneets equation \eqref{ke}.\\

We can indeed redo that in three dimensions, on $\delta \mathbb{Z}\times\delta \mathbb{Z}\times\delta \mathbb{Z}$. Taking the same rates in all directions as before with 3-dimensional ``force'' $f$, the diffusion approximation now reads
\begin{align}
	\label{serv}
	\partial_tn_t = \delta^2\;{\pmb \nabla} \cdot {\mathbf j}
\end{align}
with in Cartesian coordinates $(x_\ell, \ell=1,2,3)$ for ${\mathbf j}=\sum_\ell j_\ell \mathbf{\hat{x}_\ell}$,
\begin{equation}\label{genk}
	j_\ell=B\left(\frac{\partial n}{\partial x_\ell}+ \beta g_\ell\,(1 + n)n\right)
\end{equation}
for $g_\ell := \frac{\partial U}{\partial x_\ell} - f_\ell$.

Moving finally to the setup for the Kompaneets equation we enter frequency space by assuming that $n_t=n(t,\omega), g=g(\omega), D=D(\omega)$ with frequency $\omega = c \,\sqrt{\sum_\ell x_\ell^2}$ for speed of light $c$.  This means that we rewrite \eqref{serv}
in spherical coordinates, with $\omega$ as radial variable: 
\begin{equation}\label{genke}
	\omega^2\frac{\partial n}{\partial t}(t,\omega) = c\,\delta^2\, \frac{\partial }{\partial\omega}\left\{\omega^2 B(\omega)\left(c\,\frac{\partial n}{\partial \omega}(t,\omega)+ \beta \,g(\omega)\big(1 + n(t,\omega)\big)n(t,\omega)\right)\right\}
\end{equation}

That is an extended Kompaneets equation, to be compared with \eqref{ke}, where the energy change and the driving combine into $g(\omega) := c\frac{\partial U}{\partial \omega}(\omega) - f(\omega)$.

Making the choices
\begin{align}\label{cho}
	c^2\delta^2\,B(\omega)=\frac{k_B T}{m_ec^2 } n_e\sigma_T\,c\,\;\omega^2,\qquad g(\omega) = \hbar\,c
\end{align}
the above equation \eqref{genke} becomes exactly the one of Kompaneets 
\eqref{ke}. 

Note that $n_e\sigma_Tc= \tau^{-1}$ is the average collision rate, 
as before.  That shows that the full structure of the Kompaneets equation is obtained as the diffusion approximation to a master equation with stimulated 
emission, and this holds whenever the limiting activity and drift obey 
\eqref{cho}.  Justifications for the choices \eqref{cho} come from the physical 
nature of the process considered in the Kompaneets equation.  The photon energy 
is $U(\omega) =\hbar \omega$ and there is no driving $f\equiv 0$, making 
indeed $g=c\frac{\partial U}{\partial \omega} = \hbar c$. In order to understand the 
first equality in \eqref{cho}, we note that $c^2\delta^2\,B(\omega)$ appears as 
the diffusion constant $D(\omega)$ in \eqref{genke}. The shift in frequency for 
a photon undergoing Compton scattering determines that diffusion constant as 
the conditional average squared shift
\begin{equation}\label{di}
	D(\omega) = \left\langle \frac{(\omega' - \omega)^2}{2\tau} \,\bigg|\,\omega\,\right\rangle 
\end{equation}
this shift follows from \eqref{shift}.

In the low-temperature regime where Compton scattering is relevant, under the 
assumption that the electrons and the photons are of comparable energy much 
less than $m_ec^2$, most of the momentum is carried by the electrons, meaning 
$|\mathbf{p}| \gg \omega / c$. Hence we can replace the term between the brackets in the 
denominator of \eqref{shift} by unity, and only retain the first term in the numerator
\begin{equation}\label{lin}
	\omega' - \omega \approx \frac{\mathbf{p}\cdot(\mathbf{\hat{n}'} - \mathbf{\hat{n}})}{m_e c} \omega
\end{equation}

We can  assume the square of projection of the scattering vector $\mathbf{\hat{n}'}-\mathbf{\hat{n}}$ on the 
momentum vector $\mathbf{p}$ to average out to a constant of magnitude 1, which we 
will hereafter ignore.  Continuing then the calculation for \eqref{di} yields
\begin{equation}
	D \propto \left\langle \frac{|\mathbf{p}|^2}{2m_e^2c^2} \frac{\omega^2}{\tau} 
	\right\rangle = \frac{1}{m_ec^2} \left\langle \frac{|\mathbf{p}|^2}{2m_e} \right\rangle 
	\frac{\omega^2}{\tau} \propto \frac{k_B T}{m_ec^2 } \frac{\omega^2}{\tau}
\end{equation}
where the temperature $T$ gives the average kinetic energy of the electron 
distribution.  We thus recover the first equality in \eqref{cho}.

A crucial property of the Doppler effect that we used to compute \eqref{di} and 
to arrive at \eqref{cho} is that the shift in frequency is linear in the 
frequency itself, as seen in \eqref{lin}. On the other hand, because the shift 
in the electron's energy equals the same expression, were we to consider the 
opposite situation of an electron in a photon bath (in contact with other matter at temperature $T$), the average squared energy 
shift would be proportional to the square of the momentum instead, hence only 
linear in the energy.  Together with the electronic density of states going as 
the square root of the energy, this allows us to write down the electronic 
version of the Kompaneets equation immediately
\begin{equation}\label{ek}
	E^{1/2} \,\frac{\partial f}{\partial t}(E,t)=  b\,\frac{\partial 
	}{\partial E}E^{3/2}\left\{k_B T \frac{\partial f}{\partial E}(E,t) + f(E,t)\right\}.
\end{equation}
for some rate $b \propto  c\sigma_T\,\frac{U_\gamma}{m_e c^2}$, where $U_\gamma$ is the energy density of the photon gas. We neglect the fermionic nature of the electron, since we presume non-degeneracy of the electron gas. For verification, a derivation of this 
equation can be found in \citep{electronkompaneets}. For applications to highly dense states of fermionic matter, the Pauli exclusion is significant and departures from \eqref{ek} are expected. In such a regime however, the long-range Coulomb interactions become relevant and one must be very careful on how to perform the diffusion approximation. 

As we have seen here and noted in \cite{paper}, the description given by \eqref{meq} involves the photon occupation only and the electron bath is integrated out, remaining present only via the bath temperature. Similarly, the dynamics \eqref{tr} effectively treats the electron bath via temperature, mobility and possible driving $f$. In that sense, the description provided by the Boltzmann equation, which explicitly describes the electrons and how they modify the photon distribution via Compton effect, is one level finer.

\chapter{Extensions and generalized Kompaneets equations}\label{6}

In this chapter, we shall turn our attention to generalizations of the Kompaneets equation. These generalizations usually come in three ways, by relaxing the condition of equilibrium to the electron bath, by considering other sources of interactions or by going further in the diffusion approximation (what we have been calling a relativistic extension). This discussion was made in Section 6 of our work \cite{paper}, but here we shall give a more complete discussion to that appearing in this reference. Finally, we will close this chapter by looking at the example of a less standard generalization, proposed by \cite{arca} and that effectively changes the equilibrium distribution of this (now modified) Kompaneets equation.  

\section{Relaxation of the equilibrium condition}

The relaxation of the equilibrium condition to the electron bath was pointed out in \cite{barbosa,brown2,peebles, brown}, provided that the distribution of the electrons is isotropic, but we could not find any reference which concludes the same by using the same framework as proposed by Kompaneets in 1957, \textit{i.e.}, starting from the standard relativistic kinetic equation and performing the diffusion approximation in the energy shift $\Delta$. Therefore, here we follow \cite{paper} to observe that by assuming: (i) isotropy of the distribution of the electrons, 
\begin{equation}
	f(t,\mathbf{p})\id^3\mathbf{p} = f(t,|\mathbf{p}|)\id^3\mathbf{p}
\end{equation}
and (ii) that $f$ decays faster than $|\mathbf{p}|^3$, \textit{i.e.},
\begin{equation}\label{limit}
	\lim_{|\mathbf{p}|\to \infty}|\mathbf{p}|^3f(t,|\mathbf{p}|)=0
\end{equation} 
we find Kompaneets equation \eqref{ke} with an effective temperature $$T_\text{eff}\coloneqq \frac{\langle |\mathbf{p}|^2 \rangle}{3 k_B m_e}$$ where $$n_e\,\langle |\mathbf{p}|^2\rangle =\int \id^3\mathbf{p}\, |\mathbf{p}|^2\, f(|\mathbf{p}|)$$
the details of the derivation can be found in Appendix \ref{b}.

Observe that it is not so strange to recover the equilibrium (relaxation to) Planck distribution in the diffusion approximation.  This can either mean two things: (i) that non-equilibrium features hide in higher order terms or (ii) that a formalism starting from a Boltzmann equation washout the non-equilibrium degrees of freedom of the electron bath, as photons only ``feel" the integrated bath. 

We note here that differently from \cite{barbosa, brown2,brown}, we find the necessity of requiring the asymptotic behavior \eqref{limit} of the electronic distribution function. Not surprisingly, this generalization can be also addressed starting from the manifestly covariant formalism and we refer the reader to \cite{brown} for that.

\section{(Non-)relativistic extensions and relaxation of isotropy condition}

Relativistic extensions were mentioned in great detailed in Section \ref{review} and some references to that include \cite{cooper, barbosa, itoh, itoh2, kohyama1, kohyama2, brown, kohyama3}. It is interesting to note that while most of the references work with the manifestly covariant formalism, Barbosa uses a more kinematic approach, similar to a Fokker-Planck approximation. In the occasion of Section \ref{review}, we have also mentioned that the soft condition for the photons can be relaxed to include down-Comptonization \cite{liu, zhang}, a non-relativistic extension\footnote{As before, we note that the electron bath is still treated non-relativistically $(k_BT\ll m_ec^2)$, so that this extension should no be regarded as relativistic.} that includes the regime $(\hbar\omega\gg k_BT)$.

The condition of isotropy to the distribution function of the photons can also be relaxed and was addressed by \cite{buet, pitrou}. In that case, the diffusion approximation to retrieve a Kompaneets-like equation is much more involved and one usually uses spherical harmonics or symmetric-and-trace-free (STF) tensors to express the distribution function. Such extension is adapted to recognize spectral distortions due to some anisotropy, \textit{e.g.}, polarized photons. 

\section{Bremsstrahlung, radiative Compton and Doppler shift}

As we have noted in \cite{paper}, there are also more processes for the electron-photon system that could impact the spatio-temporal dynamics of the photon occupation number. Beyond Compton scattering we could have considered contributions due to Bremsstrahlung and radiative (or double) Compton scattering, for example. In fact, in his original paper \cite{kompa}, Kompaneets already calculates contributions due to Bremsstrahlung. Further references include \citep{hu, rybicki, longair}.

Due to the radiative nature of these processes it is not true any longer that photon number is conserved and we cannot describe them by a Boltzmann-master equation. If those processes are taken into account, we can write the time evolution of the occupation number as
\begin{equation}
	\frac{\partial n}{\partial t} = \frac{\partial n}{\partial t}\bigg{|}_C + \frac{\partial n}{\partial t}\bigg{|}_{Br} + \frac{\partial n}{\partial t}\bigg{|}_{DC} 
\end{equation}
where
\[\frac{\partial n}{\partial t}\bigg{|}_C =\frac{1}{\omega^2} \frac{n_e\sigma_T 
	c}{m_e c^2}\frac{\partial }{\partial \omega}\omega^4\left\{k_B T 
\frac{\partial n}{\partial \omega}(t,\omega) + 
\hbar\left[1+n(t,\omega)\right]n(t,\omega)\right\}\]
is the change due to Compton scattering which appears in the Kompaneets equation.

The other contributions, due to Brehmsstralung and double Compton scattering are found to be \cite{kompa, blumenthalgould, lightman, hu}
\begin{align}
	&\frac{\partial n}{\partial t}\bigg{|}_{Br}= \frac{n_e \sigma_T c}{m_ec^2}g(\omega)Y  \frac{e^{-\beta\hbar\omega}}{\omega^3} \left[1 - n(t,\omega)\left(e^{\beta\hbar\omega} - 1\right)\right]\\
	&\frac{\partial n}{\partial t}\bigg{|}_{DC}= \frac{n_e \sigma_T c}{m_e c^2}\left(\frac{4\alpha \hbar^2}{3 \pi}\right) \frac{1}{\omega^3}\left[1 - n(t,\omega)\left(e^{\beta\hbar\omega} - 1\right)\right] I(t)
\end{align}
with definitions
\begin{align*}
&g(\omega)= 
	\begin{cases}
		&\ln \left(2.2\frac{k_B T}{\hbar \omega }\right) \ \ \ \ \ \ \ \ \mathrm{if} \ \ \hbar \omega \leq k_B T\\
		&\sqrt{\frac{k_B T}{\hbar \omega }}\ln(2.2) \ \ \ \ \ \ \mathrm{if} \ \ \hbar \omega > k_B T \ 
	\end{cases}\\
&Y = \frac{\alpha c^3}{8}\sqrt{\frac{(m_ec^2)^3}{2\pi k_B T}}\sum_in_iZ^2_i\\
&I(t)=\int \id \omega'\, {\omega'}^4 \left[1+ n(t,\omega')\right] n (t, \omega ') 
\end{align*}
where $\alpha$ is the fine-structure constant and $n_i$ is the number density of ions with atomic number $Z_i$.

However, as mentioned in \cite{blumenthalgould, zeldovich}, in low-density plasmas, Compton scattering is the dominant mechanism which enables energy exchange.

It is also possible to generalize the relativistic Boltzmann equation to include curved space-times due to some gravitational field. This is done for example in \cite{ kremer, bernstein}. In that case, the equation takes the following format
\begin{equation}\label{covbolgrav}
	p_1^{\mu}\frac{\partial f_1}{\partial x^\mu} - \Gamma^\sigma_{\mu\nu}p^\mu_1p^\nu_1\frac{\partial f_1}{\partial p_1^\sigma} =\int_{\mathbf{p_2},\mathbf{p'_1},\mathbf{p'_2}} \frac{\id \mathbf{p_2}}{p^0_2} \frac{\id \mathbf{p'_1}}{{p^0_1}'} \frac{\id \mathbf{p'_2}}{{p^0_2}'} \sqrt{g} W(p_1,p_2\to p'_1, p'_2)\left(f_{1'} f_{2'} - f_1 f_2\right)
\end{equation}
where we have not taken into account external forces. Above, $-g$ is the determinant of the metric tensor and $\Gamma^\sigma_{\mu\nu}$ are the Christoffel symbols of the Levi-Civita affine connection. By using this expression, one can include the study of the Kompaneets equation in a cosmological context, see for example \cite{hu,bernstein, burigana1, burigana2}. Then, the contribution of the cosmic expansion (red shift) must be taken into consideration, appearing as a convective term in the equation
\begin{equation}\label{kered}
	\frac{\partial n}{\partial t}(t,\omega)- \frac{\dot{R}}{R}\omega \frac{\partial n}{\partial \omega }(\omega, t)= \frac{1}{\omega^2}\frac{n_e\sigma_T 
		c}{m_e c^2}\frac{\partial }{\partial \omega}\omega^4\left\{k_B T 
	\frac{\partial n}{\partial \omega}(t,\omega) + 
	\hbar\left[1+n(t,\omega)\right]n(t,\omega)\right\}
\end{equation}
where $\frac{\dot{R}}{R}$ is the \textit{Hubble parameter} (sometimes also defined with $a$ instead of $R$). For more details on the above equation, the reader can refer to \cite{hu, bernstein}. This extension to the Kompaneets equation, however, does not change the equilibrium distribution of photons\footnote{This is why we still see a blackbody radiation spectrum to the CMB, despite the expansion of the universe.} and should only modify the temperature of the distribution, which now evolves with time \cite{bernstein}. 

A less standard generalization involving curved space-times is that of a turbulent red-shift (caused by a turbulent fluctuation of the metric in the background). As we mentioned in \cite{paper}, this could appear as an integrated \textit{Sachs-Wolfe effect} for a random gravitational potential field and would give an additional diffusion in frequency, but we have not seen that being carried out yet.
 
\section{Further generalizations}

As we have seen, one generalization of the Kompaneets equation is found in Chapter \ref{5}, where we showed an extension to a more general boson system with energy $U$ and possibly driven by a force $f$. We believe that such generalization may be relevant in many contexts, one example could arise in the context of solid state physics, where phonon-electron interactions may take place with a possible drive.

Being an interesting instance where non-equilibrium features effectively changes the equilibrium distribution of a Kompaneets-like equation, we briefly discuss here the generalization contained in \cite{arca}. In this work, the authors exploit the recent observations of \cite{arcade1, arcade2, edges}, where systematic deviations of the CMB spectrum is observed at low frequencies\footnote{As we have mentioned previously, we can retrieve Planck's law of the black body radiation from the stationary solution of the Kompaneets equation. The spectral density is, thus, proportional to $\propto \omega^3 n_{Eq}(\omega)$, \textit{cf.} Section \ref{struc}.}, something referred to as \textit{space roar}. Observing that such phenomenon is not well understood in the scientific community, \cite{arca} assumes the primordial plasma to be out of equilibrium and that, as logical consequence, we should expect natural departures from the Bose-Einstein equilibrium distribution, being the new equilibrium distribution the stationary solution of the following modified Kompaneets equation

\begin{align}
	\omega^2\frac{\partial n}{\partial t}(t,\omega)= \frac{n_e\sigma_T 
		c}{m_e c^2}\frac{\partial }{\partial \omega}\omega^4\bigg{\{}k_B T 
	&\frac{\partial n}{\partial \omega}(t,\omega) + 
	\hbar\left[1+n(t,\omega)\right]n(t,\omega)\bigg{\}} \nonumber \\
	& + \frac{n_e\sigma_T 
		c}{m_e c^2}\frac{\partial }{\partial \omega} \omega^4\left\{ \frac{1}{(\xi+1)}\left(\frac{\omega_0}{\omega}\right)^\xi k_B T\frac{\partial n}{\partial \omega }(t,\omega)\right\}
\end{align}
where $\omega_0$ and $\xi$ are free parameters.

Therefore, the solution of the above equation is imprinted by the non-equilibrium features of the primordial plasma. The overall effect of non-equilibrium is to promote another source of diffusion in frequency space, increasing the occupancy of photons at low frequency, which leads to a slightly modified black-body spectrum. According to \cite{arca}, when the $\xi$ parameter is taken close to $3$, the spectral density of radiation is in good agreement with the observed data from \cite{arcade1,edges}.

The modification introduced by \cite{arca} is based on statistical and kinetic arguments, in which stochastic acceleration in frequency space due to non-equilibrium dynamical activity introduces an extra source of diffusion, with diffusivity inversely proportional to the squared of photon frequency. In order to introduce this modification as a correction based on first principles we must treat the non-equilibrium degrees of freedom of the primordial plasma, understanding also how these degrees of freedom are transferred to the photons. In any case, it would be natural to expect that, not only matter (\textit{e.g}, electrons) is turbulent in the early Universe, but also the metric\footnote{Recall that Einstein's equation couples them, such that a turbulent motion of matter induces a turbulent geometry and vice versa.} and one must include the combined dynamics of a turbulent metric space, matter and radiation, which would lead to corrections in the transition rates due to turbulent gravity, for example.

Finally, as another example of less standard generalizations, it is worth mentioning here that the Kompaneets formalism is very general and useful, yielding fruitful results even in neutrino physics. In that context, \cite{suwa, wang} have recently applied the diffusion approximation as proposed by Kompaneets to a neutrino gas.

\chapter{Conclusion}\label{7}

The topic of non-equilibrium in Statistical Mechanics is a rich and vast subject. In such context, we have addressed in this work the phenomenon of relaxation to equilibrium, where a gas initially out of equilibrium relax to it upon contact with a thermal bath. We have seen that one protagonist in this study is the so-called Boltzmann equation and, in Chapter \ref{2}, we have developed in great detail the relativistic version of it. This equation is essential to derive the Kompaneets equation, a partial differential equation that models the spatio-temporal dynamics of the occupation number distribution function of a photon gas in contact with a non-relativistic, non-degenerate electron bath in thermal equilibrium. Radiation then reaches equilibrium by undergoing the process of Comptonization, \textit{i.e}, the redistribution of photon frequency due to Compton interaction with the electrons.

The standard and traditional way of deriving the Kompaneets equation was proposed and carried by Kompaneets himself in his original paper from 1957 \cite{kompa}, where he proposes to perform a diffusion approximation to the Boltzmann equation in terms of the energy shift $\Delta$. However, we have addressed and pointed out both in here and \cite{paper} that there exist some inconsistencies regarding this derivation which are repeated in many textbook and recent references \cite{katz, liu, rybicki, zhang}. As mentioned, this happens because the diffusion approximation to the standard relativistic Boltzmann equation has some particularities and, because the set up proposed by Kompaneets is very didactic and useful, many references tend to repeat what he has done, avoiding also an approach starting from the manifestly covariant Boltzmann equation as it usually requires more background from the reader.

We feel that these inconsistencies are very useful to highlight when the Boltzmann equation fails to yield a conservation that it is, from the start, constructed to respect (this is particle number conservation) and that these problems are neither fully explored nor clarified in literature in such way that we found useful to rederive the Kompaneets equation with the traditional set up, while pointing the problems that a careful reader may encounter.

Therefore, in Chapter \ref{2}, we have shown what is the correct and natural way of expressing \eqref{komp-boltz}, which clarifies what should be the correct expression of Equation (1) in \cite{kompa}. Then, when it comes to that matter, we have seen that a prefactor appears in the Boltzmann equation, called M\o ller velocity factor, and, while it is essential to the correct description of the equation, it is a common inaccuracy in many references to neglect it, e.g. \cite{electronkompaneets, chen, tong}. In Section \ref{manif-cov-bol}, we have also seen that there exists another description of the Boltzmann equation, which is written in a manifestly covariant way, that is, of course, the manifestly covariant relativistic Boltzmann equation. In Section \ref{cov-nonc}, we have shown the equivalence of both descriptions so it becomes clear that, starting from a relativistic Boltzmann equation while being consistent, the Kompaneets equation \eqref{ke} can be regarded as the non-relativistic approximation of either version of the Boltzmann equation, as they are equivalent.

Chapter \ref{3} was devoted to the study of cross sections in different inertial frames and scattering matrices. Then, we have seen how to obtain the very general expression of the full relativistic Klein-Nishina cross section \eqref{klein-nishina}. Although not so enlightening, this expression explores Lorentz invariance and, as such, holds in any frame of reference, explaining the reason why we refer to it sometimes as the frame-independent expression of the Klein-Nishina cross section. Being one of the sources of inconsistency, we have seen that is very important to have a cross section defined in this precise way and the reason, as we mentioned, is simple: in the lab frame, where the gas mixture is being observed, we see electrons and photons with any possible velocity (this is of course compatible with using the Maxwell-Boltzmann distribution for the electrons), therefore, we need an expression of the cross section which accounts for collisions having any kind of possible momenta combination and not only collisions in which the electron is initially at rest (this is the cross section evaluated in the rest frame of the electron, as we have seen).

In Chapter \ref{4}, we have shown how to perform consistently the diffusion approximation to the Boltzmann equation, retrieving the well-known Kompaneets equation. In the first section, we started from the standard (or covariant) version of the kinetic equation and employed the traditional set up as proposed by Kompaneets. There, we have also addressed the two inconsistencies found in this framework, that is the subject of Section \ref{probcons}, where we thoroughly discussed the importance of expressing the cross section in a frame-independent way, while also accounting for the M\o ller velocity. As we mentioned in \cite{paper}, these inaccuracies are traditionally solved by employing Kompaneets' indirect reasoning \citep{kompa}, invoking the strong (but correct) assumption that \eqref{kwi} should have the form of a continuity equation and that the current should vanish for the Bose-Einstein distribution. However, we have also observed in Section \ref{struc} that assuming the Bose-Einstein distribution as the equilibrium solution to the differential equation may fail in a non-equilibrium context. 

As we observed \cite{paper}, this procedure fixes the value of $I_1(x)$, while $I_2(x)$ is calculated using the Thomson differential cross section \eqref{thomson} and evaluating the integral in the exactly same manner as we do (but neglecting the M\o ller velocity). This indirect procedure is remarkable as it relies on computing correctly $I_2(x)$ from a wrong setup. We believe this feature to be just a mathematical coincidence, \textit{i.e.}, there is no way to know from the start, within Kompaneets' set up of 1957, that $I_2(x)$ given by \eqref{seci} is the correct value. Also remarkably is that we could not find any reference which mentions these problems. The other part of Chapter \ref{4} follows more closely the work of \cite{brown} and it is devoted to the derivation of the Kompaneets equation while starting from the manifestly covariant formalism.

Looking for extensions to the famous equation, we have proposed in Chapter \ref{5} a new model, where a Kramers-Moyal expansion of suitable chosen transition rates of a random walk in reciprocal space of photons yields the Kompaneets equation from a totally different set up as ever consider in literature. We have seen then that it is also possible to generalize this equation to more general boson systems, where they can possibly be under the influence of a driving as well. Further generalizations are found in Chapter \ref{6}, where we have examined the interesting hypothesis proposed by \cite{arca}, which yields the Kompaneets equation with an extra correction due to non-equilibrium features in the primordial plasma.

Being one of the few examples where the diffusion approximation to the Boltzmann equation can be done in great detail, the Kompaneets equation is subject of great interest in Statistical Mechanics and we hope that our work serves, not only to clarify a number of issues encountered in literature, but also to guide a new reader along the vast literature that concerns this equation. Finally, we believe that the extension we propose may serve as a point of departure to further generalizations and we hope that it finds fruitful grounds even beyond systems we considered.

\appendix
\chapter{Relativistic relative velocity and M\o ller velocity}\label{a}

This Appendix will be devoted to the discussion about relativistic relative velocity and the M\o ller velocity which appears in the Boltzmann equation. Here we will follow mostly \cite{kremer} and \cite{mirco}.

\begin{figure}[H]
	\centering
	\includegraphics[width=.7\linewidth]{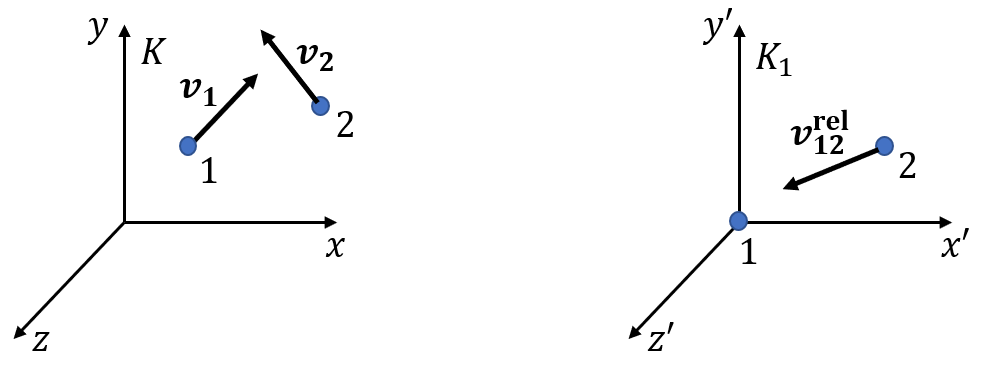}
	\caption{Frames $K$ and $K_1$. In frame $K$, particles $1$ and $2$ have velocities given by $\mathbf{v_1}$ and $\mathbf{v_2}$, respectively, while in frame $K_1$ (which is the rest frame of $1$), particle $2$ has velocity $\mathbf{v^{\text{rel}}_{12}}$.}
	\label{relativeframes}
\end{figure}

Suppose we have two particles, particle $1$ and $2$, which have respective velocities given by $\mathbf{v_1}$ and $\mathbf{v_2}$ as measured in the frame $K$. Suppose that we move to the rest frame of $1$, which we denote by $K_1$ (see Figure \ref{relativeframes}), in this new frame, the velocity of $1$ is zero and $2$ have velocity given by $\mathbf{v^{\text{rel}}_{12}}$, that is the relative velocity of $2$ with respect to $1$.

As we have seen in Chapter \ref{2}, the transformation matrix between two frames that are moving with velocity $\mathbf{v}$ along the $x$-direction with respect to each other is given by \eqref{lorentztx}
\begin{equation}
	\Lambda_x=\begin{pmatrix}
		\gamma_v & -\gamma_v\frac{|\mathbf{v}|}{c} & 0 & 0\\
		-\gamma_v\frac{|\mathbf{v}|}{c} & \gamma_v & 0 & 0\\
		0 & 0 & 1 & 0\\
		0 & 0& 0&1
	\end{pmatrix}
\end{equation}

\begin{figure}[H]
	\centering
	\includegraphics[width=.5\linewidth]{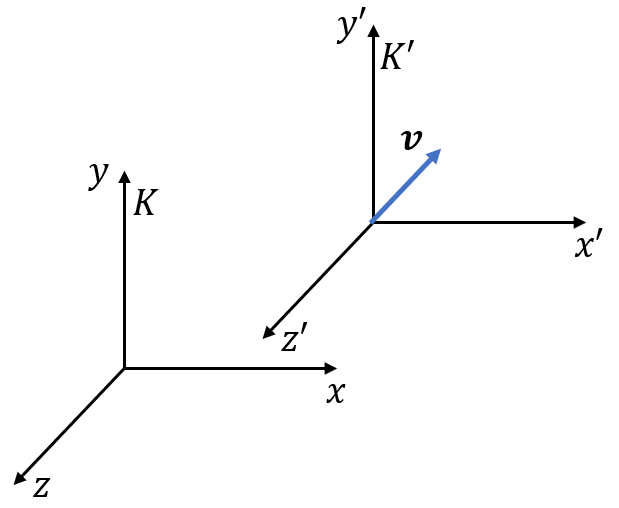}
	\caption{Frames $K$ and $K'$. Frame $K'$ is moving with velocity $\mathbf{v}$ relative to $K$.}
	\label{diffarb}
\end{figure}

Similar expressions exist if the movement is along $y$- or $z$-direction (note that there is nothing special about the $x$-axis so we could have relabeled it to work as $y$- or $z$-axis and the transformation would be found easily). If now one of the frames is moving with arbitrary velocity $\mathbf{v}$ given by
\[\mathbf{v} = (v^1,v^2,v^3)\] 
we can similarly find the transformation matrix. First let us observe that above implies 
\begin{align*}
{ct'}={x^0}'= \gamma_v\left(x^0 - \frac{v^1}{c}x^1\right); \ \ {x^1}'=\gamma_v\left(x^1 - \frac{v^1}{c}x^0\right); \ \ {x^2}'={x^2}; \ \ 	
\end{align*}
where we used that $|\mathbf{v}|=v^1$ in the case of a movement in the direction of $x$. For a movement in an arbitrary direction as seen in Figure \ref{diffarb} below
\noindent it is easy to see that we should replace
\begin{equation}
	{x^0}'= \gamma_v\left(x^0 - \frac{v^1}{c}x^1\right) \to \gamma_v\left(x^0 - \frac{\mathbf{v}\cdot \mathbf{x}}{c}\right) 
\end{equation}
because now we can decompose the movement for each direction. The spatial coordinates, in turn, can be decomposed in
\begin{equation}
\mathbf{x} = \mathbf{x_{\parallel}} + \mathbf{x_{\perp}}
\end{equation}
\textit{i.e.}, parallel and perpendicular directions of motion according to the vector $\mathbf{v}$. So that, being more specific, we can write
\begin{align*}
&\mathbf{x_{\parallel}}=(\mathbf{x}\cdot\hat{\mathbf{v}})\hat{\mathbf{v}}=\frac{(\mathbf{x}\cdot\mathbf{v})}{|\mathbf{v}|^2}\mathbf{v}\\
&\mathbf{x_{\perp}} = \mathbf{x} - \mathbf{x_{\parallel}}
\end{align*}
Lorentz contractions do not affect distances perpendicular to $\mathbf{v}$ so that
\[\mathbf{x'_{\perp}}=\mathbf{x_{\perp}}\implies \mathbf{x'_{\perp}}=\mathbf{x} - \frac{(\mathbf{x}\cdot\mathbf{v})}{|\mathbf{v}|^2}\mathbf{v} \]
relates the perpendicular component of $\mathbf{x'}$ as measured in the frame $K'$ (primed vectors) with respect to quantities measured in $K$ (unprimed).

Now, the parallel component can be found with the similar reasoning we have done for the time-component, yielding
\[\mathbf{x'_{\parallel}}=\gamma_v\left(\mathbf{x_\parallel} - \frac{\mathbf{v}x^0}{c}\right)\] 
this gives the transformation from $K$ to $K'$
\begin{align}
&\mathbf{x'} = \gamma_v\left(\frac{(\mathbf{x}\cdot\mathbf{v})}{|\mathbf{v}|^2}\mathbf{v} - \frac{\mathbf{v}x^0}{c}\right) + \mathbf{x} - \frac{(\mathbf{x}\cdot\mathbf{v})}{|\mathbf{v}|^2}\mathbf{v} \label{xprime}\\
&{x^0}'=\gamma_v\left(x^0 - \frac{\mathbf{v}\cdot \mathbf{x}}{c}\right) 
\end{align}
or, working out the algebra and writing it in a matrix form
\begin{equation}\label{lorentzv1}
	\begin{pmatrix}
		{x^0}' \\
		{x^1}' \\
		{x^2}' \\
		{x^3}' 
	\end{pmatrix}
=\Lambda_v
	\begin{pmatrix}
	{x^0} \\
	{x^1} \\
	{x^2} \\
	{x^3}
\end{pmatrix}
\end{equation}
where
\begin{equation}\label{lorentzv}
\Lambda_v=\begin{pmatrix}
		\gamma_v & -\gamma_v\frac{v^1}{c} & -\gamma_v\frac{v^2}{c} & -\gamma_v\frac{v^3}{c}\\
		-\gamma_v\frac{v^1}{c}& 1 + (\gamma_v - 1)\frac{v^1v^1}{|\mathbf{v}|^2} &  (\gamma_v - 1)\frac{v^1v^2}{|\mathbf{v}|^2} &  (\gamma_v - 1)\frac{v^1v^3}{|\mathbf{v}|^2}\\
		-\gamma_v\frac{v^2}{c}&  (\gamma_v - 1)\frac{v^2v^1}{|\mathbf{v}|^2} &1 + (\gamma_v - 1)\frac{v^2v^2}{|\mathbf{v}|^2} &  (\gamma_v - 1)\frac{v^2v^3}{|\mathbf{v}|^2}\\
		-\gamma_v\frac{v^3}{c}&  (\gamma_v - 1)\frac{v^3v^1}{|\mathbf{v}|^2} &  (\gamma_v - 1)\frac{v^3v^2}{|\mathbf{v}|^2} &1 + (\gamma_v - 1)\frac{v^3v^3}{|\mathbf{v}|^2}
	\end{pmatrix}
\end{equation}
is the matrix for a Lorentz transformation between frames moving with respect to each other with an arbitrary velocity $\mathbf{v}$.

Now, we identify frame $K'$ with $K_1$, that is the rest frame of $1$. Hence, if $x_{2}$ are the space-time coordinates of $2$ given in $K$, while $x'_2$ is the space-time coordinates of $2$ as measured in $K'=K_1$, the rest frame of $1$, we can write
\begin{alignat}{2}
&\frac{\id\mathbf{x_2}}{\id t} = \mathbf{v_2}\ \ \ &&(\mathrm{velocity \ of \ 2 \ measured \ in \ }K)\label{vel2}\\
&\frac{\id \mathbf{x'_2}}{\id t'} = \mathbf{v^{\text{rel}}_{12}}\ \ \ &&(\mathrm{velocity \ of \ 2 \ measured \ in \ }K'=K_1)\label{vel2p}
\end{alignat}
using Equation \eqref{xprime} (after rearranging it slightly) for the rule in which coordinates transforms between frames and observing that $K_1$ is moving with velocity $\mathbf{v_1}$ with respect to $K$ we can write
\begin{align}
	&\mathbf{x'_2} = \mathbf{x_2} - \mathbf{v_1}t + (\gamma_{v_1}-1)\frac{\mathbf{v_1}}{|\mathbf{v_1}|^2}\left((\mathbf{x_2}\cdot\mathbf{v_1}) - |\mathbf{v_1}|^2 t\right)\\
	&t'=\gamma_{v_1}\left(t - \frac{\mathbf{v_1}\cdot \mathbf{x_2}}{c^2}\right) 
\end{align}

Now, looking at relations \eqref{vel2} and \eqref{vel2p}, while using above equations, enables the expression (after some algebra) 
\begin{equation}\label{relvel}
	\mathbf{v^{\text{rel}}_{12}} = \frac{1}{\gamma_{v_1}(1 - \mathbf{v_1}\cdot\mathbf{v_2}/c^2)}\left[ \mathbf{v_2} - \mathbf{v_1} + (\gamma_{v_1}-1)\frac{\mathbf{v_1}}{|\mathbf{v_1}|^2}\left( \mathbf{v_1}\cdot\mathbf{v_2} -|\mathbf{v_1}|^2 \right)\right]
\end{equation}
that is the relativistic relative velocity of particle $2$ with respect to $1$ for arbitrary particle velocities. The modulus of the relative velocity can be very easily calculated following \cite{kremer} and observing that in frame $K_1$ the four-momenta of $1$ and $2$ are given by
\[p'_1 =(m_1c, 0 ) \ , \ \ \ \ p'_2 = (\gamma_{v^{\text{rel}}_{12}}m_2c, \gamma_{v^{\text{rel}}_{12}} m_2\mathbf{v^{\text{rel}}_{12}}) \]
where $m_1(m_2)$ is the mass\footnote{Although we consider here massive particles, the discussion we present is very general, working also for photons and electrons, for example. However, we note that, in this case, we can only work in the rest frame of the electron.} of particle $1 (2)$, so that
\begin{equation}\label{gammarel}
	p'_1 \cdot p'_2 = \gamma_{v^{\text{rel}}_{12}} m_1m_2c^2 = \frac{m_1m_2c^2}{\sqrt{1 - \frac{|\mathbf{v^{\text{rel}}_{12}}|^2}{c^2}}}
\end{equation}
of course that the inner product is Lorentz invariant, so that we can write more generally
\begin{equation}\label{dotrel}
	p_1 \cdot p_2 = \gamma_{v^{\text{rel}}_{12}} m_1m_2c^2 = \frac{m_1m_2c^2}{\sqrt{1 - \frac{|\mathbf{v^{\text{rel}}_{12}}|^2}{c^2}}} \implies |\mathbf{v^{\text{rel}}_{12}}| = c \sqrt{1 - \frac{m^2_1m^2_2c^4}{(p_1 \cdot p_2)^2}}
\end{equation}
this is relation \eqref{dotrel1} which appeared in Chapter \ref{2} and which shows the Lorentz invariance of the modulus of the relative velocity.

On the other hand, if we take $p_1$ and $p_2$ given in frame $K$, we have
\begin{align*}
p_1 = (\gamma_{v_1}m_1c,\gamma_{v_1}m_1\mathbf{v_1}) \ ; \ \ \ p_2 = (\gamma_{v_2}m_2c,\gamma_{v_2}m_2\mathbf{v_2})
\end{align*}
and, thus,
\begin{equation}\label{gammaonetwo}
p_1\cdot p_2 = \gamma_{v_1}\gamma_{v_2} m_1m_2c^2 \left(1 - \frac{\mathbf{v_1}\cdot\mathbf{v_2}}{c^2}\right)
\end{equation}
replacing that in \eqref{dotrel} we have
\begin{align}
	|\mathbf{v^{\text{rel}}_{12}}| &= \frac{c}{1 - (\mathbf{v_1}\cdot\mathbf{v_2})/c^2}\sqrt{\left(1- \frac{(\mathbf{v_1}\cdot\mathbf{v_2})}{c^2}\right)^2  - (\gamma_{v_1}\gamma_{v_2})^{-2}} \nonumber\\
	&= \frac{c}{1 - (\mathbf{v_1}\cdot\mathbf{v_2})/c^2}\sqrt{\left(1 - \frac{(\mathbf{v_1}\cdot\mathbf{v_2})}{c^2}\right)^2 - \left(1 - \frac{|\mathbf{v_1}|^2}{c^2}\right)\left(1 - \frac{|\mathbf{v_2}|^2}{c^2}\right)} \label{dotrel2}
\end{align}

We observe that the square root argument can be rewritten in a more suggestive way
\begin{align*}
\left(1 - \frac{(\mathbf{v_1}\cdot\mathbf{v_2})}{c^2}\right)^2 - \left(1 - \frac{|\mathbf{v_1}|^2}{c^2}\right)\left(1 - \frac{|\mathbf{v_2}|^2}{c^2}\right) = \frac{1}{c^2}\bigg{\{}|\mathbf{v_1}|^2 + |\mathbf{v_2}|^2 - &2\mathbf{v_1}\cdot\mathbf{v_2} - \\
 &\frac{1}{c^2}\left(|\mathbf{v_1}|^2|\mathbf{v_2}|^2 - \left(\mathbf{v_1}\cdot\mathbf{v_2}\right)^2\right)\bigg{\}}
\end{align*}
which simplifies to
\begin{equation}\label{sqrtarg}
\left(1 - \frac{(\mathbf{v_1}\cdot\mathbf{v_2})}{c^2}\right) - \left(1 - \frac{|\mathbf{v_1}|^2}{c^2}\right)\left(1 - \frac{|\mathbf{v_2}|^2}{c^2}\right)=\frac{1}{c^2}\left\{(\mathbf{v_1} - \mathbf{v_2})^2 - \frac{1}{c^2}\left(\mathbf{v_1}\times \mathbf{v_2}\right)^2\right\}
\end{equation}
where we have used the identity
\[(\mathbf{a}\times \mathbf{b})^2 = |\mathbf{a}|^2|\mathbf{b}|^2 - (\mathbf{a}\cdot\mathbf{b})^2\]

Substituting \eqref{sqrtarg} in \eqref{dotrel2} we can write
\begin{equation}\label{dotrel3}
		|\mathbf{v^{\text{rel}}_{12}}| = \frac{1}{1 - (\mathbf{v_1}\cdot\mathbf{v_2})/c^2}v_{M12}
\end{equation}
where we have defined the \textit{M\o ller velocity}
\begin{equation}
	v_{M12}\coloneqq \sqrt{(\mathbf{v_1} - \mathbf{v_2})^2 - \frac{1}{c^2}\left(\mathbf{v_1}\times \mathbf{v_2}\right)^2}
\end{equation}

Since from \eqref{gammarel} and \eqref{gammaonetwo} we have
\begin{equation*}
	 \gamma_{v_1}\gamma_{v_2}\left(1 - \frac{\mathbf{v_1}\cdot\mathbf{v_2}}{c^2}\right) = \gamma_{v^{\text{rel}}_{12}}
\end{equation*}
it is readily seen that, using \eqref{dotrel3}, we have
\begin{equation}
	\gamma_{v_1}\gamma_{v_2}\frac{v_{M12}}{|\mathbf{v^{\text{rel}}_{12}}|} = \gamma_{v^{\text{rel}}_{12}} \implies v_{M12} = \frac{\gamma_{v^{\text{rel}}_{12}}}{\gamma_{v_1}\gamma_{v_2}}|\mathbf{v^{\text{rel}}_{12}}|
\end{equation}
which demonstrates Equation \eqref{gammarelat} in Chapter \ref{2}.

A final relation can be found if we observe that \eqref{gammaonetwo} implies 
\[\left(1 - \frac{\mathbf{v_1}\cdot\mathbf{v_2}}{c^2}\right) = \frac{p_1\cdot p_2}{p^0_1 p^0_2}\]
which, in turn, gives
\begin{equation}\label{rel-mol-4-ap}
	v_{M12} = |\mathbf{v^{\text{rel}}_{12}}|\frac{p_1\cdot p_2}{p^0_1 p^0_2}
\end{equation}
after using \eqref{dotrel3}. This is nothing more than Equation \eqref{rel-mol-4} we use in Chapter \ref{2}.

\chapter{Kompaneets' integrals}\label{b}

This Appendix is the integral reproduction of Appendix C of our work \cite{paper}.
\section{Integral $I_1(x)$}
We need to compute 
\begin{eqnarray}
	I_1(x) = c\int \id^3\mathbf{p} \ \id\Omega\left(1 -\frac{ \mathbf{v}}{c}\cdot\mathbf{\hat{n}}\right)\frac{\id \sigma}{\id \Omega}(\mathbf{p},\mathbf{\hat{n}}, \Omega)  f_\text{Eq}(|\mathbf{p}|) \Delta
\end{eqnarray}

In what follows we omit the dependencies on the variables for simplicity. The leading order is the second on the electron momenta and the expansion yields
\begin{align}
	\frac{16 \pi}{3 \sigma_T}\left(1 -\frac{ \mathbf{p}}{\gamma m_ec}\cdot\mathbf{\hat{n}}\right)\frac{\id \sigma}{\id \Omega}\Delta &=\frac{x\mathbf{p}\cdot (\mathbf{\hat{n}'}- \mathbf{\hat{n}})}{m_ec}(1+\cos^2\theta) -\frac{x^2k_BT}{m_ec^2}(1-\cos\theta)(1 + \cos^2\theta)\nonumber \\ \nonumber
	+\frac{x}{(m_ec)^2}&\bigg\{(1+2\cos\theta - \cos^2\theta)(\mathbf{p}\cdot \mathbf{\hat{n}})^2 + (3 - 2\cos\theta + 5\cos^2\theta)(\mathbf{p}\cdot \mathbf{\hat{n}'})^2\;\\
	&- 4(\mathbf{p}\cdot \mathbf{\hat{n}})(\mathbf{p}\cdot \mathbf{\hat{n}'})(1+\cos^2\theta)\bigg\} \label{delta1}
\end{align}
where we used that $\mathbf{p}= \gamma m_e\,\mathbf{v}$.

We first compute the integral over $\mathbf{p}$ in Cartesian coordinates, and then over the solid angle. Since the distribution is isotropic, the first parcel yields a zero contribution, \textit{i.e.},
\begin{equation}
	\label{linearint}
	\int \id^3\mathbf{p}\, \mathbf{p}\cdot (\mathbf{\hat{n}'}- \mathbf{\hat{n}})f_\text{Eq}(|\mathbf{p}|) =0
\end{equation}
the integral over the momentum in the second parcel is readily done, yielding 
\begin{equation*}
	\int \id^3\mathbf{p}\, f_\text{Eq}(|\mathbf{p}|) = n_e
\end{equation*}
where we used \eqref{maxwell}.  Observe moreover
\begin{align*}
	&(\mathbf{p}\cdot \mathbf{\hat{n}})^2 = p^2_xn^2_x + p^2_yn^2_y +  p^2_zn^2_z + \ \mathrm{cross \ terms \ in \ coordinates \ \ (similarly \ for \ } \mathbf{\hat{n}'})\\
	&(\mathbf{p}\cdot \mathbf{\hat{n}})(\mathbf{p}\cdot \mathbf{\hat{n}'}) = p^2_xn_xn'_x + p^2_yn_yn'_y +  p^2_zn_zn'_z + \ \mathrm{similar \ to \ above}
\end{align*}

Cross terms in the coordinates yield zero contribution for the same reason as \eqref{linearint}. Squared terms give a similar contribution, being
\begin{align}
	& \int \id^3\mathbf{p}\,(\mathbf{p}\cdot \mathbf{\hat{n}})^2 f_\text{Eq}(|\mathbf{p}|)= I \times {\mathbf{\hat{n}}}^2 = I\nonumber\\
	& \int \id^3\mathbf{p}\,(\mathbf{p}\cdot \mathbf{\hat{n}'})^2 f_\text{Eq}(|\mathbf{p}|)= I \times {\mathbf{\hat{n'}}}^2 = I \nonumber\\
	& \int \id^3\mathbf{p}\,(\mathbf{p}\cdot \mathbf{\hat{n}})(\mathbf{p}\cdot \mathbf{\hat{n}'}) f_\text{Eq}(|\mathbf{p}|)= I \times  \mathbf{\hat{n}}\cdot\mathbf{\hat{n}'} = I \cos \theta\nonumber\\
	&\mathrm{with} \ \ I = \int \id^3 \mathbf{p} \, p^2_x \,f_\text{Eq}(|\mathbf{p}|) = n_e m_e \, k_BT \label{xint}
\end{align}

Using all that in $I_1(x)$ gives
\begin{equation*}
	I_1(x)= c\frac{3\sigma_T}{16 \pi}\frac{n_e k_B\textbf{}T}{m_ec^2}\left(-x^2 + 4x\right) \left\{2\pi \int^1_{-1}\id \cos\theta (1-\cos\theta)(1+\cos^2\theta)\right\}
\end{equation*}
and since $ \int^1_{-1}dy (1-y)(1+y^2) = 8/3$,
we find as desired
\begin{equation}
	I_1(x)= \frac{n_e\sigma_T c\,k_BT}{m_ec^2}\;x(4-x)
\end{equation}

\section{Integral $I_2(x)$}
We need to compute 
\begin{eqnarray}
	I_2(x) = c\int \id^3\mathbf{p} \id\Omega\left(1 -\frac{ \mathbf{v}}{c}\cdot\mathbf{\hat{n}}\right)\frac{\id \sigma}{\id \Omega}(\mathbf{p},\mathbf{\hat{n}}, \Omega)  f_\text{Eq}(|\mathbf{p}|) \Delta^2
\end{eqnarray}

The expansion up to second order in the electron momentum yields
\begin{align}
	\frac{16 \pi}{3 \sigma_T}\left(1 -\frac{ \mathbf{p}}{\gamma m_ec}\cdot\mathbf{\hat{n}}\right)\frac{\id \sigma}{\id \Omega}\Delta^2 =\left(\frac{x}{m_ec}\right)^2(\mathbf{p}\cdot (\mathbf{\hat{n}'} - \mathbf{\hat{n}}))^2(1+\cos^2\theta) \label{delta2}
\end{align}	
using the exact same strategy as before, we have 
\begin{align*}
	& \int \id^3\mathbf{p}\,(\mathbf{p}\cdot (\mathbf{\hat{n}'} - \mathbf{\hat{n}}))^2 f_\text{Eq}(|\mathbf{p}|)= I \times (\mathbf{\hat{n}'} - \mathbf{\hat{n}})^2 = 2I(1-\cos\theta)\\
	&\mathrm{with} \ \ I = \int \id^3\mathbf{p} \, p_x^2 \, f_\text{Eq}(|\mathbf{p}|) = n_e m_e\,k_BT
\end{align*}

This yields for $I_2(x)$
\begin{align}
	I_2(x) &=c\frac{3\sigma_T}{16 \pi}\frac{n_ek_BT}{m_ec^2}2x^2 \left\{2\pi \int^1_{-1}d\cos\theta (1-\cos\theta)(1+\cos^2\theta)\right\}\nonumber\\
	&= \frac{n_e\sigma_T c\,k_BT}{m_ec^2}2x^2
\end{align}
as desired.

\section{Nonequilibrium case}
From now on, throughout this appendix only, we denote $|\mathbf{p}|=p$ for simplicity. In the same spirit as before we define
\[
x \coloneqq \frac{\hbar \omega}{\epsilon},\quad \Delta \coloneqq \frac{\hbar (\omega'-\omega)}{\epsilon}
\]
where $\epsilon$ is the characteristic energy of the electron bath, $\epsilon \coloneqq \frac{\langle p^2 \rangle }{3 m_e}$. We do the expansion of the photon distribution, which gives the same as \eqref{pdist1}. The expansion in the electron distribution, however, gives to the leading order
\begin{align}
	\label{distrelect}
	&f(p') = f(p) - \frac{m_e\epsilon f'(p)}{p}\Delta + \left(-\frac{(m_e\epsilon)^2 f'(p)}{p^3} + \frac{(m_e\epsilon)^2 f''(p)}{p^2}\right)\frac{\Delta^2}{2} \\
	&\mathrm{with} \ f'(p)=\frac{\partial f}{\partial p} \ \mathrm{and \ so \ on} \nonumber
\end{align}
where we use the first assumption, that the electron distribution is isotropic, $f(\mathbf{p}) = f(p)$,
with normalization
\begin{eqnarray}
	\int \id^3 \mathbf{p}\,f(t,p)= n_e
\end{eqnarray}

For the equilibrium case we have, of course, $\epsilon = k_BT$ and $f(t,p)=f_\text{Eq}(p)$ as in \eqref{maxwell}. One checks that in this case
\begin{align*}
	&-\frac{m_e\epsilon f_\text{Eq}'(p)}{p}=f_\text{Eq}(p)\\
	&\frac{(m_e\epsilon)^2f_\text{Eq}''(p)}{p^2} - \frac{(m_e\epsilon)^2f_\text{Eq}'(p)}{p^3}=f_\text{Eq}(p)
\end{align*}

Plugging into \eqref{komp-boltz} reads to the leading order
\begin{align}
	\partial_t n = \, & \partial_xnI_1\left(\Delta, f\right) + \frac{\partial_{xx} n}{2}I_2\left(\Delta^2, f\right) + n(1+n)I_3\left(\Delta, \frac{f'}{p}\right)\nonumber\\
	\label{gekomp}
	&+ \partial_x n(1+n)I_4\left(\Delta^2, \frac{f'}{p}\right) + \frac{n(1+n)}{2}\left(I_5\left(\Delta^2, \frac{f''}{p^2}\right) + I_6\left(\Delta^2, \frac{f'}{p^3}\right)\right)
\end{align}
where
\begin{align*}
	&I_1\left(\Delta, f\right)= c\int \id^3\mathbf p\,\id\Omega\left(1 -\frac{ \mathbf{v}}{c}\cdot\mathbf{\hat{n}}\right) \frac{\id \sigma}{\id \Omega}\Delta f \\
	&I_2\left(\Delta^2, f\right) = c\int \id^3\mathbf p\,\id\Omega\left(1 -\frac{ \mathbf{v}}{c}\cdot\mathbf{\hat{n}}\right) \frac{\id \sigma}{\id \Omega}\Delta^2 f\\
	&I_3\left(\Delta, \frac{f'}{p}\right)=-c(m\epsilon)\int \id^3\mathbf p\,\id\Omega\left(1 -\frac{ \mathbf{v}}{c}\cdot\mathbf{\hat{n}}\right) \frac{\id \sigma}{\id \Omega}\Delta \frac{f'}{p}  \\
	&I_4\left(\Delta^2, \frac{f'}{p}\right)=-c(m_e\epsilon) \int \id^3\mathbf p\,\id\Omega\left(1 -\frac{ \mathbf{v}}{c}\cdot\mathbf{\hat{n}}\right) \frac{\id \sigma}{\id \Omega}\Delta^2 \frac{f'}{p}\\
	&I_5\left(\Delta^2, \frac{f''}{p^2}\right)=c(m_e\epsilon)^2 \int \id^3\mathbf p\,\id\Omega\left(1 -\frac{ \mathbf{v}}{c}\cdot\mathbf{\hat{n}}\right) \frac{\id \sigma}{\id \Omega}\Delta^2 \frac{f''}{p^2}\\
	&I_6\left(\Delta^2, \frac{f'}{p^3}\right)=-c(m_e\epsilon)^2\int \id^3\mathbf p\,\id\Omega\left(1 -\frac{ \mathbf{v}}{c}\cdot\mathbf{\hat{n}}\right) \frac{\id \sigma}{\id \Omega}\Delta^2 \frac{f'}{p^3}
\end{align*}

First and second integral are computed in the exact same way as before and give
\begin{align}
	\label{i1}
	&I_1\left(\Delta, f\right)= \frac{n_e\sigma_T c}{(m_ec)^2}\frac{\langle p^2 \rangle }{3}x(4-x)\\
	\label{i2}
	&I_2\left(\Delta^2, f\right) = \frac{n_e\sigma_T c}{(m_ec)^2}\frac{\langle p^2\rangle}{3}2x^2
\end{align}

The other integrals are calculated integrating by parts and using the second assumption $\lim_{p\to \infty}p^3f(p)=0$.

To compute $I_4$ we move to spherical coordinates $\id^3\mathbf{p} = p^2\id p\, \id \Omega_p $.  Using \eqref{delta2} the integral over the electron momentum becomes
\begin{align*}
	\int (\mathbf{p}\cdot (\mathbf{\hat{n}'} - \mathbf{\hat{n}}))^2\frac{f'}{p}\id^3 \mathbf{p}& = \int \id\Omega_p |\mathbf{\hat{n}'}-\mathbf{\hat{n}}|^2\cos^2\zeta\int_0^\infty \frac{p^2f'}{p}p^2\id p\\
	&=\int \id\Omega_p|\mathbf{\hat{n}'}-\mathbf{\hat{n}}|^2\cos^2\zeta \left\{p^3f\bigg{|}^\infty_0 -3\int_0^\infty p^2f\id p\right\} \\
	&=-3\int \int_0^\infty  \id\Omega_p \id p\, |\mathbf{\hat{n}'}-\mathbf{\hat{n}}|^2\cos^2\zeta\, p^2f\\
	&=-3\int \id^3 \, \mathbf{p}\,\, (\mathbf{\hat{p}}\cdot (\mathbf{\hat{n}'} - \mathbf{\hat{n}}))^2f
\end{align*}
where we used \eqref{limit},  $\zeta$ for the angle between $(\mathbf{\hat{n}}-\mathbf{\hat{n}'})$ and $\mathbf{\hat{p}}=\mathbf{p}/p$. The last integral is computed in the very same fashion as we did for the equilibrium case.  We find
\begin{equation*}
	\int \id^3\, \mathbf{p}\,\, (\mathbf{\hat{p}}\cdot (\mathbf{\hat{n}'} - \mathbf{\hat{n}}))^2f = 2n_e(1-\cos\theta)\frac{1}{3}
\end{equation*}
which after integrating the solid angle gives $I_4$,
\begin{eqnarray}
	\label{i4'}
	I_4\left(\Delta^2, \frac{f'}{p}\right) =  c(m_e\epsilon)\left(\frac{x}{m_ec}\right)^2\sigma_T 2 n_e
\end{eqnarray}

Looking at expressions \eqref{i1}, \eqref{i2} and \eqref{i4'} motivates introducing 
\begin{align*}
	&\epsilon = k_BT_\text{eff} \\
	&\langle p^2\rangle = 3m_ek_BT_\text{eff}
\end{align*}
observe that these definitions are compatible with the equilibrium case. We get now
\begin{align}
	&I_1\left(\Delta, f\right)=  \frac{n_e\sigma_T c\,k_BT_\text{eff}}{m_ec^2}x(4-x)\\
	&I_2\left(\Delta^2, f\right) = \frac{n_e\sigma_T c\, k_B T_\text{eff}}{m_ec^2}2x^2\\
	&I_4\left(\Delta^2, \frac{f'}{p}\right) =  \frac{n_e\sigma_T c \,k_B T_\text{eff}}{m_ec^2}2x^2
\end{align}
the other integrals are quite similar, to yield
\begin{align}
	&I_3\left(\Delta, \frac{f'}{p}\right)=\frac{n_e\sigma_T c\,k_BT_\text{eff}}{m_ec^2}x(4-x)\\
	&I_5\left(\Delta^2, \frac{f''}{p^2}\right) + I_6\left(\Delta^2, \frac{f'}{p^3}\right)= \frac{n_e\sigma_T c\, k_B T_\text{eff}}{m_ec^2}2x^2
\end{align}

We substitute the values back in \eqref{gekomp} to find
\begin{equation}
	\omega^2\frac{\partial n}{\partial t}(t,\omega)=    \frac{n_e\sigma_T c}{m_ec^2}\frac{\partial }{\partial \omega}\omega^4\left\{k_BT_\text{eff} \frac{\partial n}{\partial \omega}(t,\omega) + \hbar\left[1+n(t,\omega)\right]n(t,\omega)\right\}
\end{equation}
which is the Kompaneets equation but with an effective (kinetic) temperature
\begin{equation*}
	T_\text{eff}\coloneqq \frac{\langle p^2\rangle }{3m_ek_B}.
\end{equation*}

\chapter{Continuum expansion of the transition rates}\label{c}

This Appendix will follow closely the derivation proposed by \cite{brown}. We shall also follow convention of Section \ref{covdiff} of using natural units $\hbar=c=k_B=1$ in order to simplify notation. Let us begin by looking at the transition rates appearing in the Boltzmann equation as we have defined in Chapter \ref{4}
\begin{align}
	&W(k'\to k)=\frac{1}{4(2\pi)^2} \int\frac{\id \mathbf{p}}{2E}\frac{\id \mathbf{p'}}{2E'}M^{\text{KN}}(p,k'\to p', k) \delta^{(4)}(p + k' - p' - k)f_{Eq}(\mathbf{p})\label{nonisrate1}\\
	&W(k\to k')= \frac{1}{4(2\pi)^2}\int\frac{\id \mathbf{p}}{2E}\frac{\id \mathbf{p'}}{2E'}M^{\text{KN}}(p,k\to p', k') \delta^{(4)}(p + k - p' - k') f_{Eq}(\mathbf{p})\label{nonisrate2}
\end{align}
where we have already relabel the electron momenta in first equation while using dynamical reversibility
\[M^{\text{KN}}(p ',k\to p, k')=M^{\text{KN}}(p,k'\to p', k)\]

Recall that the isotropic transition rates are defined as 
\begin{align}
	&\overline{W}(\omega '\to \omega)=\frac{1}{\omega\omega'} \int\id \Omega\,  W(k'\to k)\label{in41}\\
	&\overline{W}(\omega \to \omega')= \frac{1}{\omega\omega'}\int\id \Omega\, W(k\to k')\label{out41}
\end{align}
with the transition amplitude given by \eqref{Mmand222}.

Let us then use the same trick we have been using many times, that is \eqref{3to4delta}, to rewrite the primed electron momentum measure as
\[\frac{\id \mathbf{p'}}{2E'} = \int_{{p^0}'} \id^4 p'\delta({p'}^2 - m_e^2)\]
now we integrate the outgoing electron momenta $p'$, using the four-delta to write
\begin{align*}
	&p' = p+k'-k \ \mathrm{(conservation \ imposed \ by \  \delta^{(4)} )}\\
	&{p'}^2 - m_e^2 = 2p(k'-k) - 2kk'
\end{align*}
yielding for \eqref{in41} 
\begin{align}
	\overline{W}(\omega '\to \omega)=\frac{1}{4(2\pi)^2\omega\omega'} \int\id \Omega\, \int\frac{\id \mathbf{p}}{2E}f_{Eq}(\mathbf{p})M^{\text{KN}}(p,k'\to p', k)\delta(2p(k'-k) - 2kk') \label{in42}
\end{align}
similarly, for \eqref{out41} we have
\begin{equation}
\overline{W}(\omega\to \omega')=\frac{1}{4(2\pi)^2\omega\omega'} \int\id \Omega\, \int\frac{\id \mathbf{p}}{2E}f_{Eq}(\mathbf{p})M^{\text{KN}}(p,k\to p', k')\delta(2p(k-k') - 2k'k) \label{out42}
\end{equation}

The goal now, as in Appendix \ref{b}, is to carry an expansion up to second order in the electron momenta $|\mathbf{p}|$, that is the diffusion approximation to the transition rates \eqref{in42} and \eqref{out42}. In fact, it will be more convenient this time to carry the expansion in terms of the electron velocity $\mathbf{v} = \mathbf{p}/E$ instead of momentum. We point out that, since $\mathbf{p}$ and $\mathbf{v}$ are of same order, there is no harm in doing that. From now on, we shall require the usual assumptions that are needed for the Kompaneets equation (check Chapter \ref{4}).

We first expand the Klein-Nishina transition amplitude up to second order in electron velocity, which yields (see \cite{brown})
\begin{align}
	M^{\text{KN}}(p,k\to p', k') = 12\pi m^2_e\sigma_T\big{\{} (1+\cos^2\theta) -2|\mathbf{v}|&(1-\cos\theta)\cos\theta(\cos\alpha + \cos\alpha')\nonumber\\ 
	&+ |\mathbf{v}|^2(1-\cos\theta)\left(\cos \alpha + \cos\alpha'\right)^2\big{\}}
\end{align}
where  $\alpha,\alpha'$ are the angles of the incoming electron with incoming and outgoing photons, respectively, as we have defined in Chapter \ref{3}. Since this expression is symmetric on the photon labels\footnote{A relabel of the photon momenta, $k\to k'$ would give same scattering angle $\theta$ and interchange $\alpha \leftrightarrow \alpha'$, thus not changing the expression.}, the same expansion holds for $M^{\text{KN}}(p,k'\to p', k)$.

On the other hand, the argument of the one-dimensional Dirac delta can be written as
\begin{align*}
	&p(k'-k)= E\{(\omega'-\omega)- |\mathbf{v}| \ \omega'\cos\alpha' + |\mathbf{v}| \ \omega\cos\alpha\}\\
	&p(k-k')=E\{(\omega-\omega') - |\mathbf{v}| \ \omega\cos\alpha + |\mathbf{v}| \ \omega'\cos\alpha'\}\\
	&kk'= \omega\omega' - \omega\omega'\cos\theta
\end{align*}

Hence
\begin{align*}
	&\delta(2p(k'-k)-2kk')= \frac{1}{2E}\delta\left((\omega' - \omega) - \frac{\omega\omega'}{E}(1-\cos\theta) - |\mathbf{v}|(\omega'\cos\alpha' - \omega\cos\alpha)\right) \ \ \ (*)\\
	&\delta(2p(k-k')-2k'k)= \frac{1}{2E}\delta\left((\omega - \omega') - \frac{\omega\omega'}{E}(1-\cos\theta) - |\mathbf{v}|(\omega\cos\alpha - \omega'\cos\alpha')\right) \ \ \ (**)
\end{align*}
where the delta function identity
\[\delta(ax) = \frac{1}{|a|}\delta(x)\]
has been used.

The energy shift, $\omega-\omega'$, is a small quantity, while $\omega$ being of the order as the electron energy makes $\omega'$ also of same order as the electron energy, that is $O(|\mathbf{v}|^2)$. Thus, second parcel in the expressions above is of order $|\mathbf{v}|^4$ while last parcel is of order $|\mathbf{v}|^3$.

Recalling that we divide by $\omega\omega'$, it is sufficient to expand the delta in orders up to $|\mathbf{v}|^6$. This will give all contributions up to $|\mathbf{v}|^2$ in the rates. Thus, performing the (formal) (see \cite{impa}) expansion of the delta function yields up to order $|\mathbf{v}|^6$
\begin{align}
	(*)=\frac{1}{2E}\bigg{\{}& \delta(\omega' - \omega) - |\mathbf{v}| (\omega' \cos\alpha' - \omega\cos\alpha)\delta'(\omega' -\omega) \ +\nonumber \\
	&\frac{|\mathbf{v}|^2}{2}(\omega'\cos\alpha'-\omega\cos\alpha)^2\delta''(\omega'-\omega) - \frac{\omega\omega'}{E}(1-\cos\theta)\delta'(\omega'-\omega)\bigg{\}}\\
	(**)= \frac{1}{2E}\bigg{\{}& \delta(\omega - \omega') - |\mathbf{v}|(\omega \cos\alpha - \omega'\cos\alpha')\delta'(\omega -\omega') \ + \nonumber\\
	&\frac{|\mathbf{v}|^2}{2}(\omega\cos\alpha-\omega'\cos\alpha')^2\delta''(\omega-\omega')- \frac{\omega\omega'}{E}(1-\cos\theta)\delta'(\omega-\omega')\bigg{\}}
\end{align}
where the primes indicate differentiation with respect to $\omega' - \omega$.

It is possible to simplify above expressions if we substitute the dependence on the angles $\alpha$ and $\alpha'$ by their average under the electron distribution, observing that this distribution is isotropic. In fact, this is equivalent as performing the integrals as we have done previously in Appendix \ref{b}, that approach is more direct, but now, since we have many terms, it will be more convenient to proceed by observing that integrals which involve $\cos\alpha$ (and similarly for $\alpha'$) are linear in the electron momentum and, thus, according to the discussion in Appendix \ref{b} this will always yield zero contribution, \textit{i.e.}
\[\int\frac{\id \mathbf{p}}{2E}f_{Eq}(\mathbf{p})\cos\alpha =\int\frac{\id \mathbf{p}}{2E}f_{Eq}(\mathbf{p})\mathbf{\hat{p}}\cdot\mathbf{\hat{n}} = 0 \] 

Similarly, for terms depending on $\cos^2\alpha$ we have
\[\cos^2\alpha = \left(\frac{\mathbf{p}}{|\mathbf{p}|}\cdot\mathbf{\hat{n}}\right)^2 = \frac{1}{|\mathbf{p}|^2}\bigg{\{} p_x^2n^2_x + p_y^2n^2_y + p_z^2n^2_z + \ \mathrm{cross \ terms}\bigg{\}}\]
cross terms will not contribute for the same reason as before and we can write
\[\int\frac{\id \mathbf{p}}{2E}f_{Eq}(\mathbf{p})\cos^2\alpha = I_{x}n^2_x + I_{y}n^2_y + I_zn^2_z\]
with
\[I_{x}=\int\frac{\id \mathbf{p}}{2E}f_{Eq}(\mathbf{p}) \frac{p^2_x}{|\mathbf{p}|^2}\]
and similarly for $y,z$. Note that this is just one way of representing a similar integral to that appearing in \eqref{xint}. On the other hand, observe that
\[I_x + I_y + I_z =\int\frac{\id \mathbf{p}}{2E}f_{Eq}(\mathbf{p}) \eqqcolon I_f \]
and, since the distribution is isotropic,
\[I_x=I_y=I_z \implies I_x = \frac{1}{3}I_f\]
thus, since $\mathbf{\hat{n}}$ is unitary, we can finally write
\[\int\frac{\id \mathbf{p}}{2E}f_{Eq}(\mathbf{p})\cos^2\alpha = \int\frac{\id \mathbf{p}}{2E}f_{Eq}(\mathbf{p})\frac{1}{3}\]
of course that the exact same result holds for $\cos^2\alpha'$.

Analogously, for $\cos\alpha\cos\alpha'$ we have
\[\int\frac{\id \mathbf{p}}{2E}f_{Eq}(\mathbf{p})\cos\alpha\cos\alpha' = I_{x}n_xn'_x + I_{y}n_yn '_y + I_zn_zn'_z = \int\frac{\id \mathbf{p}}{2E}f_{Eq}(\mathbf{p})\frac{\cos\theta}{3} \]

Now, if we would make the change in the integrand
\[\int\frac{\id \mathbf{p}}{2E}f_{Eq}(\mathbf{p})\bullet \to \int\frac{\id \mathbf{p}}{2E}f_{Eq}(\mathbf{p})g(\mathbf{p})\bullet\]
the results are the same for any function $g$ which is isotropic ($g(\mathbf{p})=g(|\mathbf{p}|)$)\footnote{The identification of such functions $g$ will depend on which term of the expansion we are looking.}, \textit{i.e.}
\begin{align*}
&\int\frac{\id \mathbf{p}}{2E}f_{Eq}(\mathbf{p})g(|\mathbf{p}|)\cos\alpha = 0\\
&\int\frac{\id \mathbf{p}}{2E}f_{Eq}(\mathbf{p})g(|\mathbf{p}|)\cos^2\alpha^{(')} = \int\frac{\id \mathbf{p}}{2E}f_{Eq}(\mathbf{p})g(|\mathbf{p}|)\frac{1}{3}\\
&\int\frac{\id \mathbf{p}}{2E}f_{Eq}(\mathbf{p})\cos\alpha\cos\alpha' = \int\frac{\id \mathbf{p}}{2E}f_{Eq}(\mathbf{p})g(|\mathbf{p}|)\frac{\cos\theta}{3}	
\end{align*}
 all this only means that, if isotropy holds, we can substitute the cosines by their average inside the sign of the integral. If we denote their average under the electron distribution as $\langle \bullet \rangle_p$ we can schematically write
\begin{align*}
	&\langle \cos\alpha\rangle_p = \langle \cos\alpha'\rangle_p = 0\\
	&\langle \cos^2\alpha \rangle_p = \langle \cos^2\alpha'\rangle_p = \frac{1}{3}\\
	&\langle \cos\alpha\cos\alpha'\rangle_p = \frac{1}{3}\cos\theta
\end{align*}

Hence, up to order $|\mathbf{v}|^6$ 
\begin{align}
	\langle &M^{\text{KN}}(p,k'\to p', k)\delta(2p(k'-k) - 2kk')\rangle_p=\nonumber \\
	&\frac{12\pi m^2_e\sigma_T}{2E}\bigg{\{}\left[1+\cos^2\theta + \frac{2|\mathbf{v}|^2}{3}\left(1-\cos^2\theta\right)\right] \delta(\omega' - \omega) +  \frac{2|\mathbf{v}|^2}{3}\cos\theta\left(1-\cos^2\theta\right)\left(\omega'-\omega\right)\delta'(\omega'-\omega) \nonumber \\
	&\ \ \ \ \ +\frac{|\mathbf{v}|^2}{6}(\omega^2+\omega'^2-2\omega\omega'\cos\theta)(1+\cos^2\theta)\delta''(\omega'-\omega) - \frac{\omega\omega'}{E}(1 -\cos\theta)(1+\cos^2\theta)\delta'(\omega'-\omega)\bigg{\}}  \\
	\langle &M^{\text{KN}}(p,k\to p', k')\delta(2p(k-k') - 2k'k)\rangle_p=\nonumber \\
	&\frac{12\pi m^2_e\sigma_T}{2E}\bigg{\{}\left[1+\cos^2\theta + \frac{2|\mathbf{v}|^2}{3}\left(1-\cos^2\theta\right)\right] \delta(\omega - \omega') +  \frac{2|\mathbf{v}|^2}{3}\cos\theta\left(1-\cos^2\theta\right)\left(\omega-\omega'\right)\delta'(\omega-\omega') \nonumber \\
	&\ \ \ \ \ +\frac{|\mathbf{v}|^2}{6}(\omega^2+\omega'^2-2\omega\omega'\cos\theta)(1+\cos^2\theta)\delta''(\omega-\omega') - \frac{\omega\omega'}{E}(1 -\cos\theta)(1+\cos^2\theta)\delta'(\omega-\omega')\bigg{\}} 
\end{align}

We can still simplify expressions above slightly by observing the formal delta function properties
\begin{align*}
&z\delta'(z)= - \delta(z)\\
&z^2\delta''(z) = 2\delta(z)
\end{align*}
where the primes now denote integration with respect to $z\in \mathbb{R}$. These properties only make sense inside the sign of the integral and it is very straightforward to check them using a smooth test function and integrating by parts. For example, first property follows (suppose $a,b>0$)
\begin{align*}
\int_{-a}^{b} f(z) [z\delta'(z)]\id z&= [zf(z)\delta(z)]\bigg{|}_{-a}^b - \int_{-a}^{b}[f(z) + zf'(z)]\delta(z)]\id z\\
&=\int_{-a}^{b} f(z) [-\delta(z)]\id z
\end{align*}
similarly we can check second property.

Using these two properties in the expansion
\begin{align}
	\langle M^{\text{KN}}&(p,k'\to p', k)\delta(2p(k'-k) - 2kk')\rangle_p=\nonumber \\
	&\frac{12\pi m^2_e\sigma_T}{2E}\bigg{\{}\left[1+\cos^2\theta + \frac{2|\mathbf{v}|^2}{3}\left(1-\cos\theta\right)\left(1-\cos^2\theta\right) + \frac{|\mathbf{v}|^2}{3}\right] \delta(\omega' - \omega) \nonumber \\
	&\ \ \ \ \ \ \ \ +\frac{|\mathbf{v}|^2\omega\omega'}{3}(1-\cos\theta)(1+\cos^2\theta)\delta''(\omega'-\omega) - \frac{\omega\omega'}{E}(1 -\cos\theta)(1+\cos^2\theta)\delta'(\omega'-\omega)\bigg{\}} \label{pavin} \\
	\langle M^{\text{KN}}&(p,k\to p', k')\delta(2p(k-k') - 2k'k)\rangle_p=\nonumber \\
	&\frac{12\pi m^2_e\sigma_T}{2E}\bigg{\{}\left[1+\cos^2\theta + \frac{2|\mathbf{v}|^2}{3}\left(1-\cos\theta\right)\left(1-\cos^2\theta\right) + \frac{|\mathbf{v}|^2}{3}\right] \delta(\omega - \omega') \nonumber \\
	&\ \ \ \  \ \ \ \ +\frac{|\mathbf{v}|^2\omega\omega'}{3}(1-\cos\theta)(1+\cos^2\theta)\delta''(\omega-\omega') - \frac{\omega\omega'}{E}(1 -\cos\theta)(1+\cos^2\theta)\delta'(\omega-\omega')\bigg{\}} \label{pavout}
\end{align}

Let us recall now that the rates, up to second order in the electron momentum, are given by
\begin{align}
	& \overline{W}(\omega '\to \omega)=\frac{1}{4(2\pi)^2\omega\omega'} \int\id \Omega\, \int\frac{\id \mathbf{p}}{2E}f_{Eq}(\mathbf{p}) \langle M^{\text{KN}}(p,k'\to p', k)\delta(2p(k'-k) - 2kk')\rangle_p \label{in43}\\
	&\overline{W}(\omega\to \omega')=\frac{1}{4(2\pi)^2\omega\omega'} \int\id \Omega\, \int\frac{\id \mathbf{p}}{2E}f_{Eq}(\mathbf{p}) \langle M^{\text{KN}}(p,k\to p', k')\delta(2p(k-k') - 2k'k)\rangle_p \label{out43}
\end{align}
there are three types of integrals over the solid that must be done, these are
\begin{align*}
&\int \id \Omega (1+\cos^2\theta) =\frac{16\pi}{3}\\
&\int \id \Omega (1-\cos\theta)(1-\cos^2\theta)= \frac{8\pi}{3}\\
&\int \id \Omega (1 -\cos\theta)(1+\cos^2\theta)=\frac{16\pi}{3}
\end{align*}

By using these integrals while also simplifying the prefactors in \eqref{in43} and \eqref{out43} results
\begin{align}
	&\overline{W}(\omega '\to \omega)=m_e^2\sigma_T \int\frac{\id \mathbf{p}}{E^2}f_{Eq}(\mathbf{p})\left\{\left(1 + \frac{7}{4}\frac{|\mathbf{v}|^2}{3}\right)\frac{\delta(\omega'-\omega)}{\omega\omega'} - \frac{\delta'(\omega'-\omega)}{E} + \frac{|\mathbf{v}|^2}{3}\delta''(\omega'-\omega) \right\}\\
	& \overline{W}(\omega\to \omega')=m_e^2\sigma_T\int\frac{\id \mathbf{p}}{E^2}f_{Eq}(\mathbf{p})\left\{\left(1 + \frac{7}{4}\frac{|\mathbf{v}|^2}{3}\right)\frac{\delta(\omega-\omega')}{\omega\omega'} - \frac{\delta'(\omega-\omega')}{E} + \frac{|\mathbf{v}|^2}{3}\delta''(\omega-\omega') \right\}
\end{align}

The electron momentum integral can be calculated using the Maxwell-Boltzmann distribution \eqref{maxwell}. However, we must note that, differently than before, these integrals are non-trivial and we must expand the energy up to second order in electron momentum also (for a detailed derivation, valid even for more general isotropic distributions, we invite the reader to \cite{brown}). Here we state the result
\begin{align*}
	&\int\frac{\id \mathbf{p}}{E^3}f_{Eq}(\mathbf{p})=\frac{n_e}{m^3_e}\\
	&\int\frac{\id \mathbf{p}}{E^2}\frac{|\mathbf{v}|^2}{3}f_{Eq}(\mathbf{p})=\frac{n_e}{m^3_e}T\\
	&\int\frac{\id \mathbf{p}}{E^2}f_{Eq}(\mathbf{p})=\frac{n_e}{m^2_e}
\end{align*}
giving
\begin{align}
	& \overline{W}(\omega '\to \omega)=\frac{n_e\sigma_T}{m_e}\left\{\left(m_e + \frac{7}{4}T\right)\frac{\delta(\omega'-\omega)}{\omega\omega'} - \delta'(\omega'-\omega) + T\delta''(\omega'-\omega) \right\} \label{in44}\\
	&\overline{W}(\omega\to \omega')=\frac{n_e\sigma_T}{m_e}\left\{\left(m_e + \frac{7}{4}T\right)\frac{\delta(\omega-\omega')}{\omega\omega'} - \delta'(\omega-\omega') + T\delta''(\omega-\omega') \right\} \label{out44}
\end{align}
these are the rates we have seen in Section \ref{covdiff}.

Before plugging these rates in the Boltzmann equation, we must make the final observation that first parcels in \eqref{in44} and \eqref{out44} will not contribute to the time evolution of the distribution function, since in this case, the \textit{in} term equals the \textit{out} term (in fact, the first term, which is proportional to the delta function itself, is only important to provide the correct ``normalization" of the transition rate). 

Let us then return to the Boltzmann equation \eqref{bol-kompc3}, where we use \eqref{in44} and \eqref{out44} to write
\begin{align}
	\frac{\partial n}{\partial t}(t, \omega) =\frac{n_e\sigma_T}{m_e}\int{\omega'}^2\id\omega'\{&\left( - \delta'(\omega'-\omega) + T\delta''(\omega'-\omega) \right))n(t, \omega')\left(1+n(t, \omega )\right) \nonumber \\
	&-\left( - \delta'(\omega-\omega') + T\delta''(\omega-\omega'))n(t, \omega)\left(1+n(t, \omega')\right) \right)\}
\end{align}
where we already have disregarded the first component of the transition rates.

Let us make the following definitions in order to solve the integral
\begin{alignat*}{2}
	&x\coloneqq\omega' - \omega\\
	&u(x)\coloneqq n(t, x+\omega)\left(1+n(t, \omega)\right); \ \ \ 	&&H(x)\coloneqq u(x)-v(x)\\
	&v(x)\coloneqq n(t, \omega)\left(1+n(t, x+\omega)\right);	&&G(x)\coloneqq u(x)+v(x)
\end{alignat*}

Using the shift $x$ in the rates
\begin{align*}
	&\overline{W}(\omega '\to \omega)=\frac{n_e\sigma_T}{m_e}\left\{\left(m_e + \frac{7}{4}T\right)\frac{\delta(x)}{\omega(\omega+x)} - \delta'(x) + T\delta''(x) \right\}\\
	&\overline{W}(\omega\to \omega')=\frac{n_e\sigma_T}{m_e}\left\{\left(m_e + \frac{7}{4}T\right)\frac{\delta(x)}{\omega(\omega+x)} + \delta'(x) + T\delta''(x) \right\}
\end{align*}
where we have used the properties of the delta function and its derivatives
\begin{align*}
&\delta(-x)=\delta(x)\\
& \delta'(-x)=-\delta'(x)\\
& \delta''(-x)=\delta''(x)
\end{align*}

Similarly, the Boltzmann equation can be easily rewritten in terms of the new definitions, yielding
\begin{equation}\label{bolwnewdef}
	\omega^2\frac{\partial n}{\partial t}(t, \omega)=    \frac{n_e\sigma_T\omega^2}{m_e}\int^\infty_{-\omega}(x+\omega)^2dx\left\{-\delta'(x)G(x) + T\delta''(x)H(x)\right\}
\end{equation}
it is now a matter of straightforward calculation. Integrals give
\begin{align*}
	&I_G = -\int^\infty_{-\omega}\id x\, (x+\omega)^2\delta'(x)G(x)=\int^\infty_{-\omega}\id  x\,\delta(x)\frac{\id}{\id x}\left[(x+\omega)^2G(x)\right] = 2\omega G(0) + \omega^2G'(0)\\
	&I_H= \int^\infty_{-\omega}\id x\,(x+\omega)^2\delta''(x)H(x)=\int^\infty_{-\omega}\id x\, \delta(x)\frac{\id^2}{\id x^2}\left[(x+\omega)^2H(x)\right]=4\omega H'(0) + \omega^2H''(0)
\end{align*}
replacing the defined functions:
\begin{align*}
	&I_G=4\omega\left[1+n(t,\omega)\right]n(t,\omega) + \omega^2\left[1+2n(t,\omega)\right]\frac{\partial n}{\partial \omega}(t,\omega)\\
	&I_H= 4\omega \frac{\partial n}{\partial \omega}(t,\omega) + \omega^2\frac{\partial^2 n}{\partial \omega^2}(t,\omega)
\end{align*}
substituting that back in \eqref{bolwnewdef} while performing standard manipulations leave us with the Kompaneets equation in natural units  
\begin{equation}
	\omega^2\frac{\partial n}{\partial t}(t, \omega)=    \frac{n_e\sigma_T}{m_e}\frac{\partial }{\partial \omega}\omega^4\left\{T \frac{\partial n}{\partial \omega}(t,\omega) + \left[1+n(t,\omega)\right]n(t,\omega)\right\}
\end{equation}

\chapter{Discrete expansion of the transition rates}\label{d}

This Appendix is the integral reproduction of Appendix A of our work \cite{paper}.

The transition rates \eqref{tr}  can be expanded from
\begin{align*}
	&w(x,x\pm\delta)= \left(1+ n \pm\delta n' +\frac{\delta^2}{2}n''\right)\left(B \pm\frac{\delta}{2}B' + \frac{\delta^2}{8}B''\right)\exp\left\{-\frac{\beta}{2}\left(\pm\delta (U'-f) +\frac{\delta^2}{2}(U''-f')\right)\right\}\\
	&w(x\pm\delta, x)= \left(1+n\right)\left(B \pm\frac{\delta}{2}B' + \frac{\delta^2}{8}B''\right)\exp\left\{\frac{\beta}{2}\left(\pm\delta (U'-f) +\frac{\delta^2}{2}(U''-f')\right)\right\}
\end{align*}

An expansion up to second order in $\delta$ yields
\begin{align}
	&w(x,x\pm\delta)= A(x) \pm \frac{\delta}{2} C(x) + \frac{\delta^2}{2} E(x)\\
	&w(x\pm \delta, x)=A(x) \pm \frac{\delta}{2} F(x) + \frac{\delta^2}{2} G(x)
\end{align}
with short-hands
\begin{align*}
	&A(x) = B(1+n)\\
	&C(x) = 2Bn' - (\beta B g - B')(1+n)\\
	&E(x) = \left(\frac{1}{4}(\beta^2B g^2+B'')-\frac{1}{2}(\beta g B' + \beta g' B)\right)(1+n) - (\beta B g - B')n' + Bn'' \\
	&F(x) = (\beta B g + B')(1+n)\\
	&G(x) = \left(\frac{1}{4}(\beta^2Bg^2 + B'')+\frac{1}{2}(\beta g B' +\beta g'B)\right)(1+n)\\
	&g(x)\coloneqq U'(x)-f(x)
\end{align*}

That gives to leading order in the master equation
\begin{equation}
	\partial_tn = \delta^2\left\{(G(x)-E(x))n(x) + F(x) \partial_xn(x) + A (x) \partial_{xx} n(x)\right\}
\end{equation}
substituting the short-hands we get \eqref{mm}.

\newpage

\bibliographystyle{apacite}
\bibliography{thesisrefs}
\newpage
\thispagestyle{empty}
\sffamily
\begin{textblock}{191}(113,-11)
{\color{blueline}\rule{160pt}{5.5pt}}
\end{textblock}
\begin{textblock}{191}(168,-11)
{\color{blueline}\rule{5.5pt}{59pt}}
\end{textblock}
\begin{textblock}{183}(-24,-11)
\textblockcolour{}
\flushright
\fontsize{7}{7.5}\selectfont
\textbf{DEPARTMENT OF PHYSICS AND ASTRONOMY}\\
\textbf{INSTITUTE FOR THEORETICAL PHYSICS}\\
Celestijnenlaan 200D box 2415\\
3001 LEUVEN, BELGI\"{E}\\
tel. +32 16 32 72 32\\
fys.kuleuven.be/itf
\end{textblock}
\begin{textblock}{191}(154,-7)
\textblockcolour{}
\includegraphics*[height=16.5truemm]{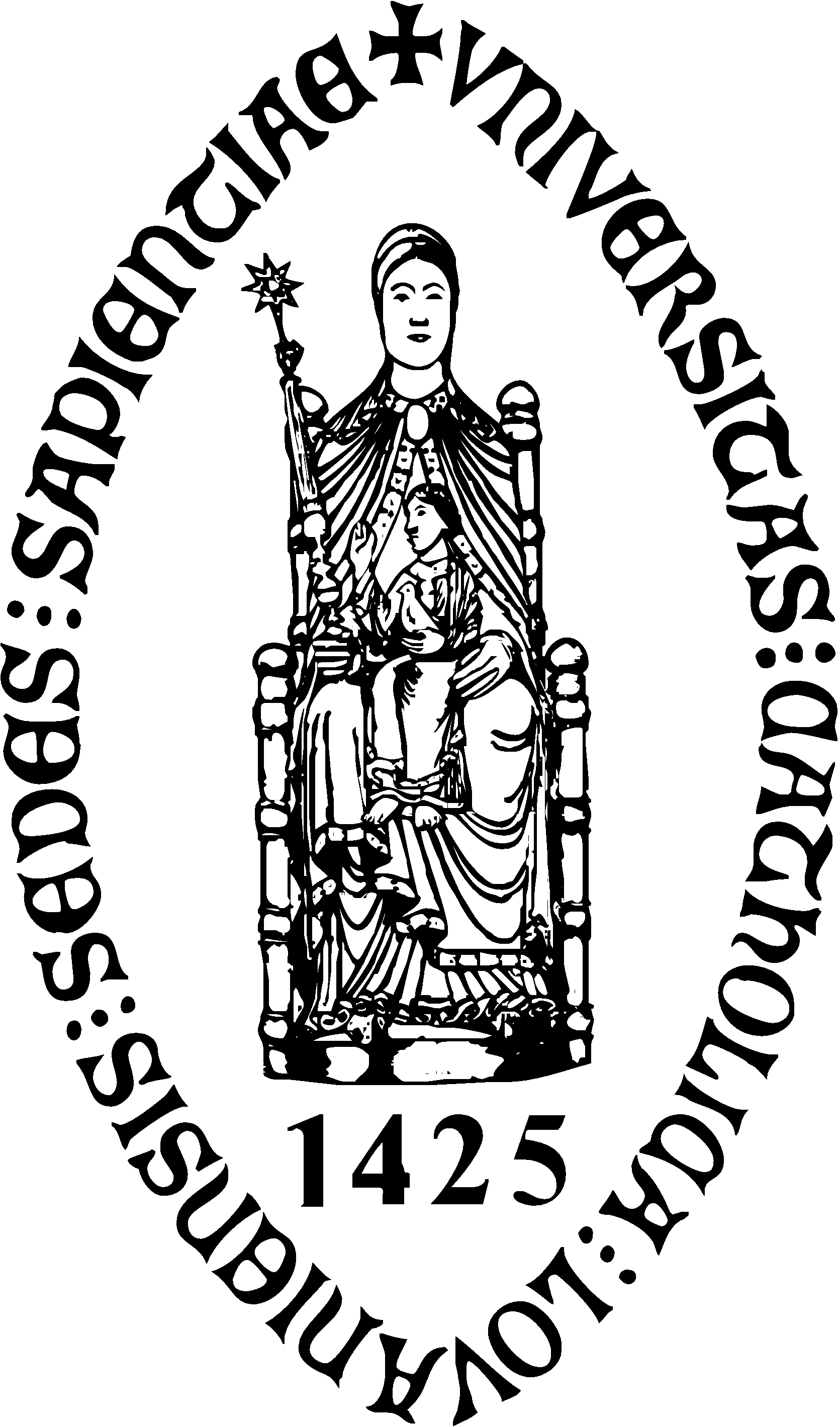}
\end{textblock}
\begin{textblock}{191}(-20,235)
{\color{bluetitle}\rule{544pt}{55pt}}
\end{textblock}
\end{document}